\documentclass[12pt,a4paper]{article}
\usepackage[utf8]{inputenc}
\usepackage{graphicx,amsmath,amssymb,mathtools,tikz}
\usetikzlibrary{cd, calc}

\usepackage[maxnames=10, sorting=none]{biblatex}
\usepackage{hyperref}
\usepackage{array}
\usepackage{float}

\numberwithin{equation}{section}

\newcommand{\VI}{\scalebox{.6}{VI}}
\newcommand{\V}{\scalebox{.6}{V}}
\newcommand{\IV}{\scalebox{.6}{IV}}
\newcommand{\III}{\scalebox{.6}{III}}
\newcommand{\II}{\scalebox{.6}{II}}
\newcommand{\I}{\scalebox{.6}{I}}
\newcommand{\ri}{\mathrm{i}}
\newcommand{\D}{\scriptscriptstyle{D}}

\newcolumntype{M}[1]{>{\centering\arraybackslash}m{#1}}

\title{Bilinear tau forms of quantum Painlev\'e equations
\\
and $\mathbb{C}^2/\mathbb{Z}_2$ blowup relations in SUSY gauge theories}
\author{Giulio Bonelli\thanks{bonelli@sissa.it},\qquad Anton Shchechkin\thanks{shch145@gmail.com},\qquad Alessandro Tanzini\thanks{tanzini@sissa.it}\\
{\small
SISSA, via Bonomea 265, 34136 Trieste, Italy}
\\ {\small INFN, Section of Trieste, 
via Valerio 2, 34127 Trieste, Italy}\\ {\small  IGAP, via Beirut 2, 34151 Trieste, Italy}}

\date{}

\begin{document}

\maketitle

\begin{abstract}
We derive bilinear tau forms of the canonically quantized Painlev\'e equations, thereby relating them to those previously obtained from the $\mathbb{C}^2/\mathbb{Z}_2$ blowup relations for the $\mathcal{N}=2$ supersymmetric gauge theory partition functions on a general $\Omega$-background. 
We fully fix the refined Painlev\'e/gauge theory dictionary by formulating the proper equations for the quantum nonautonomous Painlev\'e Hamiltonians. 
We also describe the quantum Painlev\'e symmetries at the level of tau functions and, as a byproduct of this analysis, obtain several $\mathbb{C}^2/\mathbb{Z}_2$ blowup relations including those in nontrivial holonomy sectors of the gauge theory.
\end{abstract}

\begin{quote}
\tableofcontents
\end{quote}

\setcounter{section}{-1}
\section{Introduction}
\label{sec:intro}

This paper is a companion to \cite{BST25}, where the $\mathbb{C}^2/\mathbb{Z}_2$ blowup equations \cite{BMT1106,BMT1107,BBFLT11} for tau functions in a general $\Omega$-background were supposed to be quantum Painlev\'e equations in bilinear form. In this paper we explicitly relate these blowup equations to 
the canonically quantized Painlev\'e dynamics in the symmetry approach of Hajime Nagoya \cite{N04,N09,N12} (see also \cite{JNS08,NY12}). 
More specifically:
\begin{itemize}
    \item We derive bilinear tau forms of the canonically quantized Painlev\'e equations from the quantum Hamiltonian (Heisenberg) formalism, thereby relating them to the $\mathbb{C}^2/\mathbb{Z}_2$ blowup equations of \cite{BST25}.
    \item We fix $\epsilon$-corrections to the dictionary between  quantum Painlev\'e parameters and SUSY gauge theory masses, thus completing refinement of the Painlev\'e/gauge theory correspondence. In this refined correspondence, solutions of the quantum Painlev\'e equations are expressed in terms of the corresponding SUSY gauge theory partition functions, building on the seminal paper \cite{GIL12}.
    \item We relate the symmetry structures on both sides of the refined Painlev\'e/gauge theory correspondence. This gives us several $\mathbb{C}^2/\mathbb{Z}_2$ blowup relations incuding those in nontrivial holonomy sectors of the gauge theories. In the weak-coupling regime of the gauge theories these relations were expected, via the AGT correspondence \cite{AGT09}, from
    the representation theory of the super Virasoro algebra in the Ramond sector, as in \cite{BS16b}.
\end{itemize}

The canonical quantization of the classical Painlev\'e equations, viewed as (nonautonomous) Hamiltonian systems, encounters the standard coordinate-momentum operator-ordering problem for the Hamiltonians. In Nagoya's approach, the appropriate ordering is fixed by requiring preservation of the extended affine Weyl group  symmetries of the classical Painlev\'e equations (also called B\"acklund transformation group). Taking these nonautonomous quantum Painlev\'e Hamiltonians as functions on the Heisenberg trajectories, we obtain second-order differential equations in time for them. These are quantum analogs of the so-called sigma forms of the Painlev\'e equations \cite{JM81}, \cite{Forrester_book}. We refer to such equations, together with an appropriate commutation relation on the Hamiltonian time derivatives in place of the canonical commutation relation, as the \textit{Hamiltonian form}. Further, for each quantum Hamiltonian we define two tau functions, rather than a single (isomonodromic) tau function in the classical case.
These two tau functions are related by a first-order bilinear equation. Then, from the commutation relation and the second-order equation of the Hamiltonian form, we obtain third- and fourth-order bilinear equations for the tau functions. Altogether, these three equations for the tau functions constitute the \textit{tau form} of a given quantum Painlev\'e equation, whereas in the classical case only the equation of order four remains nontrivial. We also show that these tau forms are equivalent to the original Heisenberg dynamics.

On the other hand, in \cite{BS14} the $\mathbb{C}^2/\mathbb{Z}_2$ blowup relations for the $\mathcal{N}=2$ $D=4$ SUSY $SU(2)$ partition functions with $N_f=4$ fundamental massive hypermultiplets were obtained.
In the case of the self-dual $\Omega$-background, these blowup relations were rewritten
as a fourth-order bilinear differential equation for a tau function given by the Zak transform of these partition functions.
This equation was identified with the tau form of the classical Painlev\'e VI, thereby obtaining a proof of the Painlev\'e/gauge theory correspondence conjecture in this case (formulated in \cite{GIL12}). Following the approach of \cite{BGM17}, in \cite{BST25} we rewrote the $N_f=4$ blowup relations of \cite{BS14} for a general $\Omega$-background as a system of three bilinear equations for tau functions defined as a noncommutative Zak transform of the partition functions. 
Starting from this "deformed" Painlev\'e VI, we then followed the well-known classical Painlev\'e coalescence limits, thereby obtaining analogous "deformed" systems for all Painlev\'e differential equations.
It is precisely these bilinear equations of order $1,3$, and $4$ that we identify in this paper with the tau forms of the corresponding canonically quantized Painlev\'e equations, recognizing in this "deformation" the alluded above canonical quantization with a precise -- simple and natural -- operator ordering prescription. 

The coalescence limits from the Painlev\'e VI equation to the Painlev\'e V and III's equations at the level of the partition functions were realized in \cite{BST25} as successive decouplings of heavy masses. Thus, in addition to the Painlev\'e VI case, we also obtained solutions of the corresponding quantum tau forms for these equations. 
The solutions we obtained are expressed as expansions around the regular singular points of the corresponding Painlev\'e equations ($0,1,\infty$ for Painlev\'e VI and $0$ for Painlev\'e V, III's). However, the main goal of \cite{BST25} was to study the quantum deformation of the asymptotic expansions of the Painlev\'e tau functions near the irregular singularity ($t=\infty$) found in \cite{BLMST16}, which occurs for all the Painlev\'e equations except Painlev\'e VI. These expansions correspond to the strong-coupling regime of the SUSY gauge theories, whereas the regular-type expansions above correspond to the weak-coupling regime. In~\cite{BST25} we presented several leading terms of these strong-coupling expansions, in particular 
for Painlev\'e IV, II, and I, where the corresponding theories are the Argyres--Douglas SCFTs \cite{BLMST16}.

Altogether, the results described above establish the refined  Painlev\'e/gauge theory correspondence. However, compared to the Hamiltonian forms, the tau form of each (quantum) Painlev\'e equation provides an additional integration constant, which in general cases effectively replaces one of the Painlev\'e equation parameters. In the classical case, such integration constants are fixed by requiring that the Hamiltonian form be satisfied. In the quantum case, we proceed in the same way: we derive expansions of the Hamiltonian from those of the corresponding tau functions and substitute them into the Hamiltonian form equation. In this way we obtain $\epsilon$-corrections to the Painlev\'e/gauge theory dictionary. These corrections explicitly match those derived in \cite{BST25} by the holomorphic anomaly approach. 

The $\mathbb{C}^2/\mathbb{Z}_2$ blowup relations (in the weak-coupling regime) were derived in \cite{BS14} via the AGT correspondence \cite{AGT09}, using the representation theory of the $\mathcal{N}=1$ super Virasoro algebra in the Neveu--Schwarz sector. This approach was further developed in \cite{BS16b} in the Ramond sector for the most degenerate case, corresponding to Painlev\'e III$_3$ and $N_f=0$ gauge theory. In the self-dual $\Omega$-background, this yielded the so-called Okamoto-like bilinear equations for the corresponding tau function and its B\"acklund transformation. In the present paper we proceed in the reverse direction: we first obtain such bilinear equations for the quantum Painlev\'e tau functions, then substitute their solutions in the form of the noncommutative Zak transform, which yields the $\mathbb{C}^2/\mathbb{Z}_2$ blowup relations in a nontrivial holonomy sector.
In other words, we pass from the trivial to the nontrivial holonomy sector of the blowup relations via the corresponding quantum Painlev\'e equation. 

Further, we obtained quantum Okamoto-like equations for the tau function and its certain nonautonomous symmetry transformation for each quantum Painlev\'e equation. Together with autonomous symmetry transformations, trivial on the tau functions, this transformation generate the whole symmetry group of the given quantum Painlev\'e equation. The autonomous symmetries are symmetries of the partition functions, thus they explicitly preserve the form of the Zak transform. The nonautonomous transformation acts on the non-commutative Zak transform parameters a shift, which can be retrieved from the Okamoto-like equations for the Zak transform solutions. 
Thus, in the weak- and the strong-coupling regime, we describe the whole symmetry group action on the tau functions on both sides of the refined Painlev\'e/gauge theory correspondence and present several $\mathbb{C}^2/\mathbb{Z}_2$ blowup relations including those in nontrivial holonomy sectors.

\paragraph{Structure of the paper.} In Section~\ref{sec:toy} we introduce the quantum Painlev\'e I equation, present its Hamiltonian and tau forms, and fix a universal definition of the tau functions in terms of the corresponding Hamiltonian. In Section~\ref{sec:QPVI} we extend this construction to the most general quantum Painlev\'e VI equation and also discuss its symmetry group.
In Section~\ref{sec:coalescence} we obtain all the remaining quantum Painlev\'e equations by following the classical coalescence limits; for each equation we describe its symmetry group and derive the corresponding Hamiltonian and tau forms, in parallel with the quantum Painlev\'e VI case.
In Section~\ref{sec:asymptotics} we compute the Hamiltonian expansions from the corresponding tau function expansions and fix the integration constant freedom in both the weak- and in the strong-coupling regimes.
In Section~\ref{sec:tau_symm} we describe the symmetries of the tau functions on both sides of the refined Painlev\'e/gauge theory correspondence and obtain  $\mathbb{C}^2/\mathbb{Z}_2$ blowup relations in in the weak- and strong-coupling regime.

\paragraph{Acknowledgements.}
We would like to thank Mikhail Bershtein, Pavlo Gavrylenko, and Oleg Lisovyy for stimulating discussions.

A preliminary version of the results contained in this paper was presented at the conferences SIDE15 and QTS-13 by A.S., who thanks the participants for their comments and discussions. Some of the results contained in this paper were also presented by G.B. in a talk at IST Lisboa, 
and by A.T. in the mathematical physics seminar of LPTHE Paris, which they thank for their kind hospitality.

The research of G.B. is partly supported by the INFN Iniziativa Specifica ST\&FI and by the PRIN project “Non-perturbative Aspects Of Gauge Theories And
Strings”. The research of A.S. and A.T. is partly supported by the INFN Iniziativa Specifica GAST and INdAM GNFM. This research is partly supported by the MIUR PRIN
Grant 2020KR4KN2 “String Theory as a bridge between Gauge Theories and Quantum
Gravity”. All the authors acknowledge funding from the EU project Caligola (HORIZON-
MSCA-2021-SE-01).

\section{Toy example: Quantum Painlev\'e I}
\label{sec:toy}
\subsection{Hamiltonian system}
\label{ssec:QPI_Ham}
\paragraph{Canonical quantization.}
The classical Painlev\'e I equation can be written as a nonautonomous Hamiltonian system with Hamiltonian 
\begin{equation}\label{HI}
H_{\I}(q,p|t)=\frac12p^2-2q^3-t q.    
\end{equation}
Imposing the canonical commutation relation between the coordinate and momentum by
\begin{equation}\label{pq_eps}
[p,q]=\epsilon
\end{equation}
for some $\epsilon\in\mathbb{C}$, and keeping the Hamiltonian \eqref{HI} unchanged, we obtain a quantization of this Hamiltonian system. The quantum Painlev\'e I dynamics of an observable $f(q,p|t)$ is then governed by the Heisenberg equation of motion
\begin{equation}\label{Heis_I}
\frac{df(q,p|t)}{dt}=\frac{\partial f(q,p|t)}{\partial t}+\frac1{\epsilon}\big[H_{\I}(q,p|t),f(q,p|t)\big]. 
\end{equation}
Applied to the canonical variables $(q,p)$, this yields the Hamilton equations, which coincide with the classical ones:
\begin{equation}\label{Hameq_I}
\dot{q}=p, \qquad \dot{p}=6q^2+t. 
\end{equation}
Note that this straightforward quantization is possible because in \eqref{HI} the coordinate and momentum are not mixed (there are no ordering ambiguities).

\paragraph{Dimension restoring.} It is convenient for us to exploit the homogeneity of the Hamiltonian system.
Introduce a scaling parameter $\kappa\in\mathbb{C}$ and rescale $q,p,t$ (which we now denote by a superscript ${}^{0}$) by
\begin{equation}
q^0=\frac{q}{\kappa^{\frac25}}, \qquad p^0=\frac{p}{\kappa^{\frac35}}, \qquad t^0=\frac{t}{\kappa^{\frac45}}.
\end{equation}
Then the commutator \eqref{pq_eps} and the Hamiltonian \eqref{HI} scale as
\begin{equation}
\epsilon(q^0,p^0)=\frac{\epsilon(q,p)}{\kappa}, \qquad
H_{\I}(q^0,p^0|t^0)=\frac{H_{\I}(q,p|t)}{\kappa^{\frac65}}.
\end{equation}
With this rescaling, the Heisenberg equation \eqref{Heis_I} acquires a factor $\kappa$ in front of the time derivatives:
\begin{equation}\label{Heis}
\kappa\frac{df(q,p|t)}{dt}=\kappa\frac{\partial f(q,p|t)}{\partial t}+\frac1{\epsilon}\big[H(q,p|t),f(q,p|t)\big], 
\end{equation}
where, for brevity, we henceforth omit the subscript $\I$ on $H$.
By construction, the resulting system is homogeneous. We denote the corresponding scaling dimensions by $[\cdot]$, so that
\begin{equation}\label{dim_I}
[q]=\frac25, \qquad [p]=\frac35, \qquad \quad [t]=\frac45, \qquad \quad [H]=\frac65, \qquad [\epsilon]=[\kappa]=1.    
\end{equation}
Because these dimensions are fractional, the scaling has a nontrivial branching, which gives rise to a cyclic $C_5$ symmetry of the Hamiltonian system generated by
\begin{equation}\label{C5_I}
q\mapsto e^{\frac{4\pi\ri}5} q, \qquad p\mapsto e^{\frac{6\pi\ri}5} p, \qquad \quad  t\mapsto e^{\frac{8\pi\ri}5} t.
\end{equation}
Finally, note that this dimension assignment rules out adding any polynomial (in $\epsilon,q,p,t$) $\epsilon$-corrections to the Hamiltonian \eqref{HI}.  

\paragraph{Equations for the Hamiltonian.}
It is well known (see, e.g., \cite{JM81}, \cite{Forrester_book}) that, in the classical case, the nonautonomous Painlev\'e Hamiltonian dynamics can be described as a second-order differential equation in $t$ for the Hamiltonian evaluated along the trajectories, i.e. for the function $H(t)=H(q(t),p(t)|t)$. Let us derive a few total time derivatives of such a function $H(t)$ in the quantum Painlev\'e I case, using the Heisenberg equation~\eqref{Heis}. We obtain 
\begin{equation}\label{Hpq_I}
\dot{H}=-q, \qquad \kappa\ddot{H}=-p, \qquad \kappa^2\dddot{H}=-6q^2-t,   
\end{equation}
which are formally identical to the ($\kappa$-rescaled) classical relations. Substituting $q$ and $p$ from the first two relations into the Hamiltonian \eqref{HI}, we obtain the desired (quantum) \textit{Hamiltonian form} equation:
\begin{equation}\label{Ham_I}
H=\frac12\kappa^2\ddot{H}^2+2\dot{H}^3+t\dot{H}.
\end{equation}
Besides this equation we can write another, third order equation for $H$. Indeed, elimination of $q$ from the third equation of \eqref{Hpq_I} by substituting the first one yields
\begin{equation}\label{prec_I}
\kappa^2\dddot{H}=-6\dot{H}^2-t.
\end{equation}
This equation serves as a precursor for the (quantum) tau form below. Note that, unlike \eqref{Ham_I}, it has a freedom of shifting $H(t)$ by a $t$-constant operator.

\subsection{Tau functions and tau form}
\label{ssec:tau_I}
\paragraph{Quantum Painlev\'e I of \cite{BST25}.}
As explained in the  Introduction, our first goal is to relate the canonically quantized Painlev\'e equations to the bilinear equations for the noncommutative tau functions presented in \cite{BST25} and called there quantum Painlev\'e equations. In the quantum Painlev\'e I case these are equations \cite[(3.10)]{BST25} and \cite[(3.19)]{BST25}:
\begin{equation}\label{QPI_tau}
D^1_{\epsilon_1,\epsilon_2}\left(\tau^{(1)},\tau^{(2)}\right)=0, \qquad
D^3_{\epsilon_1,\epsilon_2}\left(\tau^{(1)},\tau^{(2)}\right)=0,  
  \qquad D^4_{\epsilon_1,\epsilon_2}\left(\tau^{(1)},\tau^{(2)}\right)+\frac{t}8 \tau^{(1)}\tau^{(2)}=0, 
\end{equation}
where the generalized $(\epsilon_1,\epsilon_2)$- Hirota derivative of two (noncommutative) functions $\tau^{(1)}(t)$ and $\tau^{(2)}(t)$ is defined by the expansion
\begin{equation}\label{Hirota}
\tau^{(1)}(t+\epsilon_1 \Delta t)\, \tau^{(2)}(t+\epsilon_2 \Delta t)=\sum_{n=0}^{+\infty} D^n_{\epsilon_1,\epsilon_2} \left(\tau^{(1)}(t),\tau^{(2)}(t)\right)\, \frac{(\Delta t)^n}{n!}.  
\end{equation}
Note that in \cite{BST25} tau functions $\tau^{(1)}_{\textrm{\cite{BST25}}}$ and $\tau^{(2)}_{\textrm{\cite{BST25}}}$ are expressed in terms of a single tau function $\tau_{\textrm{\cite{BST25}}}(\epsilon_1,\epsilon_2|t)$ as
\begin{equation}\label{Omega_patches}
\tau^{(1)}_{\textrm{\cite{BST25}}}(t)=\tau_{\textrm{\cite{BST25}}}(2\epsilon_1,\epsilon_2{-}\epsilon_1|t), \qquad   \tau^{(2)}_{\textrm{\cite{BST25}}}(t)=\tau_{\textrm{\cite{BST25}}}(\epsilon_1{-}\epsilon_2,2\epsilon_2|t).
\end{equation}
In the present paper we do not impose this relation, and instead regard \eqref{QPI_tau} (and the analogous bilinear equations below) as equations for two \textit{independent} tau functions.
Further aspects of the dependence of tau functions on the $\Omega$-background parameters $\epsilon_1$ and $\epsilon_2$ are discussed in Secs.~\ref{sec:asymptotics}, \ref{sec:tau_symm}.

\paragraph{Quantum tau functions.}
In the classical case, the Painlev\'e I tau function is defined by $H(t)=\dot{\tau}/\tau$. To handle the ordering ambiguity of this definition in the quantum case, we introduce left and right tau functions $\tau^{(1)}$ and $\tau^{(2)}$ by
\begin{equation}\label{tau_def_ansatz}
H(t)=c^{(1)}\left(\tau^{(1)}\right)^{-1}\dot{\tau^{(1)}}=c^{(2)}\dot{\tau^{(2)}}\left(\tau^{(2)}\right)^{-1},
\end{equation}
where we fit the constants $c^{(1,2)},\, [c^{(1,2)}]=2$ in order to identify these tau functions with those appearing in \eqref{QPI_tau}. Note that \eqref{tau_def_ansatz} defines $\tau^{(1)}$ and $\tau^{(2)}$ only up to multiplication by $t$-constant operator prefactors from the left and from the right respectively.

The second equality in \eqref{tau_def_ansatz} becomes precisely the first equation in \eqref{QPI_tau} provided that $c^{(1)}/c^{(2)}=-\epsilon_1/\epsilon_2$. We take this as the first condition on these constants, and thus set $c^{(1)}=-2\epsilon_1\tilde{\kappa},\, c^{(2)}=2\epsilon_2\tilde{\kappa}$ with some constant $\tilde{\kappa}$. Furthermore, using \eqref{tau_def_ansatz}, any bilinear differential relation in $\tau^{(1)}$ and $\tau^{(2)}$ can be written in form
\begin{equation}
\tau^{(1)}\cdot F(H,\dot{H},\ddot{H},\ldots) \cdot\tau^{(2)}.    
\end{equation}
The successive $(\epsilon_1,\epsilon_2)$- Hirota derivatives of the tau functions in this form are 
\begin{align}
D^2_{\epsilon_1,\epsilon_2}\left(\tau^{(1)},\tau^{(2)}\right)&=\frac{\epsilon_2{-}\epsilon_1}{2\tilde{\kappa}}\,\tau^{(1)}\cdot \dot{H}_{\I}\cdot\tau^{(2)},\\ \label{D3_I}
D^3_{\epsilon_1,\epsilon_2}\left(\tau^{(1)},\tau^{(2)}\right)&=\frac{\epsilon_2{-}\epsilon_1}{2\tilde{\kappa}}\tau^{(1)}\cdot\left((\epsilon_1{+}\epsilon_2)\ddot{H}_{\I}-\frac1{\tilde{\kappa}}[H_{\I},\dot{H}_{\I}]\right)\cdot\tau^{(2)}.
\end{align}
Since the explicit $t$-dependence of the Hamiltonian \eqref{HI} is linear, the Heisenberg equation \eqref{Heis} implies the commutation relation
\begin{equation}\label{H_Hd_I}
[H,\dot{H}]=\epsilon\kappa \ddot{H}.  
\end{equation}
Therefore, the second equation in \eqref{QPI_tau} is satisfied provided that
$(\epsilon_1{+}\epsilon_2)\tilde{\kappa}=\epsilon\kappa$.
Then, the fourth order $(\epsilon_1,\epsilon_2)$- Hirota derivative yields
\begin{multline}\label{D4_I}
D^4_{\epsilon_1,\epsilon_2}\left(\tau^{(1)},\tau^{(2)}\right)=\frac{\epsilon_2^2{-}\epsilon_1^2}{2\epsilon\kappa}\,\tau^{(1)}\cdot\left((\epsilon_1^2{+}\epsilon_1\epsilon_2{+}\epsilon_2^2)\dddot{H}+\frac{3(\epsilon_1{+}\epsilon_2)^2}{4\epsilon^2\kappa^2}\left[H,[H,\dot{H}]-2\epsilon\kappa\ddot{H}\right]+\frac{3(\epsilon_2^2{-}\epsilon_1^2)}{2\epsilon\kappa}\dot{H}^2\right)\cdot\tau^{(2)}\\=\frac{(\epsilon_2^2{-}\epsilon_1^2)^2}{8\epsilon\kappa}\,\tau^{(1)}\cdot\left(\frac{\epsilon_2{-}\epsilon_1}{\epsilon_1{+}\epsilon_2}\dddot{H}+\frac{6}{\epsilon\kappa}\dot{H}^2\right)\cdot\tau^{(2)}\\=\frac{3(\epsilon_2^2{-}\epsilon_1^2)^2}{4\epsilon^2\kappa^2}\left(1-\frac{(\epsilon_2{-}\epsilon_1)\epsilon}{(\epsilon_1{+}\epsilon_2)\kappa}\right)\,\tau^{(1)}\dot{H}^2\tau^{(2)}-\frac{(\epsilon_2{-}\epsilon_1)^3(\epsilon_1{+}\epsilon_2)}{8\epsilon\kappa^3}t\,\tau^{(1)}\tau^{(2)},
\end{multline}
where for the second equality we
used \eqref{H_Hd_I} and its time derivative and for the third equality we used the precursor equation \eqref{prec_I}. The coefficient of $\dot{H}^2$ vanishes if $(\epsilon_1{+}\epsilon_2)\kappa=\epsilon(\epsilon_2{-}\epsilon_1)$, and then condition $\epsilon^4=(\epsilon_1{+}\epsilon_2)^4$ reproduces the third equation of \eqref{QPI_tau}.
Let us choose the root $\epsilon=\epsilon_1{+}\epsilon_2$, which in turn fixes $\kappa=\epsilon_2{-}\epsilon_1$. With these choices \eqref{tau_def_ansatz} becomes
\begin{equation}\label{tau_universal}
H=-2\epsilon_1(\epsilon_2{-}\epsilon_1)\left(\tau^{(1)}\right)^{-1}\frac{d\tau^{(1)}}{dt}=-2\epsilon_2(\epsilon_1{-}\epsilon_2)\frac{d\tau^{(2)}}{dt}\left(\tau^{(2)}\right)^{-1}.
\end{equation}    
We use this definition of the tau functions (with the appropriate time variable) and the above parametrizations $\epsilon=\epsilon_1{+}\epsilon_2$, $\kappa=\epsilon_2{-}\epsilon_1$ for all other quantum Painlev\'e equations as well. It is natural to refer to equations \eqref{QPI_tau} as the \textit{tau form} of the quantum Painlev\'e I equation. Finally, note that in the classical (commutative) case $\tau^{(1)}$ and $\tau^{(2)}$ coincide, and the odd $(-\epsilon_2,\epsilon_2)$- Hirota derivatives vanish, so that the first two equations of \eqref{QPI_tau} become trivial. This as well occurs for all other Painlev\'e tau forms below.

\subsection{Equivalence of the forms}
\label{ssec:equivalence_I}
In the discussion above, the trajectory $(q(t),p(t))$ can be reconstructed from the Hamiltonian function $H(t)$ via the first two relations of \eqref{Hpq_I}. However, at this point we have no evidence that \textit{any} solution $H(t)$ of the Hamiltonian form equation \eqref{Ham_I} (or of the precursor \eqref{prec_I}) necessarily reconstructs a trajectory $(q(t),p(t))$, governed by the Heisenberg equation \eqref{Heis}. A similar issue arises for the tau form: does an arbitrary solution of \eqref{QPI_tau} produce, via the definition \eqref{tau_universal}, a solution of the Hamiltonian form equation \eqref{Ham_I} (or of the precursor \eqref{prec_I})? We address these questions just below.

\paragraph{Hamiltonian form.}
Via the first two relations in \eqref{Hpq_I}, the Hamiltonian form equation \eqref{Ham_I} becomes precisely the Hamiltonian definition \eqref{HI}.
However, to recover the Heisenberg dynamics \eqref{Heis} we must also impose the commutation relation \eqref{pq_eps} between the reconstructed operators $q$ and $p$, which in terms of $H(t)$ reads
\begin{equation}\label{Hdd_Hd_I}
\kappa[\ddot{H},\dot{H}]=\epsilon. 
\end{equation}
Thus, this commutation relation should be regarded as part of the Hamiltonian form. Assuming it, we recover \eqref{H_Hd_I} and also obtain
\begin{equation}\label{H_Hdd_I}
\kappa[H,\ddot{H}]=-\epsilon(6\dot{H}^2+t), 
\end{equation}
which are precisely the ($\kappa$-rescaled) Hamilton equations \eqref{Hameq_I} and hence generate the Heisenberg dynamics \eqref{Heis}. On the other hand, taking the time derivative of \eqref{H_Hd_I} and comparing it with \eqref{H_Hdd_I} yields the precursor \eqref{prec_I}. Therefore, the Hamiltonian form equation augmented by \eqref{Hdd_Hd_I} provides all the relations needed to pass to the tau form.

\paragraph{Hamiltonian form $\Leftarrow$ Tau form.}
Assume that the tau functions $\tau^{(1)}$ and $\tau^{(2)}$ are invertible operators. The $D^1$-equation in the tau form \eqref{QPI_tau} allows us to introduce a Hamiltonian by formula \eqref{tau_universal}. Then, the $D^3$-equation in \eqref{QPI_tau} implies \eqref{H_Hd_I}. Using this commutator \eqref{H_Hd_I} and its time derivative, we then reproduce the precursor equation \eqref{prec_I} from the $D^4$-equation in \eqref{QPI_tau}.
Next we reconstruct the commutation relation \eqref{Hdd_Hd_I}. Indeed, substitute $\dddot{H}$ from the precursor \eqref{prec_I} into the time derivative of \eqref{H_Hd_I}, differentiate the resulting identity once more, and obtain
\begin{equation}
\kappa[H,\dddot{H}]+\kappa[\dot{H},\ddot{H}]=-\epsilon\left(1+6\{\dot{H},\ddot{H}\}\right).  
\end{equation}
Substituting again $\dddot{H}$ from \eqref{prec_I} and then using \eqref{H_Hd_I}, we arrive precisely at \eqref{Hdd_Hd_I}. Finally, we recover the Hamiltonian form equation \eqref{Ham_I} up to the $t$-constant operator freedom mentioned above for the precursor \eqref{prec_I}. Namely, take the anticommutator of \eqref{prec_I} with $\ddot{H}$ and, using \eqref{Hdd_Hd_I}, rearrange the terms into a total time derivative
\begin{multline}
0=\{\kappa^2\dddot{H}+6\dot{H}^2+t,\ddot{H}\}=\kappa^2\{\dddot{H},\ddot{H}\}+4\left(\ddot{H}\dot{H}^2+4\dot{H}\ddot{H}\dot{H}+4\dot{H}^2\ddot{H}\right)+2t\ddot{H}\\=
\frac{d}{dt}\big(\kappa^2\ddot{H}^2+4\dot{H}^3+2t\dot{H}-2H\big)\qquad \Longrightarrow \quad \frac12\kappa^2\ddot{H}^2+2\dot{H}^3+t\dot{H}-H=C.
\end{multline}
The $t$-constant operator $C$ defined by the latter identity commutes with $\dot{H}$ and hence with all higher derivatives, by \eqref{Hdd_Hd_I} and \eqref{H_Hd_I}. Therefore, we may redefine $H\mapsto H-C$, which restores the Hamiltonian form equation \eqref{Ham_I} while  preserving the commutation relations \eqref{H_Hd_I}, \eqref{Hdd_Hd_I}, \eqref{H_Hdd_I}.

\section{General case: Quantum Painlev\'e VI}
\label{sec:QPVI}
\subsection{Hamiltonian dynamics and its symmetries}
\label{ssec:QPVI_symm}
\paragraph{Hamiltonian.} 

According to \cite[\S 3]{N09}, the quantum Painlev\'e VI  (QPVI for brevity) Hamiltonian dynamics is defined by the Heisenberg equation \eqref{Heis}, together with the canonical commutation relation \eqref{pq_eps}, and a Hamiltonian $H_{\VI}$ in time $t$ (or, alternatively, $\ln (1{-}1/t)$), which we write as 
\begin{multline}\label{HVI}
t(t{-}1)H_{\VI}((a_i)_{i=0}^4;q,p|t)=H_{\VI}\big((a_i)_{i=0}^4;q,p|\ln (1{-}1/t)\big)=\frac16\sum_{\sigma\in S_3(0,t,1)}\big(q-\sigma(0)\big)\,p\,\big(q-\sigma(t)\big)\,p\,\big(q-\sigma(1)\big)\\-\frac{a_0{-}\kappa}2\Big(qp(q{-}1)+(q{-}1)pq\Big)-\frac{a_3}2\Big(qp(q{-}t)+(q{-}t)pq\Big)-\frac{a_4}2\Big((q{-}t)p(q{-}1)+(q{-}1)p(q{-}t)\Big)\\
+a_2(a_1{+}a_2)q+
\frac{(a_0{-}\kappa)^2{+}a_1^2{+}a_3^2{+}a_4^2}{12}(1{+}t)-\frac{a_3^2{+}a_4^2}4t-\frac{(a_4{+}a_0{-}\kappa)^2}4,
\end{multline}
where the parameters $(a_i)_{i=0}^4$ satisfy the relation 
\begin{equation}\label{D4_imrv}
a_0+a_1+2a_2+a_3+a_4=\kappa.
\end{equation}
Compared with \cite[\S 3]{N09}, we restore the dimensions as in Sec. \ref{ssec:QPI_Ham} by introducing a rescaling by a dimension-$1$ parameter $\kappa$:
\begin{equation}
\hat{q}_{\textrm{\cite{N09}}}=q, \quad \hat{p}_{\textrm{\cite{N09}}}=\frac{p}{\kappa}, \qquad t_{\textrm{\cite{N09}}}=t, \qquad 
\alpha_{i\,\textrm{\cite{N09}}}=\frac{a_i}{\kappa}\quad {\scriptstyle (i=0,\ldots 4)}, \qquad h_{\textrm{\cite{N09}}}=\frac{\epsilon}{\kappa}.
\end{equation}
This rescaling also produces
the factor $\kappa$ in front of the time derivatives in the Heisenberg equation \eqref{Heis}. For $\kappa=h_{\textrm{\cite{N09}}}^{-1}$, it translates the Hamiltonian system of \cite[\S 3]{N09} into that of \cite[Sec. 2]{N12}, with $z_{\textrm{\cite{N12}}}=t_{\textrm{\cite{N09}}}$ and $\kappa_{\textrm{\cite{N12}}}=-h_{\textrm{\cite{N09}}}^{-1}$.
The given Heisenberg dynamics with Hamiltonian \eqref{HVI} is also equivalent to the one defined by the "normal-ordered" and homogeneous Hamiltonian $H^q_{\VI}\big(\alpha|\ln (1{-}1/t)\big)$ from \cite[Sec. 2.1]{NY12}, under the dictionary
\begin{equation}\label{NY12_eps}
(\alpha_i)_{\textrm{\cite{NY12}}}=a_i, \qquad \qquad \epsilon_{1\,\textrm{\cite{NY12}}}=2\epsilon_1,\qquad \epsilon_{2\,\textrm{\cite{NY12}}}=\epsilon.    
\end{equation}
for $i=0,\ldots 4$. Below (in this section) we omit the subscript $_{\VI}$ on the Hamiltonian for brevity.

\paragraph{Symmetries.}
The QPVI symmetry group is the extended affine Weyl group $S_4 \ltimes W\left(D_4^{(1)}\right)$, acting on the coordinates $q,p,t,(a_i)_{i=0}^4$ according to \cite[Definition 2.1]{NY12}. We present this action in Table~\ref{table:QPVI_Backlund} with the corresponding diagram. 

\begin{table}[H]
\begin{tabular}{M{11cm}M{3cm}}
\vspace{-2.5cm}
\begin{tabular}{|c|c|c|c|} 
\hline
 & $q$ & $p$ &  $t$ \\
\hline
\hline
$ s_0$ & $q$ & $p - a_0 (q{-}t)^{-1}$ &  \\ 
 \cline{1-3}
 $s_1$ & $q$ & $p$ & \\ 
 \cline{1-3}
 $s_2$ & $q + a_2 p^{-1}$ & $p$  & $t$ \\
 \cline{1-3}
 $s_3$ & $q$ & $p - a_3 (q{-}1)^{-1}$ &  \\ 
 \cline{1-3}
 $s_4$ & $q$ & $p - a_4 q^{-1}$ & \\   
\hline
$\sigma_{34}$ & $1-q$ & $-p$  & $1{-}t$\\
  \hline
 $\sigma_{14}$ & $q^{-1}$ & $ -q(pq + a_2)$ &  $t^{-1}$\\
  \hline
 $\sigma_{03}$ & $q/t$ & $t p$  & $t^{-1}$\\
  \hline
 
 \end{tabular}
 &
 \begin{tikzpicture}[elt/.style={circle,draw=black!100, thick, inner sep=0pt, minimum size=2mm},scale=1.75]
			\path (-1,-1) node (a0)   [elt] {}
			(-1,1) 	node 	(a1) [elt] {}
			(0,0) node  	(a2) [elt] {}
			(1,-1) node  	(a3) [elt] {}
			(1,1) node  	(a4) [elt] {};
			
		     	\node at ($(a0.west) + (-0.2,0)$) 	{$a_0$};
		     	\node at ($(a1.west) + (-0.2,0)$) 	{$a_1$};
		     	\node at ($(a2.north) + (0,0.2)$) 	{$a_2$};
		     	\node at ($(a3.east) + (0.2,0)$) 	{$a_3$};
		     	\node at ($(a4.east) + (0.2,0)$) 	{$a_4$};
		     	
		    \draw[<->, dashed] (a3) to [bend right=40] node[fill=white]{$\sigma_{34}$} (a4);
		     \draw[<->, dashed] (a1) to [bend left=40] node[fill=white]{$\sigma_{14}$} (a4);
		     \draw[<->, dashed] (a3) to [bend left=40] node[fill=white]{$\sigma_{03}$} (a0);
		     
		     \draw[<->, dashed] (a3) to [bend left=40] node[fill=white]{$\pi_2$} (a4);
		     \draw[<->, dashed] (a1) to [bend right=40] node[fill=white]{$\pi_1$} (a4);
		     \draw[<->, dashed] (a3) to [bend right=40] node[fill=white]{$\pi_1$} (a0);
		      \draw[<->, dashed] (a1) to [bend left=40] node[fill=white]{$\pi_2$} (a0);
		   
		    \draw [black,line width=1pt] (a0) -- (a2) -- (a4) (a1) -- (a2) -- (a3);
		   
			\end{tikzpicture} 
			\\
\end{tabular}
\caption{Action of the QPVI symmetry group $S_4 \ltimes W\left(D_4^{(1)}\right)$.}
\label{table:QPVI_Backlund}
\end{table}
Here we use the standard encoding of the extended affine Weyl group by the Dynkin diagram $D_4^{(1)}$ and its automorphism group $\mathrm{Aut}\left(D_4^{(1)}\right)=S_4$. In general (for the finite and affine $ADE$ types), this encoding is as follows. To each node we assign a \textit{root variable}\footnote{Root variables are the images of the simple roots of an affine root system under the period map; see \cite[Sec.5]{S01} for details. For practical purposes, in this paper we describe Weyl group actions only at the level of the root variables.} $a_i$ and a generator $s_i$, which acts on all the root variables by
\begin{equation}\label{refl_act}
s_i(a_j) = a_j - c_{ij}a_i, 
\end{equation}
where $(c_{ij})$ is the Cartan matrix of the given Dynkin diagram. Namely, $c_{ii}=2$, and for $i\neq j$ the integer $(-c_{ij})$ is the number of edges (solid lines) between the $i$'th and $j$'th nodes.
The Weyl group associated with the Dynkin diagram is generated by these reflections $s_i$ (with $s_i^2=1$) subject to the relations
\begin{equation}
s_is_j=s_js_i \quad \textrm{if} \quad c_{ij}=0, \quad \textrm{and} \quad s_is_js_i=s_js_is_j  \quad \textrm{if} \quad c_{ij}=-1,   
\end{equation}
for all suitable pairs of distinct nodes.
This Weyl group can be further extended by the automorphisms of the Dynkin diagram. Concretely, the transposition of the $i$'th and $j$'th nodes corresponds to a generator $\sigma_{ij}$, which permutes the assigned root variables and conjugates the assigned reflections, i.e.
\begin{equation}
\sigma_{ij}(a_i) = a_j \qquad \Leftrightarrow \qquad \sigma_{ij} s_i \sigma_{ij}^{-1} = s_j. 
\end{equation}  
We depict these permutations by dashed arrows. In the diagram accompanying Table~\ref{table:QPVI_Backlund}, we also indicate the action of the automorphisms $\pi_1=\sigma_{14}\sigma_{34}\sigma_{04}\sigma_{34}$ and $\pi_2=\sigma_{34}\sigma_{14}\sigma_{04}\sigma_{14}$. They generate the Klein subgroup $C_2^2\subset S_4$, which will appear just below. 

\paragraph{Extended affine Weyl group structure.}
Let $X$ be a finite connected Dynkin diagram of $ADE$ type, and let $X^{(1)}$ be its affine extension. The affine diagram $X^{(1)}$ is obtained from $X$ by adding one extra node, which is customarily labeled by $0$. The stabilizer of this node in the affine Dynkin diagram automorphism group $\mathrm{Aut}\left(X^{(1)}\right)$ is the automorphism group $\mathrm{Aut}(X)$ of the finite diagram. Moreover,
\begin{equation}\label{aut_decomp}
\mathrm{Aut}\left(X^{(1)}\right)=\mathrm{Aut}(X)\ltimes(P_X/Q_X),
\end{equation}
where $P_X/Q_X$ is a finite abelian group arising as a quotient of two lattices. In the case at hand, $\mathrm{Aut}(D_4)=S_3$, while $P_{D_4}/Q_{D_4}=C_2^2$ is precisely the Klein subgroup mentioned above. 
The lattice $Q_X$ (the root lattice) appears in the standard semidirect-product decomposition of the affine Weyl group:
\begin{equation}
W\left(X^{(1)}\right)=W(X)\ltimes Q_X,  \quad \textrm{with} \quad Q_X=\bigotimes_{i\neq0}T^{\mathbb{Z}}_{c_{i*}}, \quad \textrm{where} \quad T_{c_{i*}}(a_j)=a_j+c_{ij}\kappa. 
\end{equation}
The other, weight-lattice group $P_X=\bigotimes\limits_{i\neq0}T^{\mathbb{Z}}_i$ is generated by the elementary shifts $T_i$ defined by
\begin{equation}\label{elem_transl_root}
T_i(a_j)=a_j+\delta_{ij}\kappa, \quad j\neq0.
\end{equation}
Note that the root variable $a_0$ is a linear combination $a_0=-\sum_{i\neq0}d_ia_i$ (recall \eqref{D4_imrv}), where marks $d_i$ are determined from the affine Cartan matrix (which has corank $1$) by $c_{0*}=-\sum_{i\neq0}d_ic_{i*}$. 
The weight-lattice group $P_X$ appears in the decomposition
\begin{equation}\label{Pdecomp}
(P_X/Q_X)\ltimes W\left(X^{(1)}\right)=W(X)\ltimes P_X \quad \xRightarrow[\textrm{\eqref{aut_decomp}}]{\mathrm{Aut}(X)\ltimes} \quad \mathrm{Aut}\left(X^{(1)}\right)\ltimes W\left(X^{(1)}\right)=\big(\mathrm{Aut}(X)\ltimes W(X)\big)\ltimes P_X,
\end{equation}
where the group in the latter parentheses is the extended finite Weyl group associated with $X$. 

\paragraph{Symmetry group structure for QPVI.}
The above description of the extended affine Weyl group for $X=D_4$, viewed as the QPVI symmetry group, controls its $t$-dependence properties. In particular, the subgroup $W(D_4)\ltimes P_{D_4}$ is a $t$-preserving one. Moreover, the extended finite Weyl group $\mathrm{Aut}(D_4)\ltimes W(D_4)$ consists of \textit{autonomous symmetries}, i.e. symmetries whose $(q,p)$-transformation formulas have no explicit time dependence. Contrary, one checks directly that every elementary translation $T_i$ is nonautonomous.\footnote{Strictly speaking, this does not prove maximality of the finite Weyl group as an autonomous subgroup; moreover, these expectations may fail for special choices of parameters.} Consequently, the Hamiltonian one-form $H_{\VI}((a_i);q,p,t)\,dt$ is invariant under any element $\sigma w$, with $\sigma\in \mathrm{Aut}(D_4)$ and $w\in W(D_4)$, namely 
\begin{equation}\label{Ham_inv}
H((a_i);q,p,t)\,dt= H\big((\sigma w(a_i));\sigma w(q),\sigma w(p)|\sigma(t)\big)\, d\sigma(t), 
\end{equation}
also thanks to the specific choice of the $q,p$-independent part of the Hamiltonian \eqref{HVI}.

Let us substitute $t=t_0+\kappa t_{\textrm{aut}}$ with a constant shift $t_0$ into the Heisenberg equation \eqref{Heis}, thus removing the factor $\kappa$ in front of the time derivatives. Sending then $\kappa\rightarrow0$, we obtain an autonomous limit of the QPVI Hamiltonian in the new time $t_{\textrm{aut}}$. Concretely, in the expression \eqref{HVI} for the Hamiltonian, $t$ is replaced by $t_0$ and $\kappa$ is set to $0$; the latter applies to the relation \eqref{D4_imrv} for the root variables as well. Simultaneously, the lattice subgroup $P_{D_4}$ acts trivially on the root variables in the limit since all of their translations are proportional to $\kappa$, however its action on $(q,p)$ remains nontrivial. Generally, the $(q,p)$-transformations in the limit remain those of  Table~\ref{table:QPVI_Backlund}
with $t$ replaced by $t_0$, while
the parameter $t_0$ transforms in the same way as $t$ in that table. The new time  $t_{\textrm{aut}}$ is still preserved by $W(D_4)\ltimes P_{D_4}$, while $\sigma_{34}(t_{\textrm{aut}})=-t_{\textrm{aut}}$ and $\sigma_{14}(t_{\textrm{aut}})=-t_{\textrm{aut}}/t_0^2$. 

\paragraph{Masses and their invariants.}
The finite Weyl group $W(D_n), \, n\geq2$ admits a standard realization as the group of signed permutations with an even number of sign changes. In particular, $W(D_4)$ acts by such signed permutations on four "basis-vector variables" $(m_f)_{f=1}^4$ related to the root variables by
\begin{equation}\label{PVI_masses}
a_0=\kappa{-}m_1{-}m_3, \quad a_1=m_4{-}m_2,
\quad a_2=m_1{-}m_4, \quad a_3=m_2{+}m_4,
\quad a_4=m_3{-}m_1.
\end{equation}
These \textit{masses} $m_f$ arise naturally on the gauge-theory side of the Painlev\'e/gauge theory correspondence; in the present paper this will appear in Sec.~\ref{sec:asymptotics}. In particular, we use them in Sec.~\ref{sec:coalescence} to take coalescence limits reflecting the renormalization-group flow of the corresponding gauge theories; see \cite[Sec. 5]{BST25} for details. 
It is also convenient to express the parameters in the Hamiltonian and tau forms of the (quantum) Painlev\'e equations in terms of Weyl-group invariant combinations of the masses. 
For QPVI, we use the basic invariants of the ring of $W(D_4)$-invariant polynomials, which can be chosen as elementary symmetric polynomials in the masses $(m_f)_{f=1}^4$ or in their squares:
\begin{equation}
\mathbb{C}[m_1,m_2,m_3,m_4]^{W(D_4)}=\mathbb{C}\left[w_2^{[4]},w_4^{[4]},e_4^{[4]},w_6^{[4]}\right],    
\end{equation}
where we denote
\begin{equation}\label{elem_def}
e_k^{[N]}\left((m_f)_{f=1}^{N}\right)=\sum_{1\leq f_1\leq \ldots \leq f_k\leq N}m_{f_1} \ldots m_{f_k} \quad \textrm{and} \quad w_{2k}^{[N]}=e_k^{[N]}\left((m_f^2)_{f=1}^N\right) \qquad \textrm{for} \quad k=1,\ldots N. 
\end{equation}
The action of $S_4=\mathrm{Aut}(D_4)$ on the masses $(m_f)_{f=1}^4$ produces some their linear transformations. Note that the only nontrivial element of $S_4$ that preserves $t=0$, that is $\sigma_{13}=\sigma_{14}\sigma_{34}\sigma_{14}$, acts on the set of masses simply by $m_2\mapsto-m_2$.

\paragraph{Possible $\epsilon$-corrections.} Finally, let us speculate on the (non)uniqueness of the Hamiltonian expression \eqref{HVI}. We would like to consider possible $\epsilon$-corrections that are polynomial in all variables. The first condition we would like to impose is homogeneity, that is such corrections should have the same dimension as the Hamiltonian. Recall that the dimensions are
\begin{equation}\label{dim_VI}
[q]=0, \qquad [p]=1, \qquad \quad [t]=0, \qquad \quad  [a_i]=1,\, {\scriptstyle i=0,\ldots4} \qquad \quad [H]=2.
\end{equation}
Then we see that the possible nontrivial corrections are of the form $\epsilon p\, \mathbb{C}[q,t]$ or $\epsilon^2 \mathbb{C}[q,t]$. The second condition we would like to impose is invariance under the $W(D_4)$-reflections listed in Table~\ref{table:QPVI_Backlund}; for these reflections, the $(q,p)$-transformation formulas do not involve ordering ambiguities. It is then straightforward to check that there are no nontrivial corrections of the above types.  On the other hand, one may allow linear $\epsilon$-corrections to the root variables of the Hamiltonian system, provided they remain compatible with the relation \eqref{D4_imrv}.

\subsection{Hamiltonian and tau forms}
\label{ssec:QPVI_Ham_tau}
\paragraph{Equations for the Hamiltonian.}
Following the QPI considerations of  Sec.~\ref{sec:toy}, we consider the Hamiltonian \eqref{HVI} written in the time variable $\ln(1{-}1/t)$ and evaluated along trajectories of the Heisenberg equation \eqref{Heis}, i.e. function  
$H\big(\ln(1{-}1/t)\big)=H\big(q(t),p(t)|\ln(1{-}1/t)\big)$. First, since the explicit $t$-dependence of the Hamiltonian \eqref{HVI} is linear, from \eqref{Heis} we have the commutation relation (c.f. \eqref{H_Hd_I})
\begin{equation}\label{H_Hd_VI}
[H,H^\backprime]=\kappa\epsilon\left(H^{\backprime\backprime}-(2t{-}1) H^{\backprime}\right),
\end{equation}
where the backprime $^\backprime$ denotes the total derivative with respect to $\ln(1{-}1/t)$.
Applying the Heisenberg equation \eqref{Heis} repeatedly, we compute $H^\backprime,H^{\backprime\backprime},H^{\backprime\backprime\backprime}$ as polynomials in $q,p,t,\kappa,\epsilon,(m_f)_{f=1}^4$. Eliminating $H^{\backprime\backprime\backprime}$ in favor of lower derivatives yields the precursor equation for the tau form (c.f. \eqref{prec_I}) 
\begin{multline}\label{prec_VI}
\kappa^2\Big(H^{\backprime\backprime\backprime}-2(2t{-}1)H^{\backprime\backprime}+\big(1{+}2t(t{-}1)\big)H^\backprime\Big)+6(H^\backprime)^2-2(2t{-}1)\left\{H,H^\backprime\right\}+2t(t{-}1)H^2\\+\frac13\big(w_2^{[4]}{-}2\epsilon^2\big)\left((2t{-}1)t(t{-}1)H-2(1{+}t(t{-}1)H^\backprime)\right)-2\left(e_4^{[4]}(1{-}t){+}\sigma_{34}\big(e_4^{[4]}\big)t\right)t(t{-}1)=0, 
\end{multline} 
where we used the $W(D_4)$- mass invariants $w_2^{[4]},e_4^{[4]}$, and $\sigma_{34}\big(e_4^{[4]}\big)=\frac12e_4^{[4]}{-}\frac14w_4^{[4]}{+}\frac1{16}(w_2^{[4]})^2$, defined in  \eqref{elem_def}.
We also obtain the second-order QPVI Hamiltonian form equation (c.f. \eqref{Ham_I}), which reads
\begin{multline}\label{Ham_VI}
t(t{-}1)\left(H^{(\backprime\backprime)}\right)^2+4\left(H^{(\backprime)}_0H^{(\backprime)}_1H^{(\backprime)}_0-H^{(\backprime)}_1H^{(\backprime)}_0H^{(\backprime)}_1\right)\\-\big(w_2^{[4]}{-}6\epsilon^2\big)t(t{-}1)\left\{H^{(\backprime)}_0,H^{(\backprime)}_1\right\}
+4t^2(t{-}1)^2\left(e_4^{[4]}H^{(\backprime)}_1-\sigma_{34}\big(e_4^{[4]}\big)H^{(\backprime)}_0\right)
\\=\left(w_6^{[4]}-w_2^{[4]}e_4^{[4]}+4\Big(e_4^{[4]}{+}\sigma_{34}\big(e_4^{[4]}\big)\Big)\epsilon^2-\frac14\big(w_2^{[4]}{-}2\epsilon^2\big)^2\epsilon^2\right)t^3(t{-}1)^3, 
\end{multline}
where we introduce the linear combinations
\begin{multline}\label{Hbas_VI}
H_0^{(\backprime)}=tH^\backprime-\left(H-\frac16\big(w_2^{[4]}{-}2\epsilon^2\big)\right)t(t{-}1), \qquad \qquad H_1^{(\backprime)}=(t{-}1)H^\backprime-\left(H+\frac16\big(w_2^{[4]}{-}2\epsilon^2\big)\right)t(t{-}1), \\  H^{(\backprime\backprime)}=\kappa\left(H^{\backprime\backprime}-(2t{-}1)H^\backprime\right). \mspace{200mu}    
\end{multline}
Note that, unlike the precursor equation \eqref{prec_VI}, this second-order equation also involves the highest-dimension mass invariant $w_6^{[4]}$. Equivalently, one may view \eqref{Ham_VI} as the time integral of \eqref{prec_VI}, with $w_6^{[4]}$ appearing as an integration constant.  
To carry this out, we need the remaining two commutations relations among $H,H^\backprime, H^{\backprime\backprime}$, in addition to \eqref{H_Hd_VI}. Differentiating \eqref{H_Hd_VI} produces the left-hand side of \eqref{prec_VI}, and hence we can write
\begin{equation}\label{H_Hdd_VI}
[H,H^{\backprime\backprime}]=\kappa\epsilon\Big(H^{\backprime\backprime\backprime}-2(2t{-}1)H^{\backprime\backprime}+\big(1{+}2t(t{-}1)\big)H^\backprime\Big)\Big|_{\textrm{\eqref{prec_VI}}}. 
\end{equation}
Differentiating this relation once more and using \eqref{prec_VI}, \eqref{H_Hd_VI}, \eqref{H_Hdd_VI}, we then obtain the commutator $[H^{\backprime\backprime},H^\backprime]$. In the notation \eqref{Hbas_VI}, the resulting commutation relations take the form
\begin{align}\label{H2_H0}
&\left[H^{(\backprime\backprime)},H_0^{(\backprime)}\right]=2\epsilon\left(\big(H_0^{(\backprime)}\big)^2-\left\{H^{(\backprime)}_0,H^{(\backprime)}_1\right\}-\frac12\big(w_2^{[4]}{-}2\epsilon^2\big)t(t{-}1)H_0^{(\backprime)}+e_4^{[4]}t^2(t{-}1)^2\right),\\ \label{H2_H1}
&\left[H^{(\backprime\backprime)},H_1^{(\backprime)}\right]=2\epsilon\left(\big(H_1^{(\backprime)}\big)^2-\left\{H^{(\backprime)}_0,H^{(\backprime)}_1\right\}-\frac12\big(w_2^{[4]}{-}2\epsilon^2\big)t(t{-}1)H_1^{(\backprime)}+\sigma_{34}\big(e_4^{[4]}\big)t^2(t{-}1)^2\right).
\end{align}
Finally, integrating the precursor equation \eqref{prec_VI} modulo the  commutation relations among $H,H^\backprime, H^{\backprime\backprime}$ yields the Hamiltonian form equation \eqref{Ham_VI}, with a $t$-constant operator $C$ in place of $w_6^{[4]}$. This operator $C$ is $W(D_4)$-invariant and central, i.e. it commutes with $H$, and hence with all time derivatives of $H$.

\paragraph{Tau form.}
To obtain the QPVI tau form, we follow the QPI construction of Sec.~\ref{ssec:tau_I}. First, we introduce tau functions $\tau^{(1)},\tau^{(2)}$ by the universal formula \eqref{tau_universal} with the time variable $t$ replaced by $\ln(1{-}1/t)$. Then the first $(\epsilon_1,\epsilon_2)$-Hirota derivative vanishes automatically, while the commutation relation \eqref{H_Hd_VI} yields the third-order Hirota equation: 
\begin{equation}\label{D13_VI}
D^1_{\epsilon_1,\epsilon_2}\left(\tau^{(1)},\tau^{(2)}\right)=0, \qquad D^3_{\epsilon_1,\epsilon_2}\left(\tau^{(1)},\tau^{(2)}\right)=\epsilon\, (2t{-}1) D^2_{\epsilon_1,\epsilon_2}\left(\tau^{(1)},\tau^{(2)}\right), 
\end{equation}
where the $(\epsilon_1,\epsilon_2)$- Hirota derivative \eqref{Hirota} is taken with respect to $\ln(1{-}1/t)$ rather than $t$. Recall that, starting from the QPI case, we set $\kappa=\epsilon_2{-}\epsilon_1$ and $\epsilon=\epsilon_1{+}\epsilon_2$. Using \eqref{H_Hd_VI} together with its time derivative and, finally, the precursor \eqref{prec_VI}, we obtain the fourth-order Hirota equation:
\begin{multline}\label{D4_VI}
D^4_{\epsilon_1,\epsilon_2}\left(\tau^{(1)},\tau^{(2)}\right)+2(2t{-}1)\epsilon_1\epsilon_2\big(D^2_{\epsilon_1,\epsilon_2}\left(\tau^{(1)},\tau^{(2)}\right)\big)^{\backprime}+t(t{-}1)(\epsilon_1\epsilon_2)^2\left(\tau^{(1)}\tau^{(2)}\right)^{\backprime\backprime}\\-\left(\big(1{+}t(t{-}1)\big)\left(\frac16(w_2^{[4]}{-}2\epsilon^2){+}\epsilon_1\epsilon_2{+}6\epsilon^2\right)-5\epsilon^2\right)D^2_{\epsilon_1,\epsilon_2}\left(\tau^{(1)},\tau^{(2)}\right)-\frac1{12}\left(w_2^{[4]}{-}2\epsilon^2\right)t(t{-}1)(2t{-}1)\epsilon_1\epsilon_2\left(\tau^{(1)}\tau^{(2)}\right)^\backprime\\-\frac14\left(e_4^{[4]}-\frac14\left(2e_4^{[4]}{+}w_4^{[4]}{-}\frac14(w_2^{[4]})^2\right)t\right)t(t{-}1)\tau^{(1)}\tau^{(2)}=0.
\end{multline}
Conversely, assuming that $\tau^{(1)}$ and $\tau^{(2)}$ are invertible, one can recover the commutation relation \eqref{H_Hd_VI} and the precursor equation \eqref{prec_VI} from the tau form equations \eqref{D13_VI}, \eqref{D4_VI}, exactly as in Sec.~\ref{ssec:equivalence_I} for the QPI case. 
The tau form equations \eqref{D13_VI}, \eqref{D4_VI} are related to the blowup equations \cite[(2.33) under (2.13),(2.14)]{BST25} (where they are also referred to as the QPVI equation) by
\begin{equation}\label{conn_VI}
\tau^{(1)}=t^{\frac{w_2^{[4]}{-}2\epsilon^2}{12\epsilon_1(\epsilon_2{-}\epsilon_1)}}(1{-}t)^{-\frac{w_2^{[4]}{-}2\epsilon^2+6\big(e_2^{[4]}{+}e_1^{[4]}\epsilon{+}\epsilon^2\big)}{24\epsilon_1(\epsilon_2{-}\epsilon_1)}}\tau^{(1)}_{\textrm{\cite{BST25}}}, \quad \tau^{(2)}=t^{\frac{w_2^{[4]}{-}2\epsilon^2}{12\epsilon_2(\epsilon_1{-}\epsilon_2)}}(1{-}t)^{-\frac{w_2^{[4]}{-}2\epsilon^2+6\big(e_2^{[4]}{+}e_1^{[4]}\epsilon{+}\epsilon^2\big)}{24\epsilon_2(\epsilon_1{-}\epsilon_2)}}\tau^{(2)}_{\textrm{\cite{BST25}}}.
\end{equation} 
Finally, note that the tau functions $\tau^{(1)}$ and $\tau^{(2)}$ are invariant under the action of the extended finite Weyl group $S_3\ltimes W(D_4)$: this follows from the Hamiltonian symmetry \eqref{Ham_inv} together with their definition \eqref{tau_universal}.

\paragraph{Reconstruction of the Heisenberg dynamics.}
Reconstructing the Heisenberg dynamics \eqref{Heis} from the Hamiltonian equations in the QPVI case is considerably more subtle than in the QPI case. First, using \eqref{HVI} and the explicit expressions for $H^\backprime,H^{\backprime\backprime}$ as polynomials in $q,p,t,\kappa,\epsilon,(m_f)_{f=1}^4$, we can express the coordinate $q$ in terms of the variables \eqref{Hbas_VI} (c.f. \cite[(D.7)]{BST25})
\begin{multline}\label{q_VI}
\left(\frac{H_0^{(\backprime)}{-}H_1^{(\backprime)}}{t(t{-}1)}{-}\frac{w_2^{[4]}{-}2\epsilon^2}2{-}\Big(m_2{-}\frac{e_1^{[4]}}2\Big)\Big(m_2{+}\epsilon{-}\frac{e_1^{[4]}}2\Big)\right)\left(\frac{H_0^{(\backprime)}{-}H_1^{(\backprime)}}{t(t{-}1)}{-}\frac{w_2^{[4]}{-}2\epsilon^2}2{-}\Big(m_4{-}\frac{e_1^{[4]}}2\Big)\Big(m_4{+}\epsilon{-}\frac{e_1^{[4]}}2\Big)\right)\times q\\
=\left(\frac{H_0^{(\backprime)}{-}H_1^{(\backprime)}}{t(t{-}1)}{-}\frac{w_2^{[4]}{-}2\epsilon^2}2{-}\Big(m_2{+}\epsilon{-}\frac{e_1^{[4]}}2\Big)\Big(m_4{+}\epsilon{-}\frac{e_1^{[4]}}2\Big)\right)\left(\frac{H_0^{(\backprime)}}{t(t{-}1)}+(m_1{-}\epsilon)(m_3{-}\epsilon)\right)\\
-\frac12(m_1{+}m_3{-}\epsilon)\left(\frac{H^{(\backprime\backprime)}}{t(t{-}1)}-(m_1{+}m_3{-}\epsilon)\big((m_1{-}\epsilon)(m_3{-}\epsilon){+}m_2m_4\big)\right)
\end{multline}
and, via this expression, the momentum $p$ (c.f. \cite[(D.8)]{BST25})
\begin{equation}\label{p_VI}
(m_1{+}m_3{+}\epsilon)p=\frac1{q{-}1}\left(\frac{H_1^{(\backprime)}}{t(t{-}1)}{+}\Big(m_2{+}m_4{-}\frac{e_1^{[4]}}2\Big)\Big(m_2{+}m_4{-}\epsilon{-}\frac{e_1^{[4]}}2\Big)\right)-\frac1{q}\left(\frac{H_0^{(\backprime)}}{t(t{-}1)}{-}m_1(m_1{+}\epsilon)\right).
\end{equation}
Our goal is to reconstruct the Heisenberg dynamics \eqref{Heis} with Hamiltonian \eqref{HVI} from the Hamiltonian-form equation \eqref{Ham_VI} together with the commutation relations \eqref{H_Hd_VI}, \eqref{H2_H0}, \eqref{H2_H1}. In general, this proceeds through the following steps (cf. Sec.~\ref{ssec:equivalence_I} in the QPI case):
\begin{enumerate}
    \item Define $q$ and $p$ in terms of $H$ and its derivatives (for QPVI, one may use \eqref{q_VI} and \eqref{p_VI}).
    \item Verify the canonical commutation relation \eqref{pq_eps} for the resulting $q,p$.
    \item Express the Hamiltonian (in our case, \eqref{HVI}) in terms of these $q$ and $p$.
    \item Verify the Heisenberg (Hamilton) equations \eqref{Heis} for $q$ and $p$.
\end{enumerate}
Note that Hamilton’s equations can, in fact, be replaced by the commutation relation between $H$ and its derivative, since the adjoint action is a derivation. Indeed, \eqref{H_Hd_VI} can be rewritten as
\begin{equation}\label{H_Hd_Hameq}
[H,\dot{H}]=\kappa\epsilon (\dot{H})^\backprime.
\end{equation}
This relation, together with its time derivative, may be used in place of Hamilton equations. The only remaining point is to check that $\dot{H}$ and $(\dot{H})^\backprime$, when expressed via the reconstructed $q,p$, have no explicit dependence on the time variable. Such time-independence holds for all (quantum) Painlev'e equations, due to the explicit linear $t$-dependence of their Hamiltonians in appropriate time variables. In the QPVI case, this amounts to the statement that the polynomials $H_0^{(\backprime)}$, $H_1^{(\backprime)}$, and $H^{(\backprime\backprime)}$, divided by $t(t{-}1)$, are $t$-independent, in agreement with \eqref{q_VI} and \eqref{p_VI}. Note also that the $t$-constant central operator $C$ introduced above is automatically central in the Heisenberg skew field $\mathbb{C}(\langle p,q\rangle)/\big([p,q]=\epsilon\big)$. Hence $C$ must be a finite Weyl group mass invariant (here, a $W(D_4)$-invariant) of the appropriate dimension (in our case, $[C]=6$).  

Unfortunately, when attempting to carry out these steps for QPVI, we encountered computational difficulties: the commutators of the resulting expressions become too cumbersome for us to control directly. Nevertheless, we believe the reconstruction should be feasible in principle, since we were able to implement the same procedure not only for QPI but also for all quantum Painlev'e equations in the next section. Resolving this issue appears to be related to finding suitable combinations of $H$ and its derivatives for which the Hamiltonian form can be established as a second-order equation together with a single commutation relation that enforces the canonical commutation relation \eqref{pq_eps} for $q$ and $p$.
Another important point in the reconstruction is the requirement that certain combinations of $H$ and its derivatives be invertible. In particular, this concerns the bracketed factors multiplying $q$ in \eqref{q_VI}. Below, we do not address invertibility issues explicitly, and we tacitly assume that all operators we invert are invertible in the generic situation. This expectation is motivated by the classical case, where vanishing denominators correspond to additional algebraic constraints on the dynamical variables. We hope to return to a more rigorous study of Hamiltonian forms elsewhere.

\section{Quantum Painlev\'e coalescence}
\label{sec:coalescence}
\setcounter{subsection}{-1}
\subsection{The general scheme of the coalescence}

\begin{figure}[H]
\begin{center}
\begin{tikzcd}[row sep=3em, column sep=3em]
\mathrm{QPVI} \arrow[r] & \mathrm{QPV}  \arrow[r]  \arrow[d] & \mathrm{QPIII}_1 \arrow[r]  \arrow[d] & \mathrm{QPIII}_2 \arrow[r]  \arrow[d] & \mathrm{QPIII}_3 \\
& \mathrm{QPIV} \arrow[r] & \mathrm{QPII} \arrow[r] & \mathrm{QPI} &
\end{tikzcd}
\end{center}
\caption{Quantum Painlev\'e coalescence diagram}
	        
         \label{fig:QPainleve_coalescence}              
\end{figure}
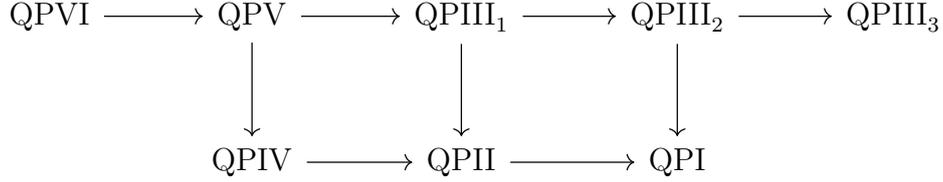 

In this section, we obtain a quantum Painlev\'e coalescence (Fig. \ref{fig:QPainleve_coalescence})starting from the QPVI equation presented in  Sec.~\ref{sec:QPVI}, actually following the same limiting procedures as in the classical Painlev\'e coalescence scheme. More precisely, we use the limits of \cite[App. D]{BST25}, adjusted to normalizations of the present paper, and successively derive Hamiltonian systems for all quantum Painlev\'e equations. As explained in the Introduction, this limiting procedure at the level of tau forms was already carried out in \cite[Sec. 3.1,3.2]{BST25}; here we 
lift this coalescence to the level of
quantum Hamiltonian systems. 

The results for the quantum Painlevé equations presented below are parallel to those obtained for the toy model QPI (Sec.~\ref{sec:toy}) and for the most sophisticated case QPVI (Sec.~\ref{sec:QPVI}). For each equation, we provide the following:
\begin{enumerate}
    \item Limiting procedure(s) in terms of $q,p,t$ and mass parameters, together with the resulting quantum Hamiltonian, and a dictionary relating it to the Hamiltonians of \cite{NY12}, \cite{N12}. 
    \item An extended affine Weyl symmetry group of the Hamiltonian system and its structure. 
    \item A Hamiltonian form equation, a tau form precursor equation, and commutation relations for the total time derivatives of the Hamiltonian.
    \item A tau form and its equivalence to the Hamiltonian form as well as a relation of the tau form equations with those in \cite{BST25}.
    \item Establishment of the Hamiltonian form and its equivalence to the original Hamiltonian system.
\end{enumerate}

\subsection{Quantum Painlev\'e V}
\label{ssec:QPV}
\paragraph{Hamiltonian.}
Making the autonomous rescaling (canonical transformation)
\begin{equation}\label{lim_VI_V}
 q_{\VI}=\frac{q_{\V}}{m_4}, \qquad \quad p_{\VI}=p_{\V}m_4, \qquad \qquad t_{\VI}=\frac{t_{\V}}{m_4},
 \end{equation}
 and then sending $m_4\rightarrow\infty$, we obtain that the QPVI Heisenberg dynamics \eqref{Heis} with Hamiltonian~\eqref{HVI} degenerates to the Heisenberg dynamics defined by the Hamiltonian
\begin{multline}\label{HV}
 H_{\V}((a_i)_{i=1}^3;q,p|\ln t)=t H_{\V}((a_i)_{i=1}^3;q,p|t)=pq(q{-}t)p+\frac12\Big(qp(q{-}t)+(q{-}t)pq\Big)+\frac{a_1{+}a_3}2\{p,q\}\\+a_2tp+a_1q-\frac{a_1{-}2a_2{-}a_3}4t+\frac{(a_1{+}a_3)^2}4,
 \end{multline}
 where we introduce the root variables $(a_i)_{i=0}^3$ by
\begin{equation}
a_0=\kappa{-}m_1{-}m_3, \quad a_1=m_1{-}m_2, \quad a_2=m_3{-}m_1, \quad a_3=m_1{+}m_2 \quad \Rightarrow \quad \sum_{i=0}^3a_i=\kappa.  
\end{equation}
More precisely, we have
\begin{equation}
H_{\VI}\big(\ln(1{-}t_{\VI}^{-1})\big)-\frac{w_2^{[4]}{-}2\epsilon^2}{12}(2{-}t)=-H_{\V}(\ln t_{\V})+O(m_4^{-1}), \qquad  d\ln(1{-}t^{-1}_{\VI})=-d\ln t_{\V}\left(1-\frac{t_{\V}}{m_4}\right)^{-1}.
\end{equation}
Under the limiting procedure \eqref{lim_VI_V}, the dimensions \eqref{dim_VI} induce the following dimensions for the QPV variables:
\begin{equation}\label{dim_V}
[q]=1, \qquad [p]=0, \qquad \quad [t]=1, \qquad \quad  [a_i]=1,\, {\scriptstyle i=0,\ldots3} \qquad \quad [H(\ln t)]=2.
\end{equation}

The resulting Heisenberg dynamics with Hamiltonian \eqref{HV} is
equivalent to that defined by the homogeneous Hamiltonian $H^q_{\V}\big(\alpha|\ln t\big)$ in \cite[Sec. 2.2]{NY12} under the dictionary
\begin{equation}
q_{\textrm{\cite{NY12}}}=q/t, \qquad \qquad 
p_{\textrm{\cite{NY12}}}=tp  
\end{equation}
together with \eqref{NY12_eps} and the additional permutation $a_i\mapsto a_{3-i},\, {\scriptstyle i=0,\ldots 3}$. The same dynamics is also equivalent to that of
\cite[Sec. 3]{N12} under a special dimension \eqref{dim_V} rescaling, i.e.
\begin{equation}\label{5to9}
q_{\textrm{\cite{N12}}}=\frac{q_{\textrm{\cite{NY12}}}}{(\epsilon_2)^{[q]}_{\textrm{\cite{NY12}}}}, \quad
p_{\textrm{\cite{N12}}}=\frac{p_{\textrm{\cite{NY12}}}}{(\epsilon_2)^{[p]}_{\textrm{\cite{NY12}}}}, \qquad z_{\textrm{\cite{N12}}}=\frac{t_{\textrm{\cite{NY12}}}}{(\epsilon_2)^{[t]}_{\textrm{\cite{NY12}}}} \qquad  (\alpha_i)_{\textrm{\cite{N12}}}=\frac{(\alpha_i)_{\textrm{\cite{NY12}}}}{(\epsilon_2)^{[a_i]}_{\textrm{\cite{NY12}}}} \quad \Rightarrow \quad \kappa_{\textrm{\cite{N12}}}=\frac{(\epsilon_1)_{\textrm{\cite{NY12}}}}{(\epsilon_2)_{\textrm{\cite{NY12}}}}{-}1. 
\end{equation}

\paragraph{Symmetries.}
The QPV symmetry group is the extended affine Weyl group $\mathrm{Dih}_4 \ltimes W\left(A_3^{(1)}\right)$, acting on the coordinates $q,p,t,(a_i)_{i=0}^3$ according to \cite[Definition 2.6]{NY12}. We present this action in Table~\ref{table:QPV_Backlund} with the corresponding diagram, following the encoding described in the QPVI case (see Sec.~\ref{ssec:QPVI_symm}).
\begin{table}[H]
\begin{tabular}{M{10cm}M{4cm}}
\vspace{-4.5cm}
\begin{tabular}{ |c|c|c|c|c| } 
\hline
 & $q$ & $p$  & $t$ \\
\hline
\hline
 $s_0$ & $q$ & $p-a_0 (q{-}t)^{-1}$ &  \\ 
 \cline{1-3}
 $s_1$ & $q + a_1 p^{-1}$ & $p$ &
  \\ 
 \cline{1-3}
 $s_2$ & $q$ & $p-a_2 q^{-1}$ & $t$\\
 \cline{1-3}
 $s_3$ & $q + a_3(p{+}1)^{-1}$ & $p$ &  \\
 \cline{1-3}
 $\pi$ & $-tp$ & $ q/t-1$ & \\
\hline
$\sigma_{13}$ & $-q$ & $-p-1$ & $-t$\\
\hline

\end{tabular}
&
\begin{tikzpicture}[elt/.style={circle,draw=black!100,thick, inner sep=0pt,minimum size=2mm},scale=1.25]
			\path 	(-1,-1) 	node 	(a0) [elt] {}
			(-1,1) 	node 	(a1) [elt] {}
			(1,-1) node  	(a3) [elt] {}
			(1,1) node  	(a2) [elt] {};
			
		     	\node at ($(a0.west) + (-0.2,-0.1)$) 	{$a_0$};
		     	\node at ($(a1.west) + (-0.2,0.1)$) 	{$a_1$};
		     	\node at ($(a2.east) + (0.2,0.1)$) 	{$a_2$};
		     	\node at ($(a3.east) + (0.2,-0.1)$) 	{$a_3$};

		    \draw[<-, dashed] (a0) to [bend left=40] node[fill=white]{$\pi$} (a1);
		     \draw[<-, dashed] (a1) to [bend left=40] node[fill=white]{$\pi$} (a2);
		    \draw[<-, dashed] (a2) to [bend left=40] node[fill=white]{$\pi$} (a3);
		    \draw[<-, dashed] (a3) to [bend left=40] node[fill=white]{$\pi$} (a0);
			\draw[<->, dashed] (a3) to [bend right=0] node[fill=white]{$\sigma_{13}$} (a1);	   
			\draw [black,line width=1pt] (a0) -- (a1);
			\draw [black,line width=1pt] (a1) -- (a2);
			\draw [black,line width=1pt] (a2) -- (a3);
			\draw [black,line width=1pt] (a3) -- (a0);
			\end{tikzpicture} 
			\\
\end{tabular}
\caption{Action of the QPV symmetry group $\mathrm{Dih}_4\ltimes W\left(A_3^{(1)}\right)$.}
\label{table:QPV_Backlund}
\end{table}
For the dihedral group, we have the following decomposition:
\begin{equation}
\mathrm{Dih}_4=\mathrm{Aut}\left(A_3^{(1)}\right)=\underbrace{C_2\langle\sigma_{13}\rangle}_{\mathrm{Aut}(A_3)}\ltimes \underbrace{C_4\langle\pi\rangle}_{P_{A_3}/Q_{A_3}}.    
\end{equation}
We reproduce the considerations of Sec.~\ref{ssec:QPVI_symm} for the QPVI symmetry group in the present QPV setting: 
\begin{itemize}
    \item The subgroup $C_4\ltimes W\left(A_3^{(1)}\right)$ is the $t$-preserving subgroup of the full group.
    \item The extended finite Weyl group $C_2\ltimes W(A_3)$ is the subgroup of the autonomous symmetries. Under this subgroup, the one-form $H_{\V}((a_i);q,p,t)dt$ is invariant by \eqref{Ham_inv} ($\forall\sigma w: \sigma\in C_2, \, w\in W(A_3)$).
    \item  Under the autonomization limit $t=t_0+\kappa t_{\textrm{aut}}, \kappa\rightarrow0$, the Hamiltonian \eqref{HV} becomes autonomous, with $t$ replaced by the constant $t_0$, while the lattice group $P_{A_3}$ acts on the root variables trivially.  
    \item The action of the finite Weyl group $W(A_3)$ on the mass variables $(m_f)_{f=0}^3$ is realized as $W(D_3)$, i.e. the group of signed permutations with an even number of sign changes. For the ring of $W(D_3)$-invariant polynomials, we take the basic invariants
 $w_2^{[3]},e_3^{[3]},w_4^{[3]}$ (recall the notation \eqref{elem_def}). 
    \item There are no nontrivial polynomial $\epsilon$-corrections to the Hamiltonian \eqref{HV} that simultaneously preserve the dimensions \eqref{dim_V} and the $W(A_3)$-action given in Table~\ref{table:QPV_Backlund}. However, the root variables $(a_i)_{i=1}^3$ may be shifted by arbitrary linear $\epsilon$-corrections.
\end{itemize}

\paragraph{Equations for the Hamiltonian.}
Here we follow Secs.~\ref{sec:toy},\ref{ssec:QPVI_Ham_tau} for the QPI and QPVI cases, respectively. We consider the Hamiltonian \eqref{HV} in time $\ln t$ along the Heisenberg trajectories \eqref{Heis}, i.e.  
$H(t)=H_{\V}\big(q(t),p(t)|\ln t)$. The explicit linear $t$-dependence of \eqref{HV} yields the commutation relation (c.f. \eqref{H_Hd_I}, \eqref{H_Hd_VI})
\begin{equation}\label{H_Hd_V}
[H,H']=\kappa\epsilon (H''-H'),   
\end{equation}
where the prime $'$ denotes the total derivative with respect to $\ln t$. Using \eqref{Heis}, we compute $H',H'',H'''$ from \eqref{HV} as polynomials in $q,p,t,\kappa,\epsilon,(m_f)_{f=1}^3$. This yields the precursor equation (c.f. \eqref{prec_I},\eqref{prec_VI})
\begin{equation}\label{prec_V}
\kappa^2(H'''-2H''+H')+6(H')^2-2\left\{H,H'\right\}+\big(H-H'\big)t^2-\frac12\big(w_2^{[3]}{-}\epsilon^2\big)t^2+2e_3^{[3]}t=0,
\end{equation} 
and the second-order QPV Hamiltonian form equation (c.f. \eqref{Ham_I}, \eqref{Ham_VI}):
\begin{multline}\label{Ham_V}
\kappa^2t^2\Big(H''-H'\Big)^2-\left(2(H')^2+\left(H-H'{-}\frac12\big(w_2^{[3]}{-}\epsilon^2\big)\right)t^2\right)^2+4(H')^4-2\big(w_2^{[3]}{-}3\epsilon^2\big)(H')^2t^2+4e_3^{[3]}t^3H'\\=\Big(w_4^{[3]}-\frac14\big(w_2^{[3]}{-}\epsilon^2\big)^2\Big)t^4.
\end{multline}
The Hamiltonian form equation \eqref{Ham_V} depends on the $W(D_3)$- basic mass invariants $w_2^{[3]},e_3^{[3]},w_4^{[3]}$ defined in \eqref{elem_def}, whereas the precursor \eqref{prec_V} depends on the same invariants except for the highest-dimension invariant $w_4^{[3]}$. 
To integrate the precursor in $t$ and recover the Hamiltonian form equation, we complete the commutation relation \eqref{H_Hd_V} to a full set of relations among $H,H',H''$, using the precursor itself, analogously to the QPVI case. These relations can be written as
\begin{equation}\label{triple_V}
[H_\pm,H']=\pm\epsilon t H_{\pm}, \qquad [H_+,H_-]=4\epsilon t\left(4(H')^3-\big(w_2^{[3]}{-}\epsilon^2\big)t^2H'+e_3^{[3]}t^3\right),
\end{equation}
where we introduce the combinations
\begin{equation}\label{comb_V}
H_{\pm}=2(H')^2+\left(H-H'{-}\frac12\big(w_2^{[3]}{-}\epsilon^2\big)\right)t^2\pm\kappa t(H''{-}H').
\end{equation}
After integration, we obtain the Hamiltonian form equation \eqref{Ham_V}, in which $w_4^{[3]}$ is replaced by a $t$-constant central $W(D_3)$-invariant operator $C$.

\paragraph{Tau form.}
Deriving the QPV tau form is completely analogous to the QPVI case in Sec.~\ref{ssec:QPVI_Ham_tau}. We introduce the tau functions $\tau^{(1)},\tau^{(2)}$ by the universal formula \eqref{tau_universal}, with the time variable $t$ replaced by $\ln t$. Then the first $(\epsilon_1,\epsilon_2)$- Hirota derivative vanishes automatically, while the commutation relation \eqref{H_Hd_V} yields the third-order Hirota equation:
\begin{equation}\label{D13_V}
D^1_{\epsilon_1,\epsilon_2}\left(\tau^{(1)},\tau^{(2)}\right)=0, \qquad
D^3_{\epsilon_1,\epsilon_2}\left(\tau^{(1)},\tau^{(2)}\right)=\epsilon D^2_{\epsilon_1,\epsilon_2}\left(\tau^{(1)},\tau^{(2)}\right),
\end{equation}
where the $(\epsilon_1,\epsilon_2)$- Hirota derivative \eqref{Hirota} is taken with respect to $\ln t$ rather than $t$. Using \eqref{H_Hd_V} together with its time derivative and, finally, the precursor \eqref{prec_V}, we obtain the fourth-order Hirota equation (c.f. \eqref{D4_VI}):
\begin{multline}\label{D4_V}
D^4_{\epsilon_1,\epsilon_2}\left(\tau^{(1)},\tau^{(2)}\right)+2\epsilon_1\epsilon_2\big(D^2_{\epsilon_1,\epsilon_2}\left(\tau^{(1)},\tau^{(2)}\right)\big)'-\left(\frac14t^2{+}\epsilon_1\epsilon_2{+}\epsilon^2\right)D^2_{\epsilon_1,\epsilon_2}\left(\tau^{(1)},\tau^{(2)}\right)-\frac14t^2\epsilon_1\epsilon_2\left(\tau^{(1)}\tau^{(2)}\right)'\\+\frac14\left(e_3^{[3]}{-}\frac14\big(w_2^{[3]}{-}\epsilon^2\big)t\right)t\tau^{(1)}\tau^{(2)}=0.
\end{multline} 
Conversely, assuming that $\tau^{(1)}$ and $\tau^{(2)}$ are invertible, one can recover the commutation relation \eqref{H_Hd_V} and the precursor equation \eqref{prec_V} from the tau form equations \eqref{D13_V}, \eqref{D4_V}, exactly as in Sec.~\ref{ssec:equivalence_I} for the QPI case. 
The tau form equations \eqref{D13_V}, \eqref{D4_V} are related to the blowup equations \cite[(3.1),(3.2)]{BST25} (where they are also referred to as the QPV equation) by 
\begin{equation}\label{conn_V}
\tau^{(1)}=e^{\frac{e_1^{[3]}+\epsilon}{4\epsilon_1(\epsilon_2{-}\epsilon_1)}\,t}\,\tau^{(1)}_{\textrm{\cite{BST25}}}, \qquad   \tau^{(2)}=e^{\frac{e_1^{[3]}+\epsilon}{4\epsilon_2(\epsilon_1{-}\epsilon_2)}\,t}\,\tau^{(2)}_{\textrm{\cite{BST25}}}.
\end{equation}
Finally, as in the QPVI case, the tau functions $\tau^{(1)}$ and $\tau^{(2)}$ are invariant under the action of the extended finite Weyl group $C_2\ltimes W(A_3)$.

\paragraph{Hamiltonian form.}
Here we reconstruct the Heisenberg dynamics \eqref{Heis} from the equations for the Hamiltonian. First, however, we specify the starting equations, i.e. we establish the Hamiltonian form, as was done in Sec.~\ref{ssec:equivalence_I} for the QPI case. Using the last commutation relation in \eqref{triple_V}, we can rewrite the Hamiltonian form equation \eqref{Ham_V} as
\begin{equation}
\label{Ham_V_twisted}
4H_+H_-=\prod_{\varsigma_1,\varsigma_2=\pm1}\Big(2H'+\big(\varsigma_1m_1{+}\varsigma_2m_2{+}\varsigma_1\varsigma_2m_3+\epsilon\big)t\Big). 
\end{equation}
We refer to the equation \eqref{Ham_V_twisted} together with one of the first commutation relations in \eqref{triple_V} as the \textit{Hamiltonian form} of the QPV equation. Below, for definiteness, we choose the relation with the "$+$" sign. From this Hamiltonian form one can recover the remaining two commutation relations in \eqref{triple_V} and the precursor \eqref{prec_V}. Indeed, commuting \eqref{Ham_V_twisted} with $H'$ yields the commutator $[H_-,H']$ from \eqref{triple_V}. Next, commuting \eqref{Ham_V_twisted} with $H_+$ and using the commutation relation between $H_+$ and $H'$ recovers the last relation in \eqref{triple_V}. Finally, deriving $[H,H']$ and $[H,H'']$ from the full set of the commutation relations and using the identity $[H,H']'=[H,H'']$, we recover the precursor \eqref{prec_V}.

To reconstruct the Heisenberg dynamics, we follow the general scheme and the discussion at the end of Sec.~\ref{ssec:QPVI_Ham_tau}.
First, viewing  $H,H',H''$ as polynomials in $q,p,t,\kappa,\epsilon,(m_f)_{f=1}^3$, we express the momentum $p$
\begin{equation}\label{p_V}
H_+=\frac12\Big(2H'+\big(m_1{-}m_2{-}m_3+\epsilon\big)t\Big) \Big(2H'+\big({-}m_1{+}m_2{-}m_3+\epsilon\big)t\Big)\big(1+p^{-1}).
\end{equation}
Via this formula, we have an implicit expression for the coordinate $q$ in terms of $H,H',H''$:
\begin{equation}\label{q_V}
-\dot{H}=qp(p+1)+\frac12(m_1{+}m_3{-}\epsilon)(2p+1)-\frac{m_2}2. 
\end{equation}
We define $q$ and $p$ by these formulas. With these definitions, we recover the canonical commutation relation \eqref{pq_eps} by commuting \eqref{p_V} with $\dot{H}$ given by \eqref{q_V} and using the commutator $[H_+,H']$ from \eqref{triple_V}.
Next, we express $H$ and $H''$ as polynomials in the newly introduced $q,p$, in addition to \eqref{q_V}.
An expression for $H_+$ is already provided by \eqref{p_V} and \eqref{q_V}.
Substituting \eqref{p_V} into the rewritten Hamiltonian form equation \eqref{Ham_V_twisted}, we obtain
\begin{equation}
(1+p^{-1})H_-=  \frac12\Big(2H'+\big(m_1{+}m_2{+}m_3+\epsilon\big)t\Big) \Big(2H'+\big({-}m_1{-}m_2{+}m_3+\epsilon\big)t\Big).
\end{equation}
Combining this relation with \eqref{q_V}, we obtain an expression for $H$ that coincides with \eqref{HV}. We also obtain an expression for $(\dot{H})'=H''/t-\dot{H}$, which is explicitly time-independent, just as $\dot{H}$ given by \eqref{q_V}. This shows that we have indeed reconstructed the original QPV Heisenberg dynamics.

\subsection{Quantum Painlev\'e III$_1$}
\label{ssec:QPIII1}
\paragraph{Hamiltonian.}
Making the autonomous rescaling (canonical transformation)
\begin{equation}\label{lim_V_III1}
q_{\V}=q_{\III_1}, \qquad p_{\V}=p_{\III_1}, \qquad \quad t_{\V}=\frac{t_{\III_1}}{m_3},
 \end{equation}
 and then sending $m_3\rightarrow\infty$, we obtain that the QPV Heisenberg dynamics \eqref{Heis} with Hamiltonian~\eqref{HV} degenerates to the Heisenberg dynamics defined by the Hamiltonian
 \begin{equation}\label{HIII1}
 H_{\III_1}(a_1^{\pm};q,p|\ln t)=t H_{\III_1}(a_1^{\pm};q,p|t)=qp(p{+}1)q+\frac{a^+_1{+}a_1^-}2\{p,q\}+a_1^-q+t p+\frac12t+\frac{(a^+_1{+}a_1^-)^2}4,
 \end{equation}
where we introduce the root variables $(a_i^{\pm})_{i=0}^1$ by
\begin{equation}
a_0^+=\kappa{-}m_1{-}m_2, \quad a_1^+=m_1{+}m_2, \qquad a_0^-=\kappa{+}m_2{-}m_1, \quad a_1^-=m_1{-}m_2 \quad \Rightarrow \quad a_0^{\pm}+a_1^{\pm}=\kappa.    
\end{equation}
More precisely, we have
$H_{\V}(\ln t_{\V})=H_{\III_1}(\ln t_{\III_1})+O(m_3^{-1})$ and $d\ln t_{\V}=d\ln t_{\III_1}$.
Under the limiting procedure \eqref{lim_V_III1}, the dimensions \eqref{dim_V} induce the following dimensions for the QPIII$_1$ variables: 
\begin{equation}\label{dim_III1}
[q]=1, \qquad [p]=0, \qquad \quad [t]=2, \qquad \quad  [a_i^{\pm}]=1,\, {\scriptstyle i=0,1} \qquad \quad [H(\ln t)]=2.
\end{equation}

The resulting Heisenberg dynamics with Hamiltonian \eqref{HIII1} is
equivalent to that defined by the homogeneous Hamiltonian $H^q_{\III}\big(\alpha|\ln t\big)$ in \cite[Sec. 2.4]{NY12} under the dictionary
\begin{equation}
p_{\textrm{\cite{NY12}}}=p+1, \qquad \qquad
(\alpha_0)_{\textrm{\cite{NY12}}}=a_1^-,\qquad (\alpha_2)_{\textrm{\cite{NY12}}}=a_1^+
\end{equation}
together with \eqref{NY12_eps} for $\epsilon_1,\epsilon_2$. The same dynamics is also equivalent to that of
\cite[Sec. 5]{N12} under a special dimension \eqref{dim_III1} rescaling \eqref{5to9} with the permutation $\alpha_0\leftrightarrow \alpha_2$.

\paragraph{Symmetries.}
The QPIII$_1$ symmetry group is the extended affine Weyl group $C_2\ltimes \left(C_2\ltimes W\left(A_1^{(1)}\right)\right)^2$, acting on the coordinates $q,p,t,(a_i^{\pm})_{i=0}^1$.
We present this action in Table~\ref{table:QPIII1_Backlund} with the corresponding diagram, following the encoding described in the QPVI case (see Sec.~\ref{ssec:QPVI_symm}). The extended affine Weyl group symmetry $C_2\ltimes W\left(C_2^{(1)}\right)$, presented in \cite[Definition 2.16]{NY12} is the subgroup of the above symmetry group, generated by $s_0^q=s_1^-,\, s_1^q=\pi^+\pi^-\sigma,\, s_2^q=s_1^+,\, \sigma^q=\sigma$. We have verified the relations for our extended symmetry group in the quantum case. Note that the corresponding Dynkin diagram is disconnected (with two connected components).
\begin{table}[H]
\begin{tabular}{M{11cm}M{4cm}}
\vspace{0.5cm}
\begin{tabular}{|c|c|c|c|c|c|c|} 
\hline
 & $q$ & $p$ & $t$ \\
\hline
\hline
 $s_0^+$ & $\pi^+ s_1^+\pi^+(q)$ & $\pi^+ s_1^+\pi^+(p)$ & \\ 
 \cline{1-3}
 $s_1^+$ & $q + a_1^+ (p{+}1)^{-1}$ & $p$ & $t$\\ 
 \cline{1-3}
 $\pi^+$ & $ -tq^{-1} $ & $q(pq+a_1^-)/t$ &  \\
 \hline
 \hline
 $s_0^-$ & $\pi^-s_1^-\pi^-(q)$ & $\pi^-s_1^-\pi^-(p)$ & \\ 
 \cline{1-3}
 $s_1^-$ & $q + a_1^- p^{-1}$ & $p$ & $t$\\ 
 \cline{1-3}
  $\pi^- $ & $ tq^{-1}$&$-q \left((p{+}1)q + a_1^+\right)/t-1$ & \\
  \hline
  \hline
 $\sigma$ & $-q$ & $-1-p$& $-t$\\
  \hline
\end{tabular}

&

\begin{tikzpicture}[elt/.style={circle,draw=black!100,thick, inner sep=0pt,minimum size=2mm},scale=1.5]
			\path 	(-1,0.5) 	node 	(a0) [elt] {}
			(1,0.5) 	node 	(a1) [elt] {}
			(-1,-0.5) node  	(b0) [elt] {}
			(1,-0.5) node  	(b1) [elt] {};
			
		     	\node at ($(b0.west) + (-0.2,-0.1)$) 	{$a_0^-$};
		     	\node at ($(a0.west) + (-0.2,0.1)$) 	{$a_0^+$};
		     	\node at ($(a1.east) + (0.2,0.1)$) 	{$a_1^+$};
		     	\node at ($(b1.east) + (0.2,-0.1)$) 	{$a_1^-$};
		     	
		    \draw[<->, dashed] (a0) to [bend right=40] node[fill=white]{$\sigma$} (b0);
		    \draw[<->, dashed] (a1) to [bend left=40] node[fill=white]{$\sigma$} (b1);
		    \draw[<->, dashed] (a0) to [bend left=40] node[fill=white]{$\pi^+$} (a1);		    
			\draw[<->, dashed] (b0) to [bend right=40] node[fill=white]{$\pi^-$} (b1);	
		    \draw [black,line width=1pt, double distance=.07cm] (a0) -- (a1);
		    \draw [black,line width=1pt, double distance=.07cm] (b0) -- (b1);    
			\end{tikzpicture} 
			\\
\end{tabular}
\caption{Action of the QPIII$_1$ symmetry group  $C_2\ltimes \left(C_2\ltimes W\left(A_1^{(1)}\right)\right)^2$.}
\label{table:QPIII1_Backlund}
\end{table}
For this extended affine Weyl group, the outer factor $C_2\langle\sigma\rangle$ is $\mathrm{Aut}(2A_1)$, while the inner factors $C_2\langle \pi^{\pm}\rangle$ are $\mathrm{Aut}\left(A_1^{(1)}\right)$ for the corresponding $A_1$-components of the Dynkin diagram.
Despite this disconnectedness, we reproduce the considerations of Sec.~\ref{ssec:QPVI_symm} for the QPVI symmetry group in the present QPIII$_1$ setting: 
\begin{itemize}
    \item The subgroup $\left(C_2\ltimes W\left(A_1^{(1)}\right)\right)^2$ is the $t$-preserving subgroup of the full group.
    \item The extended finite Weyl group $C_2\langle\sigma\rangle\ltimes W(A_1)^2$ is the subgroup of the autonomous symmetries. Under this subgroup, the one-form $H_{\III_1}(a_1^\pm;q,p,t)dt$ is invariant by \eqref{Ham_inv} ($\forall\sigma w: \sigma\in C_2\langle\sigma\rangle, \, w\in W(A_1)^2$).
    \item  Under the autonomization limit $t=t_0+\kappa t_{\textrm{aut}}, \kappa\rightarrow0$, the Hamiltonian \eqref{HIII1} becomes autonomous, with $t$ replaced by the constant $t_0$, while the lattice group $P_{A_1}^2$ acts on the root variables trivially. 
    \item The action of the finite Weyl group $W(A_1)^2$ on the mass variables $m_1,m_2$ is realized as $W(D_2)$, i.e. the group of signed permutations with an even number of sign changes. For the ring of $W(D_2)$-invariant polynomials, we take the basic invariants
    $w_2^{[2]},e_2^{[2]}$ (recall the notation \eqref{elem_def}).
    \item Preserving the dimensions \eqref{dim_III1} and the Weyl group action does not seem to be sufficient to rule out $\epsilon$-corrections to the Hamiltonian \eqref{HIII1}. However, the ordering ambiguity of the Hamiltonian \eqref{HIII1} is given by numerical linear combinations of $\epsilon pq$ and $\epsilon q$, which can be absorbed by an appropriate redefinition of masses $m_1, m_2$.
\end{itemize}

\paragraph{Equations for the Hamiltonian.}
Here we follow Sec.~ \ref{ssec:QPV} for the QPV case as well as the previous ones. We consider the Hamiltonian \eqref{HIII1} in time $\ln t$ along the Heisenberg  trajectories \eqref{Heis}, i.e.  
$H(t)=H_{\III_1}\big(q(t),p(t)|\ln t)$. As in the QPV case, we have the commutation relation \eqref{H_Hd_V}. Moreover, using \eqref{Heis} we obtain
\begin{equation}\label{Hder_III1}
\dot{H}=p+\frac12,\qquad -\kappa t\ddot{H}=2qp(p{+}1)+(m_1{+}\epsilon)(2p+1)-m_2.   
\end{equation}
Using these expressions, we obtain the second-order 
QPIII$_1$ Hamiltonian form equation (c.f. \eqref{Ham_V})
\begin{equation}
\label{Ham_III1}
\kappa^2(H''-H')^2-4H'(H-H')H'+t^2(H-H')+4e_2^{[2]}tH'=\big(w_2^{[2]}{-}\epsilon^2\big)t^2.
\end{equation}
Furthermore, computing $H'''$ as a polynomial in $q,p,t,\kappa,\epsilon,m_1,m_2$, we obtain the precursor equation (c.f. \eqref{prec_V})
\begin{equation}\label{prec_III1}
\kappa^2(H'''-2H''+H')+6(H')^2-2\left\{H,H'\right\}-\frac12t^2+2e_2^{[2]}t=0. 
\end{equation}
The Hamiltonian form equation \eqref{Ham_III1} depends on the $W(D_2)$- basic mass invariants $w_2^{[2]},e_2^{[2]}$ defined in \eqref{elem_def}, whereas the precursor \eqref{prec_III1} depends only on $e_2^{[2]}$. 
To integrate the precursor in $t$ and recover the Hamiltonian form equation, we complete the commutation relation \eqref{H_Hd_V} to a full set of relations among $H,H',H''$, analogously to the previous cases, by imposing (c.f. \eqref{H_Hdd_I}, \eqref{Hdd_Hd_I}) 
\begin{equation}\label{triple_III1}
[H,\kappa (H''{-}H')]=\epsilon\left(2\{H,H'\}-6(H')^2+\frac12t^2-2e_2^{[2]}t\right), \qquad [\kappa H'',H']=\frac12\epsilon\left(4(H')^2{-}t^2\right).
\end{equation}
After integration, we obtain the Hamiltonian form equation \eqref{Ham_III1}, in which $w_2^{[2]}$ is replaced by a $t$-constant central $W(D_2)$-invariant operator $C$. 

\paragraph{Tau form.}
The QPIII$_1$ tau form is analogous to that for QPV in Sec.~\ref{ssec:QPV}. We introduce the tau functions $\tau^{(1)},\tau^{(2)}$ by the universal formula \eqref{tau_universal}, with the time variable $t$ replaced by $\ln t$. Then we have the same first- and third-order Hirota equations \eqref{D13_V} as for the QPV. Using \eqref{H_Hd_V} together with its time derivative and, finally, the precursor \eqref{prec_III1}, we obtain the fourth-order Hirota equation (c.f. \eqref{D4_V}):
\begin{equation}\label{D4_III1}
D^4_{\epsilon_1,\epsilon_2}\left(\tau^{(1)},\tau^{(2)}\right)+2\epsilon_1\epsilon_2\big(D^2_{\epsilon_1,\epsilon_2}\left(\tau^{(1)},\tau^{(2)}\right)\big)'-(\epsilon_1\epsilon_2{+}\epsilon^2)D^2_{\epsilon_1,\epsilon_2}\left(\tau^{(1)},\tau^{(2)}\right)+\frac14\left(e_2^{[2]}{-}\frac14t\right)t\tau^{(1)}\tau^{(2)}=0.
\end{equation}
Conversely, assuming that $\tau^{(1)}$ and $\tau^{(2)}$ are invertible, one can recover the commutation relation \eqref{H_Hd_V} and the precursor equation \eqref{prec_III1} from the tau form equations \eqref{D13_V}, \eqref{D4_III1}, exactly as in Sec.~\ref{ssec:equivalence_I} for the QPI case. 
The tau form equations \eqref{D13_V}, \eqref{D4_III1} are related to the blowup equations \cite[(3.1),(3.4)]{BST25} (where they are also referred to as the QPIII$_1$ equation) by 
\begin{equation}\label{conn_III1}
\tau^{(1)}=e^{\frac{t}{4\epsilon_1(\epsilon_2{-}\epsilon_1)}}\tau^{(1)}_{\textrm{\cite{BST25}}}, \qquad \tau^{(2)}=e^{\frac{t}{4\epsilon_2(\epsilon_1{-}\epsilon_2)}}\tau^{(2)}_{\textrm{\cite{BST25}}}.    
\end{equation}
Finally, as in the previous cases, the tau functions $\tau^{(1)}$ and $\tau^{(2)}$ are invariant under the action of the extended finite Weyl group $C_2\langle\sigma\rangle\ltimes W(A_1)^2$.

\paragraph{Hamiltonian form.}
As in the QPV case, we first establish the Hamiltonian form of the QPIII$_1$ equation and then reconstruct the Heisenberg dynamics \eqref{Heis} from it. Using the commutation relation \eqref{H_Hd_V}, we can rewrite the Hamiltonian form equation \eqref{Ham_III1} as
\begin{equation}
\label{Ham_III1_twisted}
\kappa(H''-H')(\kappa(H''-H')+4\epsilon H')-(H-H')\left(4(H')^2{-}t^2\right)+4e_2^{[2]}tH'=\big(w_2^{[2]}{-}\epsilon^2\big)t^2.
\end{equation}
We refer to the equation \eqref{Ham_III1_twisted} together with the second commutation relation in \eqref{triple_III1} as the \textit{Hamiltonian form} of the QPIII$_1$ equation.
From this Hamiltonian form, one can recover the remaining two commutation relations of the full set \eqref{triple_III1}, \eqref{H_Hd_V} and the precursor \eqref{prec_III1}. Indeed, commuting \eqref{Ham_III1_twisted} with $H'$ and $H''$ yields the commutators $[H,H']$ and $[H,H'']$, respectively. Using the identity $[H,H']'=[H,H'']$, we recover the precursor \eqref{prec_III1}.

To reconstruct the Heisenberg dynamics, we define $q$ and $p$ by formulas \eqref{Hder_III1}, i.e.
\begin{equation}\label{pq_dict_III1}
p=t^{-1}H'-\frac12, \qquad q\big(t^2-4(H')^2\big)=2t\left(\kappa(H''{-}H')+2(m_1{+}\epsilon)H'-m_2\right).
\end{equation}
Then the canonical commutation relation \eqref{pq_eps} follows from the second relation in \eqref{triple_III1}. Further, the Hamiltonian form equation \eqref{Ham_III1_twisted}, together with the definition \eqref{Hder_III1}, yields the Hamiltonian expression \eqref{HIII1}. The same definition also implies that $\dot{H}$ and $(\dot{H})'$ are explicitly time-independent. This completes the reconstruction of the original QPIII$_1$ Heisenberg dynamics.

\subsection{Quantum Painlev\'e III$_2$}
\label{ssec:QPIII2}
\paragraph{Hamiltonian.}  Making the autonomous rescaling (canonical transformation)
\begin{equation}\label{lim_III1_III2}
 q_{\III_1}=\frac{q_{\III_2}}{m_2}, \qquad p_{\III_1}=p_{\III_2}m_2, \qquad \quad  t_{\III_1}=\frac{t_{\III_2}}{m_2},
 \end{equation}
 and then sending $m_2\rightarrow\infty$, we obtain that the QPIII$_1$ Heisenberg dynamics \eqref{Heis} with Hamiltonian~\eqref{HIII1} degenerates to the Heisenberg dynamics defined by the Hamiltonian
 \begin{equation}\label{HIII2}
 H_{\III_2}(a_1;q,p|\ln t)=t H_{\III_2}(a_1;q,p|t)=qp^2q+\frac{a_1}2\{p,q\}+t p-q+m_1^2,
 \end{equation}
 where we introduce the root variables $a_0,a_1$ by
\begin{equation}
a_0=\kappa-2m_1, \qquad a_1=2m_1  \quad \Rightarrow \quad a_0+a_1=\kappa.  
\end{equation} 
More precisely, we have
$H_{\III_1}(\ln t_{\III_1})=H_{\III_2}(\ln t_{\III_2})+O(m_2^{-1})$ and $d\ln t_{\III_1}=d\ln t_{\III_2}$.
Under the limiting procedure \eqref{lim_III1_III2}, the dimensions \eqref{dim_III1} induce the following dimensions for the QPIII$_2$ variables:
\begin{equation}\label{dim_III2}
[q]=2, \qquad [p]=-1, \qquad \quad [t]=3, \qquad \quad  [a_i]=1,\, {\scriptstyle i=0,1} \qquad \quad [H(\ln t)]=2.
\end{equation}
The negative dimension for $p$ makes no sense to discuss arbitrary polynomial $\epsilon$-corrections. However, the ordering ambiguity of the Hamiltonian \eqref{HIII2} is proportional to $\epsilon pq$, which can be absorbed by an appropriate redefinition of mass $m_1$.

The resulting Heisenberg dynamics with Hamiltonian \eqref{HIII2} is
equivalent to that defined by the homogeneous Hamiltonian $H^{D_7,q}_{\III}\big(\alpha|\ln t\big)$ in \cite[Sec. 2.5]{NY12} under the dictionary
\begin{equation}
q_{\textrm{\cite{NY12}}}=-q, \quad
p_{\textrm{\cite{NY12}}}=-p, \qquad  
t_{\textrm{\cite{NY12}}}=-t
\end{equation}
and \eqref{NY12_eps}.

\paragraph{Symmetries.}
The QPIII$_2$ symmetry group is the extended affine Weyl group $C_2 \ltimes W\left(A_1^{(1)}\right)$, acting on the coordinates $q,p,t,a_0,a_1$ according to \cite[Definition 2.21]{NY12}. We present this action in Table~\ref{table:QPIII2_Backlund} with the corresponding diagram, following the encoding described in the QPVI case (see Sec.~\ref{ssec:QPVI_symm}).
\begin{table}[H]
\begin{tabular}{M{10cm}M{4cm}}
\begin{tabular}{ |c|c|c|c|} 
\hline
 & $q$ & $p$ & $t$ \\
\hline
\hline
 $s_0$ & $q$ & $p - a_0 q^{-1} + tq^{-2}$ & $-t$\\  
 \hline
 $s_1$ & $-q-a_1p^{-1}+p^{-2}$ & $-p$ & $-t$\\  
 \hline
 $\pi$ & $-tp$ & $q/t$ & $-t$ \\
 \hline

\end{tabular}
&
\begin{tikzpicture}[elt/.style={circle,draw=black!100,thick, inner sep=0pt,minimum size=2mm},scale=1.75]
			\path 	(-1,0) 	node 	(a0) [elt] {}
			(1,0) 	node 	(a1) [elt] {};
		
		     	\node at ($(a0.west) + (-0.2,0)$) 	{$a_0$};
		     	\node at ($(a1.east) + (0.2,0)$) 	{$a_1$};

		    \draw[<->, dashed] (a0) to [bend left=40] node[fill=white]{$\pi$}(a1);
		    \draw [black,line width=1pt, double distance=.07cm] (a0) -- (a1);
			\end{tikzpicture} 
			\\

\end{tabular}
\caption{Action of the QPIII$_2$ symmetry group  $C_2\ltimes W(A_1^{(1)})$.}
\label{table:QPIII2_Backlund}
\end{table}
The factor $C_2\langle\pi\rangle$ is $\mathrm{Aut}\left(A_1^{(1)}\right)$.
We reproduce the considerations of Sec.~\ref{ssec:QPVI_symm} for the QPVI symmetry group in the present QPIII$_2$ setting: 
\begin{itemize}
    \item Unlike all other cases, the $t$-preserving subgroup of the full group is $P_{A_1}$, generated, e.g., by $\pi s_1$.
    \item The finite Weyl group $W(A_1)$ is the subgroup of the autonomous symmetries. Under this subgroup, the one-form $H_{\III_2}((a_i);q,p,t)dt$ is invariant by \eqref{Ham_inv} ($\forall\sigma w: \sigma=1, \, w\in W(A_1)$).
    \item  Under the autonomization limit $t=t_0+\kappa t_{\textrm{aut}}, \kappa\rightarrow0$, the Hamiltonian \eqref{HIII2} becomes autonomous, with $t$ replaced by the constant $t_0$, while the lattice group $P_{A_1}$ acts on the root variables trivially. 
    \item The finite Weyl group $W(A_1)$ is realized as the group that changes the sign of $m_1$. So the ring of $W(A_1)$-invariant polynomials is generated by $m_1^2$.
\end{itemize}

\paragraph{Equations for the Hamiltonian.}
Here we follow Sec.~ \ref{ssec:QPIII1} for the QPIII$_1$ case as well as the previous ones. We consider the Hamiltonian \eqref{HIII2} in time $\ln t$ along the Heisenberg trajectories \eqref{Heis}, i.e. $H(t)=H_{\III_2}\big(q(t),p(t)|\ln t)$. 
As in the QV and QPIII$_1$ cases, we have the commutation relation \eqref{H_Hd_V}. Moreover, using \eqref{Heis} we obtain
\begin{equation}\label{Hder_III2}
\dot{H}=p,\qquad -\kappa t\ddot{H}=2qp^2+2(m_1{+}\epsilon)p-1.   
\end{equation}
Using these expressions, we obtain the second-order QPIII$_2$ Hamiltonian form equation (c.f. \eqref{Ham_III1})
\begin{equation}\label{Ham_III2}
\kappa^2(H''-H')^2-4H'(H-H')H'+4m_1tH'=t^2.
\end{equation}
Furthermore, computing $H'''$ as a polynomial in $q,p,t,\kappa,\epsilon,m_1$, we obtain the precursor equation (c.f. \eqref{prec_III1})
\begin{equation}\label{prec_III2}
\kappa^2(H'''-2H''+H')+6(H')^2-2\left\{H,H'\right\}+2m_1t=0. 
\end{equation} 
If we rescale $t\mapsto t/m_1$, we see that the precursor \eqref{prec_III1} no longer depends on $m_1$, whereas the Hamiltonian form equation \eqref{Ham_III1} still depends on it through the right-hand side, via the combination $t^2/m_1^2$. To integrate the precursor in $t$ and recover the Hamiltonian form equation, we complete the commutation relation \eqref{H_Hd_V} to a full set of relations among $H,H',H''$ by imposing (c.f. \eqref{triple_III1}) 
\begin{equation}\label{triple_III2}
[H,\kappa(H''{-}H')]=\epsilon\left(2\{H,H'\}-6(H')^2-2m_1t\right), \qquad [\kappa H'',H']=2\epsilon(H')^2.
\end{equation}
After integration, we obtain the Hamiltonian form equation \eqref{Ham_III2} with right-hand side $Ct^2$, where $C$ is a $t$-constant central operator of dimension $0$. Under the above rescaling of $t$, this operator actually rescales mass $m_1$. 

\paragraph{Tau form.}
The QPIII$_2$ tau form is analogous to those for QPV, QPIII$_1$ in Secs.~\ref{ssec:QPV},\ref{ssec:QPIII1}, respectively. We introduce the tau functions $\tau^{(1)},\tau^{(2)}$ by the universal formula \eqref{tau_universal}, with the time variable $t$ replaced by $\ln t$. Then we have the same first- and third-order Hirota equations \eqref{D13_V} as for QPV, QPIII$_1$. Using  \eqref{H_Hd_V} together with its time derivative and, finally, the precursor \eqref{prec_III2}, we obtain the fourth-order Hirota equation (c.f. \eqref{D4_III1}):
\begin{equation}\label{D4_III2}
D^4_{\epsilon_1,\epsilon_2}\left(\tau^{(1)},\tau^{(2)}\right)+2\epsilon_1\epsilon_2\big(D^2_{\epsilon_1,\epsilon_2}\left(\tau^{(1)},\tau^{(2)}\right)\big)'-(\epsilon_1\epsilon_2{+}\epsilon^2)D^2_{\epsilon_1,\epsilon_2}\left(\tau^{(1)},\tau^{(2)}\right)+\frac14m_1t\tau^{(1)}\tau^{(2)}=0.   
\end{equation} 
Conversely, assuming that $\tau^{(1)}$ and $\tau^{(2)}$ are invertible, one can recover the commutation relation \eqref{H_Hd_V} and the precursor equation \eqref{prec_III2} from the tau form equations \eqref{D13_V}, \eqref{D4_III2}, exactly as in Sec.~\ref{ssec:equivalence_I} for the QPI case. 
The tau form equations \eqref{D13_V}, \eqref{D4_III2} coincide with the blowup equations \cite[(3.1),(3.5)]{BST25} (where they are also referred to as the QPIII$_2$ equation), i.e.
\begin{equation}\label{conn_III2}
\tau^{(1)}=\tau^{(1)}_{\textrm{\cite{BST25}}}, \qquad \tau^{(2)}=\tau^{(2)}_{\textrm{\cite{BST25}}}.    
\end{equation}
Finally, as in the previous cases, the tau functions $\tau^{(1)}$ and $\tau^{(2)}$ are invariant under the action of the finite Weyl group $W(A_1)$.

\paragraph{Hamiltonian form.}
Here we establish the Hamiltonian form of the QPIII$_2$ equation and reconstruct the Heisenberg dynamics \eqref{Heis} from it, exactly as in the QPIII$_1$ case. Using the commutation relation \eqref{H_Hd_V}, we can rewrite the Hamiltonian form equation \eqref{Ham_III2} as
\begin{equation}
\label{Ham_III2_twisted}
\kappa(H''-H')(\kappa(H''-H')+4\epsilon H')-4(H-H')(H')^2+4m_1tH'=t^2.
\end{equation}
We refer to the equation \eqref{Ham_III2_twisted} together with the second commutation relation in \eqref{triple_III2} as the \textit{Hamiltonian form} of the QPIII$_2$ equation.
From this Hamiltonian form, one can recover the remaining two commutation relations of the full set \eqref{triple_III2}, \eqref{H_Hd_V} and the precursor \eqref{prec_III2}, as in the QPIII$_1$ case.

To reconstruct the Heisenberg dynamics, we define $q$ and $p$ by formulas \eqref{Hder_III2}, i.e.
\begin{equation}\label{pq_dict_III2}
p=t^{-1}H', \qquad q(H')^2=-\frac12t\left(\kappa(H''{-}H')+2(m_1{+}\epsilon)H'-t\right).
\end{equation}
Then the canonical commutation relation \eqref{pq_eps} follows from the second relation in \eqref{triple_III2}. Further, the Hamiltonian form equation \eqref{Ham_III2_twisted}, together with the definition \eqref{Hder_III2}, yields the Hamiltonian expression \eqref{HIII2}. The same definition also implies that $\dot{H}$ and $(\dot{H})'$ are explicitly time-independent. This completes the reconstruction of the original QPIII$_2$ Heisenberg dynamics.
 
 \subsection{Quantum Painlev\'e III$_3$.}
 \paragraph{Hamiltonian.}
 Making the autonomous rescaling (canonical transformation)
\begin{equation}\label{lim_III2_III3}
 q_{\III_2}=q_{\III_3}, \qquad p_{\III_2}=p_{\III_3}-\frac{m_1}{q_{\III_3}}, \qquad  \quad t_{\III_2}=\frac{t_{\III_3}}{m_1},
 \end{equation}
and then sending $m_1\rightarrow\infty$, we obtain that the QPIII$_2$ Heisenberg dynamics \eqref{Heis} with Hamiltonian~\eqref{HIII2} degenerates to the Heisenberg dynamics defined by the Hamiltonian
\begin{equation}\label{HIII3}
 H_{\III_3}(q,p|\ln t)=t H_{\III_3}(q,p|t)=qp^2q-q-t q^{-1}.
\end{equation}
More precisely, we have
$H_{\III_2}(\ln t_{\III_2})=H_{\III_3}(\ln t_{\III_3})+O(m_1^{-1})$ and $d\ln t_{\III_2}=d\ln t_{\III_3}$.
Under the limiting procedure \eqref{lim_III2_III3}, the dimensions \eqref{dim_III2} induce the following dimensions for the QPIII$_3$ variables:
\begin{equation}
[q]=2, \qquad [p]=-1, \qquad \quad [t]=4,  \qquad \quad [H(\ln t)]=2.
\end{equation}
The negative dimension for $p$ and the $q^{-1}$-term in \eqref{HIII2} make no sense to discuss arbitrary polynomial $\epsilon$-corrections. However, the ordering ambiguity of the Hamiltonian \eqref{HIII3} is proportional to $\epsilon pq$, which can be absorbed by a (canonical) shift of $p$ by $\epsilon q^{-1}$ with an appropriate coefficient.

\paragraph{Symmetries.} The QPIII$_3$ symmetry group is $C_2$, the action of its generator is presented in Table~\ref{table:QPIII3_Backlund}.
\begin{table}[H]
\begin{center}
\begin{tabular}{ |c|c|c|c|c| } 
\hline
 & $q$ & $p$ & $t$ \\
\hline
\hline
 $\pi$ & $tq^{-1}$ & $q\left(\frac{\kappa}2-pq\right)/t$ & $t$\\ 
 \hline
\end{tabular}
\caption{Action of the QPIII$_3$ symmetry group $C_2$.}
\label{table:QPIII3_Backlund}
\end{center}
\end{table}

\paragraph{Equations for the Hamiltonian.}

Here we follow Sec.~\ref{ssec:QPIII2} for the QPIII$_2$ case as well as the previous ones. We consider the Hamiltonian \eqref{HIII3} in time $\ln t$ along the Heisenberg trajectories \eqref{Heis}, i.e. $H(t)=H_{\III_3}\big(q(t),p(t)|\ln t)$. 
As in the QV and QPIII$_{1,2}$ cases, we have the commutation relation \eqref{H_Hd_V}. Moreover, using \eqref{Heis} we obtain
\begin{equation}\label{Hder_III3}
\dot{H}=-q^{-1},\qquad \kappa t\ddot{H}=2p.   
\end{equation}
Using these expressions, we obtain the second-order QPIII$_3$ Hamiltonian form equation (c.f. \eqref{Ham_III2})
\begin{equation}\label{Ham_III3}
\kappa^2(H''-H')^2-4H'(H-H')H'+4tH'=0.
\end{equation}
Furthermore, computing $H'''$ as a polynomial in $q,q^{-1},p,t,\kappa,\epsilon$, we obtain the precursor equation (c.f. \eqref{prec_III2})
\begin{equation}\label{prec_III3}
\kappa^2(H'''-2H''+H')+6(H')^2-2\left\{H,H'\right\}+2t=0. 
\end{equation} 
The QPIII$_2$ and QPIII$_3$ precursors coincide after the time rescaling $t_{\III_2}= t_{\III_2}/m_1$ from the limiting procedure \eqref{lim_III2_III3}. Under this rescaling, the QPIII$_3$ Hamiltonian form equation \eqref{Ham_III3} is obtained as the $m_1\rightarrow\infty$ limit of that \eqref{Ham_III2} for QPIII$_2$. 
Moreover, the commutation relations \eqref{triple_III2} under the same time rescaling yield the corresponding relations for the QPIII$_3$ Hamiltonian:
\begin{equation}\label{triple_III3}
[H,\kappa(H''{-}H')]=\epsilon\left(2\{H,H'\}-6(H')^2-2t\right), \qquad [\kappa H'',H']=2\epsilon(H')^2.
\end{equation}
Thus, the integration constant operator $C$ arising when integrating the QPIII$_2$ precursor is precisely what distinguishes the QPIII$_2$ and QPIII$_3$ Hamiltonian-form equations: for the latter, one has $C=0$.

\paragraph{Tau form.}
As the corresponding precursors, the QPIII$_3$ tau form coincides with that for QPIII$_2$ up to the time rescaling $t_{\III_2}= t_{\III_2}/m_1$. We introduce the tau functions $\tau^{(1)},\tau^{(2)}$ by the universal formula \eqref{tau_universal}, with the time variable $t$ replaced by $\ln t$. Then we have the same first- and third-order Hirota equations \eqref{D13_V} as for QPV, QPIII$_{1,2}$, and the fourth-order Hirota equation is given by a rescaled version of \eqref{D4_III2}:
\begin{equation}\label{D4_III3}
D^4_{\epsilon_1,\epsilon_2}\left(\tau^{(1)},\tau^{(2)}\right)+2\epsilon_1\epsilon_2\big(D^2_{\epsilon_1,\epsilon_2}\left(\tau^{(1)},\tau^{(2)}\right)\big)'-(\epsilon_1\epsilon_2{+}\epsilon^2)D^2_{\epsilon_1,\epsilon_2}\left(\tau^{(1)},\tau^{(2)}\right)+\frac14t\tau^{(1)}\tau^{(2)}=0.   
\end{equation} 
Conversely, assuming that $\tau^{(1)}$ and $\tau^{(2)}$ are invertible, one can recover the commutation relation \eqref{H_Hd_V} and the precursor equation \eqref{prec_III2} from the tau form equations \eqref{D13_V}, \eqref{D4_III3}, exactly as in Sec.~\ref{ssec:equivalence_I} for the QPI case. 
The tau form equations \eqref{D13_V}, \eqref{D4_III3} coincide with the blowup equations \cite[(3.1),(3.6)]{BST25} (where they are also referred to as the QPIII$_3$ equation), i.e.
\begin{equation}\label{conn_III3}
\tau^{(1)}=\tau^{(1)}_{\textrm{\cite{BST25}}}, \qquad \tau^{(2)}=\tau^{(2)}_{\textrm{\cite{BST25}}}.    
\end{equation}

\paragraph{Hamiltonian form.}
As before, the Hamiltonian form of the QPIII$_3$ equation coincides with that of QPIII$_2$ up to the time rescaling $t_{\III_2}= t_{\III_2}/m_1$ and the limit $m_1\rightarrow\infty$. Hence it is given by the equation
\begin{equation}
\label{Ham_III3_twisted}
\kappa(H''-H')(\kappa(H''-H')+4\epsilon H')-4(H-H')(H')^2+4tH'=0,
\end{equation}
together with the second commutation relation in \eqref{triple_III3}; the remaining relations are recovered from these, exactly as in the QPIII$_2$ case. 

To reconstruct the Heisenberg dynamics, we define $q$ and $p$ by formulas \eqref{Hder_III3}.
Then the canonical commutation relation \eqref{pq_eps} follows from the second relation in \eqref{triple_III3}. Further, the Hamiltonian form equation \eqref{Ham_III3_twisted}, together with the definition \eqref{Hder_III3}, yields the Hamiltonian expression \eqref{HIII3}. The same definition also implies that $\dot{H}$ and $(\dot{H})'$ are explicitly time-independent. This completes the reconstruction of the original QPIII$_3$ Heisenberg dynamics.

\subsection{Quantum Painlev\'e IV}
\label{ssec:QPIV}
\paragraph{Hamiltonian.}
Making the autonomous canonical transformation
\begin{multline}\label{lim_V_IV}
 q_{\V}=q_{\IV}M^{\frac12}, \qquad p_{\V}=\frac{p_{\IV}}{M^{\frac12}}, \qquad \quad t_{\V}=M\left(1-\frac{t_{\IV}}{M^{\frac12}}\right)   \\ \textrm{with} \qquad 
m_f=\frac12M+\boldsymbol{m}_f,\,\, {\scriptstyle f=1,2,3} \quad \textrm{such that} \quad \sum_{f=1}^3\boldsymbol{m}_f=0,
 \end{multline}
 and then sending $M\rightarrow\infty$, we obtain that the QPV Heisenberg dynamics \eqref{Heis} with Hamiltonian~\eqref{HV} degenerates to the Heisenberg dynamics defined by the Hamiltonian 
 \begin{equation}\label{HIV}
 H_{\IV}(a_1,a_2;q,p|t)=pqp-qpq-\frac12t(pq{+}qp)-a_1p-a_2q+\frac{a_1{-}a_2}3t, 
 \end{equation}
  where we introduce the root variables $(a_i)_{i=0}^2$ by
\begin{equation}
a_0=\kappa+\boldsymbol{m}_2{-}\boldsymbol{m}_3, \qquad a_1=\boldsymbol{m}_3{-}\boldsymbol{m}_1, \qquad a_2=\boldsymbol{m_1}{-}\boldsymbol{m}_2 \quad \Rightarrow \quad \sum_{i=0}^2a_i=\kappa. 
\end{equation}
More precisely, we have
\begin{equation}
H_{\V}(\ln t_{\V})-\frac13w_2^{[3]}-\frac16e_1^{[3]}t_{\V}=-H_{\IV}M^{\frac12}+O(1), \qquad \quad  d\ln t_{\V}=-\frac{dt_{\IV}}{M^{\frac12}}\left(1-\frac{t_{\IV}}{M^{\frac12}}\right)^{-1}.  
\end{equation}
Under the limiting procedure \eqref{lim_V_IV}, the dimensions \eqref{dim_V} induce the following dimensions for the QPIV variables:
\begin{equation}\label{dim_IV}
[q]=[p]=[t]=\frac12, \qquad \quad  [a_i]=1,\, {\scriptstyle i=0,1,2} \qquad \quad [H]=\frac32.
\end{equation}
Then, the possible nontrivial polynomial $\epsilon$-corrections to the homogeneous Hamiltonian are numerical linear combinations of $\epsilon q$ and $\epsilon p$, which can be absorbed by an appropriate redefinition of masses $\boldsymbol{m}_f$. 

The resulting Heisenberg dynamics with Hamiltonian \eqref{HIV} is
equivalent to that defined by the homogeneous Hamiltonian $H^q_{\IV}(\alpha)$ in \cite[Sec. 2.3]{NY12} under the dictionary \eqref{NY12_eps}. The same dyanmics is also equivalent to that of
\cite[Sec. 4]{N12} under a special dimension \eqref{dim_IV} rescaling \eqref{5to9}.

\paragraph{Symmetries.}
The QPIV symmetry group is the extended affine Weyl group $\mathrm{Dic}_3 \ltimes W\left(A_2^{(1)}\right)$, with a further central extension, acting on the coordinates $q,p,t,(a_i)_{i=0}^2$ according to \cite[Definition 2.11]{NY12}\footnote{The dicyclic structure is not explicitly reflected in \cite[Sec. 2.3]{NY12}, although the central element does appear there.}. We present this action in Table~\ref{table:QPIV_Backlund} with the corresponding diagram, following the encoding described in the QPVI case (see Sec.~\ref{ssec:QPVI_symm}).
\begin{table}[H]
\begin{tabular}{M{11cm}M{4cm}}
\vspace{-2.5cm}
\begin{tabular}{ |c|c|c|c|c|c| } 
\hline
 & $q$ & $p$& $t$ \\
\hline
\hline
 $s_0$ & $q + a_0(p{-}q{-}t)^{-1}$ & $p + a_0 (p{-}q{-}t)^{-1}$ & \\ 
 \cline{1-3}
 $s_1$ & $q$ & $p - a_1 q^{-1}$ & \parbox{0.25cm}{\vspace{0.5cm} $t$}\\ 
 \cline{1-3}
 $s_2$ & $q +a_2 p^{-1}$ & $p$ & \\
 \cline{1-3}
 $\pi$ & $-p$ & $-(p{-}q{-}t)$ & \\ 
\hline
\hline
 $\sigma_{12}$ & $-\text{i}p$ & $-\text{i}q$ & $\text{i}t$\\
 \hline
\end{tabular}
&
\begin{tikzpicture}[elt/.style={circle,draw=black!100,thick, inner sep=0pt,minimum size=2mm},scale=1.25]
			\path 	(-1,0) 	node 	(a1) [elt] {}
			(1,0) 	node 	(a2) [elt] {}
			(0, 1.73205) 	node 	(a0) [elt] {};
		
		     	\node at ($(a0.north) + (0,0.2)$) 	{$a_0$};
		     	\node at ($(a1.west) + (-0.2,-0.1)$) 	{$a_1$};
		     	\node at ($(a2.east) + (0.2,-0.1)$) 	{$a_2$};

		    \draw[<->, dashed] (a1) to [bend left=40] node[fill=white]{$\sigma_{12}$}(a2);
		    \draw[->, dashed] (a0) to [bend right=40] node[fill=white]{$\pi$}(a1);
		    \draw[->, dashed] (a1) to [bend right=40] node[fill=white]{$\pi$}(a2);
		    \draw[->, dashed] (a2) to [bend right=40] node[fill=white]{$\pi$}(a0);
		    \draw [black,line width=1pt] (a0) -- (a1);
		    \draw [black,line width=1pt] (a1) -- (a2);
		    \draw [black,line width=1pt] (a2) -- (a0);
			\end{tikzpicture} 
			\\

\end{tabular}

\caption{Action of the QPIV symmetry group $\mathrm{Dic}_3\ltimes W\left(A_2^{(1)}\right)$.}
\label{table:QPIV_Backlund}
\end{table}
The dicyclic group $\mathrm{Dic}_3$ is a nontrivial central extension of the Dynkin diagram automorphism group $\mathrm{Aut}\left(A_2^{(1)}\right)=S_3$ by the element $\sigma_{12}^2$. This element acts trivially on the root variables, but changes the signs of $q,p,t$:
\begin{equation}\label{C2_IV}
q\mapsto -q, \qquad p\mapsto -p, \qquad \quad  t\mapsto -t.
\end{equation}
This central symmetry is analogous to the $C_5$ symmetry \eqref{C5_I} of the QPI equation and arises from the branching of the dimension scaling associated with the fractional dimensions \eqref{dim_IV}.
Under the decomposition of the dicyclic group as a semidirect product, this central extension arises in the finite automorphism 
group $\mathrm{Aut}(A_2)$, namely
\begin{equation}
\mathrm{Dic}_3=\mathrm{Aut}\left(A_2^{(1)}\right)=\underbrace{C_4\langle\sigma_{12}\rangle}_{C_2\cdot\mathrm{Aut}(A_2)}\ltimes \underbrace{C_3\langle\pi\rangle}_{P_{A_2}/Q_{A_2}}.    
\end{equation}

We reproduce the considerations of Sec.~\ref{ssec:QPVI_symm} for the QPVI symmetry group in the present QPIV setting, taking the central extension into account:
\begin{itemize}
    \item The subgroup $C_3\ltimes W\left(A_2^{(1)}\right)$ is the $t$-preserving subgroup of the full group.
    \item The extended finite Weyl group $C_4\ltimes W(A_2)$, where $C_4=C_2\cdot \mathrm{Aut}(A_2)$, is the subgroup of the autonomous symmetries. Under this subgroup, the one-form $H_{\IV}((a_i);q,p,t)dt$ is invariant by \eqref{Ham_inv} ($\forall\sigma w: \sigma\in C_4, \, w\in W(A_2)$).
    \item  Under the autonomization limit $t=t_0+\kappa t_{\textrm{aut}}, \kappa\rightarrow0$, the Hamiltonian \eqref{HIV} becomes autonomous, with $t$ replaced by the constant $t_0$, while the lattice group $P_{A_2}$ acts on the root variables trivially (in addition to the central extension). 
    \item The finite Weyl group $W(A_2)$ is realized as the permutation group of masses $(\boldsymbol{m}_f)_{f=0}^2$ subject to the constraint $\sum_f\boldsymbol{m_f}=0$. For the ring of $W(A_2)$-invariant polynomials, we take the basic invariants $\boldsymbol{e}_1\equiv0,\boldsymbol{e}_2, \boldsymbol{e}_3$, defined in \eqref{elem_def} as the elementary symmetric polynomials in the three dependent masses.
\end{itemize}

\paragraph{Equations for the Hamiltonian.}
Following the previous cases, we consider the Hamiltonian \eqref{HIV} along the Heisenberg  trajectories \eqref{Heis}, i.e.  
$H(t)=H_{\IV}\big(q(t),p(t)|t)$. As in the QPI case, we have the commutation relation \eqref{H_Hd_I}. Moreover, using \eqref{Heis} we obtain
\begin{equation}\label{Hder_IV}
-\dot{H}=qp+\boldsymbol{m}_1+\frac{\epsilon}2, \qquad -\kappa\ddot{H}=pqp+qpq+(\boldsymbol{m_1}{-}\boldsymbol{m}_3)p+(\boldsymbol{m_1}{-}\boldsymbol{m}_2)q.
\end{equation}
Using these expressions, we obtain the second-order QPIV Hamiltonian form equation (c.f. \eqref{Ham_V})
\begin{equation}\label{Ham_IV}
\kappa^2\ddot{H}^2-(H-t\dot{H})^2+4\dot{H}^3+(4\boldsymbol{e}_2{+}3\epsilon^2)\dot{H}=-4\boldsymbol{e}_3
\end{equation}
Furthermore, computing $\dddot{H}$ as a polynomial in $q,p,t,\kappa,\epsilon,(\boldsymbol{m}_f)_{f=1}^3$, we obtain the precursor equation (c.f. \eqref{prec_V})
\begin{equation}\label{prec_IV}
\kappa^2\dddot{H}+6\dot{H}^2+\big(H-t\dot{H}\big)t+2\boldsymbol{e}_2{+}\frac12\epsilon^2=0.
\end{equation} 
The Hamiltonian form equation \eqref{Ham_IV} depends on the $W(A_2)$- basic mass invariants $\boldsymbol{e}_2,\boldsymbol{e}_3$ defined in \eqref{elem_def} on masses $(\boldsymbol{m}_f)_{f=1}^3$ subject to  $\sum_f\boldsymbol{m}_f=0$. The precursor \eqref{prec_IV} depends only on $\boldsymbol{e}_2$ and not on the highest-dimension basic invariant $\boldsymbol{e}_3$.
To integrate the precursor in $t$ and recover the Hamiltonian form equation, we complete the commutation relation \eqref{H_Hd_I} to a full set of relations among $H,\dot{H},\ddot{H}$, analogously to the previous cases. These relations can be written as (c.f. \eqref{triple_V})
\begin{equation}\label{triple_IV}
[H_\pm,\dot{H}]=\pm\epsilon H_\pm, \qquad [H_+,H_-]=\epsilon\left(12\dot{H}^2+4\boldsymbol{e}_2{+}\epsilon^2\right),
\end{equation}
where we introduce the combinations (c.f. \eqref{comb_V})
\begin{equation}
H_\pm=H-t\dot{H}\pm\kappa\ddot{H}.    
\end{equation}
After integration, we obtain the Hamiltonian form equation \eqref{Ham_IV}, in which $\boldsymbol{e}_3$ is replaced by a $t$-constant central $W(A_2)$-invariant operator $C$.

\paragraph{Tau form.}
The QPIV tau form is analogous to that for QPI in Sec.~\ref{ssec:tau_I}. We introduce the tau functions $\tau^{(1)},\tau^{(2)}$ by the universal formula \eqref{tau_universal}. Then we have the same first- and third-order Hirota equations
\begin{equation}\label{D13_IV}
D^1_{\epsilon_1,\epsilon_2}\left(\tau^{(1)},\tau^{(2)}\right)=0, \qquad
D^3_{\epsilon_1,\epsilon_2}\left(\tau^{(1)},\tau^{(2)}\right)=0
\end{equation}
as for QPI, i.e. the first and second equations in \eqref{QPI_tau}.
Using \eqref{H_Hd_I} together with its time derivative and, finally, the precursor \eqref{prec_IV}, we obtain the fourth-order Hirota equation (c.f. the third equation of \eqref{QPI_tau}):
\begin{equation}\label{D4_IV}
D^4_{\epsilon_1,\epsilon_2}\left(\tau^{(1)},\tau^{(2)}\right)-\frac14t^2D^2_{\epsilon_1,\epsilon_2}\left(\tau^{(1)},\tau^{(2)}\right)-\frac14t\epsilon_1\epsilon_2\frac{d}{dt}\left(\tau^{(1)}\tau^{(2)}\right)+\frac14\left(\boldsymbol{e}_2{+}\frac{\epsilon^2}4\right)\tau^{(1)}\tau^{(2)}=0.
\end{equation}
Conversely, assuming that $\tau^{(1)}$ and $\tau^{(2)}$ are invertible, one can recover the commutation relation \eqref{H_Hd_I} and the precursor equation \eqref{prec_IV} from the tau form equations \eqref{D13_IV}, \eqref{D4_IV}, exactly as in Sec.~\ref{ssec:equivalence_I} for the QPI case. 
The tau form equations \eqref{D13_IV}, \eqref{D4_IV} coincide with the blowup equations \cite[(3.10),(3.11)]{BST25} (where they are also referred to as the QPIV equation), i.e. 
\begin{equation}\label{conn_IV}
\tau^{(1)}=\tau^{(1)}_{\textrm{\cite{BST25}}}, \qquad \tau^{(2)}=\tau^{(2)}_{\textrm{\cite{BST25}}}.    
\end{equation}
Finally, as in the previous cases, the tau functions $\tau^{(1)}$ and $\tau^{(2)}$ are invariant under the action of the extended finite Weyl group $C_4\ltimes W(A_2)$.

\paragraph{Hamiltonian form.}
Here we establish the Hamiltonian form of the QPIV equation and reconstruct the Heisenberg dynamics \eqref{Heis} from it, analogously to the QPV case of Sec.~\ref{ssec:QPV}.
Using the last commutation relation of \eqref{triple_IV}, we can rewrtite the Hamiltonian form equation \eqref{Ham_IV} as (c.f. \eqref{Ham_V_twisted})
\begin{equation}\label{Ham_IV_twisted}
H_+H_-=4\prod_{f=1}^3\left(\dot{H}{+}\boldsymbol{m}_f{+}\frac{\epsilon}2\right).
\end{equation}
We refer to the equation \eqref{Ham_IV_twisted} together with one of the first commutation relations of \eqref{triple_IV} as the \textit{Hamiltonian form} of the QPIV equation. Below, for definiteness, we choose the relation with the "$+$" sign. From this Hamiltonian form one can recover the remaining two commutation relations in \eqref{triple_IV} and the precursor \eqref{prec_IV}, analogously to the QPV case.

To reconstruct the Heisenberg dynamics, viewing $H,\dot{H},\ddot{H}$ as polynomials in $q,p,t,\kappa,\epsilon,(\boldsymbol{m}_f)_{f=1}^3$, we define the coordinate $q$ and momentum $p$:
\begin{equation}\label{pq_dict_IV}
H_+=2\left(\dot{H}{+}\boldsymbol{m}_2{+}\frac{\epsilon}2\right)q, \qquad H_-=-2p\left(\dot{H}{+}\boldsymbol{m}_3{+}\frac{\epsilon}2\right).
\end{equation}
Combining these definitions with the Hamiltonian form equation \eqref{Ham_IV_twisted}, we obtain immediately the first formula of \eqref{Hder_IV}, i.e. the expression for $\dot{H}$. Using it together with the definitions \eqref{pq_dict_IV}, we obtain the expression \eqref{HIV} and the the second formula of \eqref{Hder_IV}, i.e. the expression for $\ddot{H}$. We recover the canonical commutation relation \eqref{pq_eps} by commuting any of the definitions \eqref{pq_dict_IV} with $\dot{H}$, given by \eqref{Hder_IV}, and using the first commutators of \eqref{triple_IV}. Finally, we see that obtained $\dot{H}$ and $\ddot{H}$ are explicitly time-independent. This completes the reconstruction of the original QPIV Heisenberg dynamics.
 
\subsection{Quantum Painlev\'e II}
\paragraph{Hamiltonian.}
Making the autonomous canonical transformation
\begin{multline}\label{lim_III1_II}
q_{\III_1}=-2M\left(1-\frac{q_{\II}}{M^{\frac13}}\right), \qquad  p_{\III_1}=-1+\frac{p_{\II}}{2M^{\frac23}}, \qquad \quad  t_{\III_1}=-4M^2\left(1+\frac{t_{\II}}{2M^{\frac23}}\right) \\  \textrm{with} \qquad m_1=\boldsymbol{m}-2M, \quad m_2=\boldsymbol{m}+2M,
\end{multline}
and then sending $M\rightarrow\infty$, we obtain that the QPIII$_1$ Heisenberg dynamics \eqref{Heis} with Hamiltonian~\eqref{HIII1} degenerates to the Heisenberg dynamics defined by the Hamiltonian
\begin{equation}\label{HII}
H_{\II}(a_1;q,p|t)=\frac12p^2-qpq-\frac12t p-a_1q, 
\end{equation}
where we introduce the root variables $a_0, a_1$ by
\begin{equation}
a_0=\kappa-2\boldsymbol{m}, \qquad a_1=2\boldsymbol{m} \quad \Rightarrow \quad a_0+a_1=\kappa.
\end{equation}
More precisely, we have
\begin{equation}
H_{\III_1}(\ln t_{\III_1})-\frac{w_2^{[2]}{-}2e_2^{[2]}}4+\frac12t_{\III_1}=2H_{\II}M^{\frac23}+O\left(M^{\frac13}\right), \qquad \quad d\ln t_{\III_1}=\frac{dt_{\II}}{2M^{\frac23}}\left(1+\frac{t_{\II}}{2M^{\frac23}}\right)^{-1}. 
\end{equation}
Alternatively, making the autonomous canonical transformation
\begin{multline}\label{lim_IV_II}
q_{\IV}=\frac12M^{\frac12}\left(1+\frac{2q_{\II}}{M^{\frac13}}\right), \qquad p_{\IV}=\frac{p_{\II}}{M^{\frac16}}, \qquad \quad  t_{\IV}=-M^{\frac12}\left(1-\frac{t_{\II}}{M^{\frac23}}\right)\\
\textrm{with} \qquad \boldsymbol{m}_1=-\frac{M}{12}+\boldsymbol{m}, \quad \boldsymbol{m}_2=-\frac{M}{12}-\boldsymbol{m}, \quad 
\boldsymbol{m}_3=\frac{M}6, 
\end{multline}
and then sending $M\rightarrow\infty$, we obtain that QPIV with Hamiltonian~\eqref{HIV} degenerates to QPII with Hamiltonian~\eqref{HII}, namely
\begin{equation}
H_{\IV}-\frac{\boldsymbol{m}_3}2t_{\IV}=H_{\II}M^{\frac16}+O\left(\frac1{M^{\frac16}}\right), \qquad \quad dt_{\IV}=\frac{dt_{\II}}{M^{\frac16}}.   
\end{equation}
Under the limiting procedure \eqref{lim_III1_II} or \eqref{lim_IV_II}, the dimensions \eqref{dim_III1} or \eqref{dim_IV} respectively induce the following dimensions for the QPII variables:
\begin{equation}\label{dim_II}
[q]=\frac13, \qquad [p]=[t]=\frac23, \qquad \quad  [a_i]=1,\, {\scriptstyle i=0,1} \qquad \quad [H]=\frac43.
\end{equation}
Then, the possible nontrivial polynomial $\epsilon$-corrections to the homogeneous Hamiltonian are numerically proportional to $\epsilon q$, which can be absorbed by an appropriate redefinition of mass $\boldsymbol{m}$. 

The resulting Heisenberg dynamics with Hamiltonian \eqref{HII} is
equivalent to that defined by the homogeneous Hamiltonian $H^q_{\II}(\alpha)$ in \cite[Sec. 2.6]{NY12} under the dictionary
\eqref{NY12_eps}. The same dynamics is also equivalent to that of
\cite[Sec. 6]{N12} under a special dimension \eqref{dim_II} rescaling \eqref{5to9}.

\paragraph{Symmetries.}
The QPII symmetry group is the extended affine Weyl group $C_3\times \left(C_2\ltimes W\left(A_1^{(1)}\right)\right)$ with an additional central factor $C_3$, acting on the coordinates $q,p,t,a_0,a_1$ according to \cite[Definition 2.26]{NY12}\footnote{The central extension by $C_3$ is not described in \cite[Sec. 2.6]{NY12}}. We present this action in Table~\ref{table:QPII_Backlund} with the corresponding diagram, following the encoding described for the QPVI case (see Sec.~\ref{ssec:QPVI_symm}).
\begin{table}[H]
\begin{tabular}{M{10cm}M{4cm}}
	\begin{tabular}{ |c|c|c|c|c|c| } 
\hline
 & $q$ & $p$  & $t$ \\
\hline
\hline
 $s_1$ & $q + a_1 p^{-1}$ & $p$ & \parbox{0.25cm}{\vspace{0.5cm} $t$}\\ 
 \cline{1-3}
 $\pi$ & $-q$ & $-(p{-}2q^2{-}t)$ & \\
 \hline
 \hline
  $\sigma$ & $e^{-2\pi\ri/3}q$ & $e^{2\pi\ri/3}p$ & $e^{2\pi\ri/3}t$ \\
  \hline
\end{tabular}
&
\begin{tikzpicture}[elt/.style={circle,draw=black!100,thick, inner sep=0pt,minimum size=2mm},scale=1.75]
			\path 	(-1,0) 	node 	(a0) [elt] {}
			(1,0) 	node 	(a1) [elt] {};
		
		     	\node at ($(a0.west) + (-0.2,0)$) 	{$a_0$};
		     	\node at ($(a1.east) + (0.2,0)$) 	{$a_1$};

		    \draw[<->, dashed] (a0) to [bend left=40] node[fill=white]{$\pi$}(a1);
		    \draw [black,line width=1pt, double distance=.07cm] (a0) -- (a1);
			\end{tikzpicture} 
			\\

\end{tabular}

\caption{Action of the QPII symmetry group $C_3\times \left(C_2\ltimes W\left(A_1^{(1)}\right)\right)$.}
\label{table:QPII_Backlund}
\end{table}
The central group $C_3$ is generated by $\sigma$, which acts trivially on the root variables. This central symmetry is analogous to the $C_5$ symmetry \eqref{C5_I} of the QPI equation and to the $C_2$ symmetry \eqref{C2_IV} of the QPIV equation.
As above, it arises from the branching of the dimension scaling associated with the fractional dimensions \eqref{dim_II}. The other cyclic group $C_2$ is the affine Dynkin diagram automorphism group $\mathrm{Aut}\left(A^{(1)}\right)$, while the corresponding finite diagram automorphism group is trivial.

We reproduce the considerations of Sec.~\ref{ssec:QPVI_symm} for the QPVI symmetry group in the present QPII setting, taking the trivial central extension into account: 
\begin{itemize}
    \item The subgroup $C_2\ltimes W\left(A_1^{(1)}\right)$ is the $t$-preserving subgroup of the full group.
    \item The $C_3$-multiplied finite Weyl group $C_3\times W(A_1)$ is the subgroup of the autonomous symmetries. Under this subgroup, the one-form $H_{\II}(a_1;q,p,t)dt$ is invariant by  \eqref{Ham_inv} ($\forall\sigma w: \sigma\in C_3, \, w\in W(A_1)$).
    \item  Under the autonomization limit $t=t_0+\kappa t_{\textrm{aut}}, \kappa\rightarrow0$, the Hamiltonian \eqref{HII} becomes autonomous, with $t$ replaced by the constant $t_0$, while the lattice group $P_{A_1}$ acts on the root variables trivially (in addition to the central extension). 
    \item The finite Weyl group $W(A_1)$ is realized as the group that changes the sign of $\boldsymbol{m}$. So the ring of $W(A_1)$-invariant polynomials is generated by $\boldsymbol{m}^2$.
\end{itemize}

\paragraph{Equations for the Hamiltonian.}
Following the previous cases, we consider the Hamiltonian \eqref{HII} along the Heisenberg trajectories \eqref{Heis}, i.e.  
$H(t)=H_{\II}\big(q(t),p(t)|t)$. As in the QPI case, we have the commutation relation \eqref{H_Hd_I}. Moreover, using \eqref{Heis} we obtain
\begin{equation}\label{Hder_II}
\dot{H}=-\frac12p,\qquad -\kappa \ddot{H}=qp+\boldsymbol{m}+\frac{\epsilon}2.
\end{equation}
Using these expressions, we obtain the second-order QPII Hamiltonian form equation (c.f. \eqref{Ham_IV}, \eqref{Ham_III1})
\begin{equation}\label{Ham_II}
\kappa^2\ddot{H}^2+2t\dot{H}^2-\{H,\dot{H}\}+4\dot{H}^3=\boldsymbol{m}^2-\frac14\epsilon^2.
\end{equation}
Furthermore, computing $\dddot{H}$ as a polynomial in $q,p,t,\kappa,\epsilon,\boldsymbol{m}$, we obtain the precursor equation (c.f. \eqref{prec_IV}, \eqref{prec_III1})
\begin{equation}\label{prec_II}
\kappa^2\dddot{H}+6\dot{H}^2+2t\dot{H}-H=0. 
\end{equation}
The Hamiltonian form equation \eqref{Ham_II} depends on the $W(A_1)$-invariant mass square $\boldsymbol{m}^2$, whereas the precursor \eqref{prec_II} does not.
To integrate the precursor in $t$ and recover the Hamiltonian form equation, we complete the commutation relation \eqref{H_Hd_I} to a full set of relations among $H,\dot{H},\ddot{H}$, analogously to the previous cases, by imposing (c.f. \eqref{H_Hdd_I}, \eqref{Hdd_Hd_I}) 
\begin{equation}\label{triple_II}
[H,\kappa \ddot{H}]=\epsilon\left(H-2t\dot{H}-6\dot{H}^2\right), \qquad [\kappa \ddot{H},\dot{H}]=\epsilon\dot{H}.
\end{equation}
After integration, we obtain the Hamiltonian form equation \eqref{Ham_II}, in which $\boldsymbol{m}^2$ is replaced by a $t$-constant central $W(A_1)$-invariant operator $C$. 

\paragraph{Tau form.}
The QPII tau form is analogous to those for QPI and QPIV in Secs.~\ref{ssec:tau_I} and \eqref{ssec:QPIV}, respectively. We introduce the tau functions $\tau^{(1)},\tau^{(2)}$ by the universal formula \eqref{tau_universal}. Then we have the same first- and third-order Hirota equations \eqref{D13_IV} as for QPIV and QPI. Using \eqref{H_Hd_I} together with its time derivative and, finally, the precursor \eqref{prec_II}, we obtain the fourth-order Hirota equation (c.f. \eqref{D4_IV} and the third equation of \eqref{QPI_tau}):
\begin{equation}\label{D4_II}
D^4_{\epsilon_1,\epsilon_2}\left(\tau^{(1)},\tau^{(2)}\right)+\frac12t D^2_{\epsilon_1,\epsilon_2}\left(\tau^{(1)},\tau^{(2)}\right)+\frac14\epsilon_1\epsilon_2\frac{d}{dt}\left(\tau^{(1)}\tau^{(2)}\right)=0.    
\end{equation}
Conversely, assuming that $\tau^{(1)}$ and $\tau^{(2)}$ are invertible, one can recover the commutation relation \eqref{H_Hd_I} and the precursor equation \eqref{prec_II} from the tau form equations \eqref{D13_IV}, \eqref{D4_II}, exactly as in Sec.~\ref{ssec:equivalence_I} for the QPI case. 
The tau form equations \eqref{D13_IV}, \eqref{D4_II} coincide with the blowup equations \cite[(3.10),(3.15)]{BST25} (where they are also referred to as the QPII equation), i.e.
\begin{equation}\label{conn_II}
\tau^{(1)}=\tau^{(1)}_{\textrm{\cite{BST25}}}, \qquad \tau^{(2)}=\tau^{(2)}_{\textrm{\cite{BST25}}}.    
\end{equation}
Finally, as in the previous cases, the tau functions $\tau^{(1)}$ and $\tau^{(2)}$ are invariant under the action of the centrally extended finite Weyl group $C_3\times W(A_1)$.

\paragraph{Hamiltonian form.}
Here we establish the Hamiltonian form of the QPII equation and reconstruct the Heisenberg dynamics \eqref{Heis} from it, analogously to the previous cases. Using the commutation relation \eqref{H_Hd_I}, we can rewrite the Hamiltonian form equation \eqref{Ham_II} as
\begin{equation}\label{Ham_II_twisted}
\kappa\ddot{H}(\kappa\ddot{H}+\epsilon)+2t\dot{H}^2-2H\dot{H}+4\dot{H}^3=\boldsymbol{m}^2-\frac14\epsilon^2.
\end{equation}
We refer to the equation \eqref{Ham_II_twisted} together with the second commutation relation of \eqref{triple_II} as the \textit{Hamiltonian form} of the QPII equation.
From this Hamiltonian form, one can recover the remaining two commutation relations of the full set \eqref{triple_II}, \eqref{H_Hd_I} and the precursor \eqref{prec_II}, analogously to the previous cases.

To reconstruct the Heisenberg dynamics, we define $q$ and $p$ by formulas \eqref{Hder_II}, i.e.
\begin{equation}\label{pq_dict_II}
p=-2\dot{H}, \qquad 2q\dot{H}=\kappa\ddot{H}+\boldsymbol{m}+\frac{\epsilon}2.
\end{equation}
Then the canonical commutation relation \eqref{pq_eps} follows from the second relation in \eqref{triple_II}. Further, the Hamiltonian form equation \eqref{Ham_II_twisted}, together with the definition \eqref{Hder_II}, yields the Hamiltonian expression \eqref{HII}. The same definition also implies that $\dot{H}$ and $\ddot{H}$ are explicitly time-independent. This completes the reconstruction of the original QPII Heisenberg dynamics.

 \subsection{Limits to QPI}
 Making the autonomous canonical transformation 
 \begin{equation}
 q_{\III_2}=-4M^2\left(1-\frac{q_{\I}} {M^{\frac25}}\right), \qquad p_{\III_2}=\frac1{4M}\left(1-\frac{q_{\I}}{M^{\frac25}}+\frac{p_{\I}}{M^{\frac35}}\right), \qquad  \quad t_{\III_2}=16M^3\left(1+\frac{t_{\I}}{2M^{\frac45}}\right)
 \end{equation} 
 with $M=m_1/3$, and then sending $M\rightarrow\infty$, we obtain that QPIII$_2$ with Hamiltonian~\eqref{HIII2} degenerates to QPI with Hamiltonian~\eqref{HI} and dimensions \eqref{dim_I}, namely
\begin{equation}
H_{\III_2}(\ln t_{\III_2})-\frac{8m_1^2}9-\frac3{4m_1}t_{\III_2}=2H_{\II}M^{\frac45}+O\left(M^{\frac35}\right), \qquad \quad  d\ln t_{\III_2}=\frac{dt_{\I}}{2M^{\frac45}}\left(1+\frac{t_{\I}}{2M^{\frac45}}\right)^{-1}.    
\end{equation}
Alternatively, making the autonomous canonical transformation
\begin{equation} q_{\II}=M^{\frac13}\left(1+\frac{q_{\I}}{M^{\frac25}}\right), \qquad p_{\II}=-2M^{2/3}\left(1-\frac{q_{\I}}{M^{\frac25}}-\frac{p_{\I}}{2M^{\frac35}}\right), \qquad \quad t_{\II}=-6M^{\frac23}\left(1-\frac{t_{\I}}{6M^{\frac45}}\right)  
\end{equation}
with $M=\boldsymbol{m}/2$, and then sending $M\rightarrow\infty$, we obtain that QPII with Hamiltonian~\eqref{HII} degenerates to QPI with Hamiltonian~\eqref{HI} and dimensions \eqref{dim_I}, namely
\begin{equation}
H_{\II}+\frac1{12}t_{\II}^2+3(\boldsymbol{m}/2)^{\frac43}=H_{\I}M^{\frac2{15}}+O\left(M^{-\frac1{15}}\right), \qquad \quad  dt_{\II}=\frac{dt_{\I}}{M^{\frac2{15}}}.
\end{equation}

\section{Quantum Painlev\'e tau functions expansions}
\label{sec:asymptotics}
\setcounter{subsection}{-1}

\subsection{Overview: from classical to quantum expansions}
As already mentioned in the Introduction, in \cite{BST25} we presented quantum deformations of the Painlev\'e tau function expansions around regular and irregular singularities. Building on \cite{GIL12}, these expansions were obtained as the Zak transforms of SUSY gauge theory partition functions. In general, it was established that each Painlev\'e equation corresponds to a certain SUSY gauge theory; Table~\ref{tab:theories} summarizes the corresponding theories for all (differential) Painlev\'e equations, together with the associated regular and irregular singularities. The Painlev\'e~VI, V, and III's equations correspond to $\mathcal{N}=2$ $D=4$ SUSY $SU(2)$ gauge theories with $N_f$ hypermultiplets in the (anti-)fundamental representation \cite{GIL13}. Meanwhile, the Painlev\'e~IV, II, and I equations correspond to the Argyres--Douglas theories $H_k$ \cite{BLMST16}, where the subscript $k$ indicates the number of mass parameters. The type of singularity determines the gauge theory regime: regular singularities correspond to the weak-coupling regime, whereas irregular singularities correspond to the strong-coupling regime.
\begin{table}[H]
    \centering
    \begin{tabular}{|c|c|c|c|c|c||c|c|c|}
      \hline Equation & PVI & PV & PIII$_1$ & PIII$_2$ & PIII$_3$ & PIV & PII & PI \\ \hline
        Theory & $N_f=4$ & $N_f=3$ & $N_f=2$ & $N_f=1$ & $N_f=0$ & $H_2$ & $H_1$ & $H_0$ \\ \hline\hline
        Reg. expansion & $0,1,\infty$ & 0 & 0 & 0 & 0 & - & - & -  \\ 
        \hline
        Irr. expansion & - & $\infty$ & $\infty$ & $\infty$ & $\infty$ & $\infty$ & $\infty$ & $\infty$  \\ \hline
    \end{tabular}
    \caption{(Quantum) Painlev\'e tau function expansions at the singular points}
    \label{tab:theories}
\end{table}
In \cite{BST25}, the quantum deformations of the Painlev\'e tau function expansions were obtained as solutions of the bilinear blowup equations on $\mathbb{C}^2/\mathbb{Z}_2$ for the tau functions. These equations were called \emph{quantum Painlev\'e equations} there, since they arise as a natural deformation of the classical Painlev\'e equations in the tau form. In the previous sections we identified these deformed equations with the tau forms of the canonically quantized Painlev\'e equations. The solutions of these quantum tau forms take the form of the formal noncommutative Zak transforms of the refined partition functions, i.e.\ of partition functions in a generic $\Omega$-background (away from the self-dual point). The noncommutativity of the Zak transform originates from the canonical quantization of the Painlev\'e monodromy (or Stokes) data.

In the previous sections we also saw that, when passing from the canonically quantized Painlev\'e equations to the quantum tau forms, one typically gains an additional integration constant $C$, which effectively replaces one of the equation parameters. This is even more transparent in the classical setting: passing from the second-order Hamiltonian form equation to the third-order precursor equation introduces an integration constant. Accordingly, the irregular-type expansions in \cite{BST25} generally depend on an extra integration constant rather than on the corresponding equation parameter. In the classical case, this freedom is fixed by straightforward checks of the Hamiltonian form equations (e.g., in \cite{GIL13}, \cite{BS14} \cite{BLMST16}). Here we proceed in the same way, using the quantum Hamiltonian form equations obtained in the previous sections. Namely, we first extract several terms of the expansions of the Hamiltonian functions from the corresponding tau function expansions of \cite{BST25}. Substituting these truncated expansions into the corresponding Hamiltonian form equation, we obtain a relation between the "missing" quantum Painlev\'e parameter and the integration constant in the tau function expansion. More precisely, we determine the $\epsilon$-corrections to the classical relations. We find that, for the irregular-type expansions, these $\epsilon$-corrections (or their absence) exactly reproduce the results of \cite[Sec.~5]{BST25}. In \cite{BST25} they were obtained by identifying the preimage of the Zak transform of the irregular tau function expansions with refined partition functions computed via the holomorphic anomaly approach.

This section is organized as follows. In Sec.~\ref{ssec:small} we recall the regular-type tau function expansions of QPVI, QPV, and QPIII's from \cite{BST25}, together with the structure of the corresponding partition functions. We then derive an equation determining the successive coefficients in the corresponding regular-type expansions of the Hamiltonian functions, and verify that the Hamiltonian-form equations hold without any $\epsilon$-corrections. Analogously, in Sec.~\ref{ssec:large} we describe the general ansatz of \cite{BST25} for the irregular-type tau function expansions and obtain a general equation determining the successive coefficient in the corresponding irregular-type expansions of the Hamiltonian functions. Finally, in Sec.~\ref{ssec:fixing} we use this general equation to compute $\epsilon$-corrections relating the quantum Painlev\'e parameters to the integration constants in the irregular-type tau function expansions of \cite{BST25}.

\subsection{Quantum tau and Hamiltonian functions regular type expansions}
\label{ssec:small}
\paragraph{Tau functions as the noncommutative Zak transforms.} In \cite[Sec. 2,3]{BST25} we presented the QPVI, QPV, QPIII's tau functions around the regular singularity $t=0$. These tau functions are given by the (noncommutative) Zak transforms
\begin{equation}\label{Zak_ncmt_weak}
\tau^{[N_f]}(a,\eta;\epsilon_1,\epsilon_2|t)=\sum_{n\in\mathbb{Z}}e^{\ri n \eta }\cdot\mathcal{Z}^{[N_f]}(a+n\epsilon_2;\epsilon_1,\epsilon_2|t), 
\end{equation}
where the operator $e^{\ri\eta}$ canonically commute with the other integration constant $a$ in the sense that
\begin{equation}\label{as_comm_rel}
a e^{\ri \eta}=e^{\ri \eta} (a+\epsilon) \quad \Leftarrow \quad \ri [a,\eta]=\epsilon.    
\end{equation}
Here $\mathcal{Z}^{[N_f]}$ is the full $\mathcal{N}=2$ $D=4$ SUSY $SU(2)$ partition function 
with the corresponding (by Table \ref{tab:theories}) number $0\leq N_f\leq4$ of hypermultiplets in the (anti-) fundamental representation of the gauge group in the weak-coupling regime. Then the solutions of the quantum Painlev\'e tau forms were given there by two tau functions: $\tau^{[N_f]}(2\epsilon_1,\epsilon_2{-}\epsilon_1)$ and $\tau^{[N_f]}(\epsilon_1{-}\epsilon_2,2\epsilon_2)$ (recall \eqref{Omega_patches} for the QPI case). More precisely, for the tau forms obtained in the previous sections, the solutions of \cite{BST25} should be multiplied by prefactors, namely
\begin{equation}
\begin{aligned}
\label{tau_factor}
\tau^{(1)}_{\textrm{eq}(N_f)}=f^{[N_f]}(2\epsilon_1,\epsilon_2{-}\epsilon_1|t)\, \tau^{[N_f]}(a,\eta; 2\epsilon_1,\epsilon_2{-}\epsilon_1|t), \\ \tau^{(2)}_{\textrm{eq}(N_f)}=f^{[N_f]}(\epsilon_2{-}\epsilon_1,2\epsilon_2|t)\, \tau^{[N_f]}(a,\eta;2\epsilon_1,\epsilon_2{-}\epsilon_1|t),
\end{aligned}
\end{equation}
where the subscript $\textrm{eq}(N_f)$ marks the equation corresponding (via Table~\ref{tab:theories}) to the number $N_f$ of the hypermultiplets. These prefactors are monomials whose powers are expressed via the mass invariants \eqref{elem_def}:
\begin{multline}\label{prefactors_weak}
f^{[4]}(\epsilon_1,\epsilon_2)=\big(t(1{-}t)\big)^{\frac{w_2^{[4]}{-}2\epsilon^2}{6\epsilon_1\epsilon_2}}(1{-}t)^{-\frac{e_1^{[4]}\big( e_1^{[4]}{+}2\epsilon\big)}{4\epsilon_1\epsilon_2}}, \qquad  f^{[3]}(\epsilon_1,\epsilon_2)=e^{\frac{e_1^{[3]}+\epsilon}{2\epsilon_1\epsilon_2}\,t},  \qquad f^{[2]}(\epsilon_1,\epsilon_2)=e^{\frac{t}{2\epsilon_1\epsilon_2}}, \\  f^{[1]}(\epsilon_1,\epsilon_2)=f^{[0]}(\epsilon_1,\epsilon_2)=1,
\end{multline}
according to \eqref{conn_VI}, \eqref{conn_V}, \eqref{conn_III1}, \eqref{conn_III2}, \eqref{conn_III3}, respectively. The partition functions $\mathcal{Z}^{[N_f]}$ also depend on the $N_f$ hypermultiplet masses of the gauge theory. These masses of \cite{BST25} are precisely the masses $(m_f)_{f=1}^{N_f}$ that parametrize quantum Painlev\'e equations in the previous sections. Finally, note that in the QPVI case there are also regular singularities at $t=1$ and $t=\infty$, around which there are  solutions of the form \eqref{Zak_ncmt_weak} with $t$ replaced by $t{-}1$ and $t^{-1}$, respectively. For details see the end of \cite[Sec. 2.3]{BST25} and the further discussion in Sec.~\ref{ssec:aut_symm}. Through this section, in the QPVI case, we understand by $t$ any of these three choices of the partition function variable.

\paragraph{Structure of the partition function $\mathcal{Z}^{[N_f]}$.} Generally following \cite{AGT09}, the full partition function $\mathcal{Z}^{[N_f]}$ in the weak-coupling regime factorizes into three parts, 
\begin{equation}\label{Zstr}
\mathcal{Z}^{[N_f]}=\mathcal{Z}_{cl}^{[N_f]}\mathcal{Z}_{1-loop}^{[N_f]}\mathcal{Z}_{inst}^{[N_f]},
\end{equation}
which are given by\footnote{We use slightly modified classical and 1-loop part; see \cite[Sec. 2.1]{BST25} for details.}
\begin{enumerate}
\item The classical part is given by a monomial-type expression
\begin{equation}\label{Zcl}
	\mathcal{Z}^{[N_f]}_{cl}(a;\epsilon_1,\epsilon_2|t)=t^{\frac{\epsilon^2/4-a^2}{\epsilon_1\epsilon_2}}.
\end{equation}
\item The 1-loop part is given by a product of the double gamma functions $\gamma_{\epsilon_1,\epsilon_2}$ of \cite{NY03L}
\begin{equation}
\label{Z1loop}
\mathcal{Z}^{[N_f]}_{1-loop}(a,(m_f)_{f=1}^{N_f};\epsilon_1,\epsilon_2)=\prod\limits_{\pm}\frac{\prod\limits_{f=1}^{N_f}\exp\gamma_{\epsilon_1,\epsilon_2}(m_f\pm a-\epsilon/2)}{\exp(\gamma_{\epsilon_1,\epsilon_2}(\pm 2a))}.
\end{equation}
These double gamma functions are formally defined by
\begin{equation} \label{gammaNYdef}
\gamma_{\epsilon_1,\epsilon_2}(x):=
   \frac{d}{ds}\Bigg|_{s=0}\frac1{\Gamma(s)}\int\limits_0^{+\infty}\frac{dz}{z} z^s \frac{e^{-xz}}{(e^{\epsilon_1z}-1)(e^{\epsilon_2z}-1)},\qquad \mathrm{Re}(\epsilon_{1,2})\neq0, \quad \mathrm{Re}(x)>0,
\end{equation}
where the integral should be understood via analytic continuation in $s$ from the region $\mathrm{Re}(s)>2$ to a neighborhood of $s=0$. We refer to \cite[App. B]{BST25} for details and for further useful properties of $\gamma_{\epsilon_1,\epsilon_2}$.
\item The most important, the instanton part is given by the Nekrasov formula \cite{N02,Flume:2002az,Bruzzo:2002xf}. It has the structure of an expansion in small $t$
\begin{equation}
\label{Zinst}
\mathcal{Z}^{[N_f]}_{inst}(a,(m_f)_{f=1}^{N_f};\epsilon_1,\epsilon_2|t)=\sum_{Y^+,Y^-}\frac{\prod\limits_{f=1}^{N_f} Z_{fund}(a,m_f;\epsilon_1,\epsilon_2|Y^+,Y^-)}{Z_{vec}(a;\epsilon_1,\epsilon_2|Y^+,Y^-)} t^{|Y^+|+|Y^-|},
\end{equation}
where the sum runs over all pairs of integer partitions $Y^{+}, Y^{-}$, and $|Y^{\pm}|$ denotes the number of boxes of $Y^\pm$. The factors in each summand of $\mathcal{Z}^{[N_f]}_{inst}$ are
 \begin{equation}
   Z_{fund}(a,m;\epsilon_1,\epsilon_2|Y^+,Y^-)=\prod_{\pm}\prod_{(i,j)\in Y^{\pm}} (m\pm a-\epsilon/2+\epsilon_1 i+\epsilon_2 j),
   \end{equation}
   \vspace{-0.7cm}
    \begin{multline}
  \mspace{-20mu}Z_{vec}(a;\epsilon_1,\epsilon_2|Y^+,Y^-)=\prod_{\pm}\left(\prod_{(i,j)\in Y^{\pm}}\!\Bigl(\epsilon_2 (Y^{\pm}_i{-}j{+}1)-\epsilon_1 (\tilde{Y}^{\pm}_j{-}i)\Bigr)\prod_{(i,j)\in Y^{\pm}}\!\Bigl(\epsilon_1 (\tilde{Y}^{\pm}_j{-}i{+}1)-\epsilon_2 (Y^{\pm}_i{-}j)\Bigr)\right.  \\ \mspace{-19mu}\times  \left.\prod_{(i,j)\in Y^{\pm}}\!\Bigl(\pm 2a+\epsilon_2 (Y^{\pm}_i{-}j{+}1)-\epsilon_1 (\tilde{Y}^{\mp}_j{-}i)\Bigr)\prod_{(i,j)\in Y^{\mp}}\!\Bigl(\pm 2a+\epsilon_1 (\tilde{Y}^{\pm}_j{-}i{+}1)-\epsilon_2 (Y^{\mp}_i{-}j)\Bigr)\right),
\end{multline} 
where $\tilde{Y}$ denotes the transpose of $Y$. This instanton partition function is convergent with infinite radius of convergence for $N_f<4$ \cite{ILTy14,Arnaudo:2022ivo} and with finite radius for 
$N_f=4$ \cite{guillarmou2024conformal}.
\footnote{
Strictly speaking, for technical reasons these proofs apply to a restricted set of the $\Omega$-background parameters, but it is reasonable to expect that the results extend to more general situations.}
\end{enumerate}

All three parts of $\mathcal{Z}^{[N_f]}$ are explicitly invariant under the mass permutations as well as under $a\mapsto -a$.
They are also invariant under the exchange $\epsilon_1\leftrightarrow\epsilon_2$, although this symmetry is not manifest for the instanton part in the form above. Indeed, the exchange symmetry follows from the symmetry of the $(\mathbb{C}^*)^2$ torus action on the ADHM data.
It also follows from the AGT correspondence \cite{AGT09}, as we discuss in Sec.~\ref{ssec:aut_symm}.

\paragraph{Symmetries of the Zak transforms.}
Writing the tau functions \eqref{tau_factor} as the Zak transforms \eqref{Zak_ncmt_weak},
we obtain the regular-type solutions (expanded around $t=0$) of the Painlev\'e tau forms: 
\begin{align}\label{Zak_tau1}
&\tau^{(1)}(a,\eta|t)=\sum_{n\in\mathbb{Z}}e^{\ri n \eta}\cdot\widetilde{\mathcal{Z}}^{[N_f]}(a+n(\epsilon_2{-}\epsilon_1);2\epsilon_1,\epsilon_2{-}\epsilon_1|t), \\ \label{Zak_tau2}
&\tau^{(2)}(a,\eta|t)=\sum_{n\in\mathbb{Z}}\widetilde{\mathcal{Z}}^{[N_f]}(a+n(\epsilon_2{-}\epsilon_1);\epsilon_1{-}\epsilon_2,2\epsilon_2|t)\cdot e^{\ri n \eta},
\end{align}
where we denoted $\widetilde{\mathcal{Z}}^{[N_f]}=f^{[N_f]}\mathcal{Z}^{[N_f]}$, and in $\tau^{(2)}$ we moved $e^{\ri n\eta}$ to the right using \eqref{as_comm_rel}. Recall that, according to \eqref{tau_universal}, the tau functions $\tau^{(1)}$ and $\tau^{(2)}$ are defined up to left and right $t$-constant operator prefactors, respectively. As in the classical case, these tau functions, together with the commutation relation \eqref{as_comm_rel}, enjoy the natural shift symmetry
\begin{equation}\label{weak_shift}
\tau^{(1)}(a-\kappa,\eta|t)=e^{\ri\eta}\cdot\tau^{(1)}(a,\eta|t),  \qquad \tau^{(2)}(a-\kappa,\eta|t)=\tau(a,\eta|t)\cdot e^{\ri\eta},
\end{equation}
where, as usual, $\kappa=\epsilon_2{-}\epsilon_1$. The partition function symmetry $a\mapsto-a$ additionally requires $\eta\mapsto-\eta$, in agreement with \eqref{as_comm_rel}.
Without loss of generality, these two symmetries allow us to assume $0\leq\mathrm{Re}\,\frac{a}{\kappa}\leq\frac12$ below. The mass permutation symmetry of the tau functions is inherited from the corresponding symmetries of the partition functions.

The tau functions $\tau^{(1)}$ and $\tau^{(2)}$ are related not only by the defining relation \eqref{tau_universal} (or, the same, $D^1$-relation), but also by an involutive antiautomorphism, which we denote by $^\mathrm{T}$. We define its action on the gauge theory variables (treated formally) by
\begin{equation}\label{aut_T}
\epsilon_1\xleftrightarrow[]{\mathrm{T}}\epsilon_2, \qquad a^{\mathrm{T}}=a, \quad (\ri\eta)^{\mathrm{T}}=-\ri \eta, \qquad t^{\mathrm{T}}=t, \qquad m_f^{\mathrm{T}}=m_f,\, {\scriptstyle f=1,\ldots N_f}, 
\end{equation}
where the action on $\ri\eta$ is chosen so as to preserve the commutation relation \eqref{as_comm_rel}. 
Then, by the $\epsilon_1\leftrightarrow\epsilon_2$ symmetry of the partition functions, the Zak transforms \eqref{Zak_tau1} and \eqref{Zak_tau2} are exchanged under $^\mathrm{T}$. Further, this means that the corresponding Hamiltonian function, reconstructed from the definition \eqref{tau_universal}, is invariant under $^\mathrm{T}$. 
Apropos, the $\epsilon$-prefactors in the definition \eqref{tau_universal} are simply the products of the $\epsilon$-parameters of the corresponding tau functions.
We postpone further discussion of tau function symmetries to Sec.~\ref{ssec:aut_symm}. 

\paragraph{Rearrangement of the Zak transforms.}
For definiteness, let us consider tau function $\tau^{(1)}$ and factor out on the left the classical and the 1-loop part from \eqref{Zak_tau1}. Namely, using that the combined homogeneity factor of the classical \cite[(2.7)]{BST25} and the 1-loop \cite[(2.8)]{BST25} parts is $a$-independent, we have
\begin{equation}\label{rearr}
\tau^{(1)}(a,\eta|t)=\mathcal{Z}^{[N_f]}_{cl+1-loop}(a;\epsilon{-}\kappa,\kappa|t) \sum_{n\in\mathbb{Z}}e^{\ri n \eta}\cdot\frac{\mathcal{Z}^{[N_f]}_{cl+1-loop}\left(\frac{a}{\kappa}{+}n;\frac{\epsilon}{\kappa}{-}1,1\Big|\kappa^{N_f{-}4}t\right)}{\mathcal{Z}^{[N_f]}_{cl+1-loop}\left(\frac{a}{\kappa}{+}n\frac{\epsilon}{\kappa};\frac{\epsilon}{\kappa}{-}1,1\Big|\kappa^{N_f{-}4} t\right)}\widetilde{\mathcal{Z}}^{[N_f]}_{inst}(a{+}n\kappa;\epsilon{-}\kappa,\kappa|t),
\end{equation}
where we denoted $\mathcal{Z}_{cl+1-loop}^{[N_f]}=\mathcal{Z}_{cl}\mathcal{Z}_{1-loop}^{[N_f]}$ and $\widetilde{\mathcal{Z}}_{inst}^{[N_f]}=f^{[N_f]}\mathcal{Z}_{inst}^{[N_f]}$,
and moved to the quantum Painlev\'e notations $\epsilon=\epsilon_1{+}\epsilon_2$, $\kappa=\epsilon_2{-}\epsilon_1$ for convenience.
To compute the ratio of the 1-loop parts, we need a shift relation for the double gamma function $\gamma_{\epsilon/\kappa{-}1,1}$ defined by \eqref{gammaNYdef}. Using the Hurwitz zeta function integral representation \cite[\href{https://dlmf.nist.gov/25.11.E25}{(25.11.25)}]{DLMF} and the property \cite[\href{https://dlmf.nist.gov/25.11.E18}{(25.11.18)}]{DLMF}, we obtain
\begin{multline}
\frac{\exp\gamma_{\epsilon/\kappa{-}1,1}(x{+}n)}{\exp\gamma_{\epsilon/\kappa{-}1,1}(x{+}n\epsilon/\kappa)}=(2\pi)^{-\frac{n}2}\prod_{k=\frac12}^{|n|-\frac12}\Gamma^{\mathrm{sgn}(n)}\left(x+\Big(n{+}\frac12{-}\mathrm{sgn}(n)k\Big)+\Big(\frac12{+}\mathrm{sgn}(n)k\Big)\frac{\epsilon}{\kappa}\right)\\
=(2\pi)^{-\frac{n}2}P_{(n)}(x;1,\epsilon/\kappa)\prod_{k=\frac12}^{|n|-\frac12}\Gamma^{\mathrm{sgn}(n)}\left(x+1+\Big(\frac12{+}\mathrm{sgn}(n)k\Big)\frac{\epsilon}{\kappa}\right), \qquad n\in\mathbb{Z},
\end{multline}
where the polynomial $P_{(n)}(x;\kappa,\epsilon)$, of degree $\frac{n(n{-}1)}2$ in $x$, is defined by
\begin{equation}\label{Pn}
P_{(n)}(x;\kappa,\epsilon)=\prod_{k=\frac12}^{|n|-\frac12}\prod_{l=\frac12}^{|n-\frac12|-k-\frac12} \left(x+\Big(\frac12{+}\mathrm{sgn}(n)l\Big)\kappa+\Big(\frac12{+}\mathrm{sgn}(n)k\Big)\epsilon\right).   
\end{equation}
We then attach the resulting gamma function factors to the operator $e^{\ri\eta}$ by introducing
\begin{equation}\label{eta_t}
e^{\ri\eta(t)}=e^{\ri\eta}\cdot\prod_{\pm}\frac{\prod\limits_{f=1}^{N_f}\Gamma^{\pm1}\Big(1+\kappa^{-1}\big(m_f\pm(a{+}\epsilon/2)\big)\Big)}{\Gamma^{\pm1}\Big(1\pm\kappa^{-1}(2a{+}\epsilon)\Big) \Gamma^{\pm1}\Big(1+\kappa^{-1}\big(\epsilon\pm(2a{+}\epsilon)\big)\Big)}\,t^{\frac{2a{+}\epsilon}{\kappa}}.
\end{equation}
Dropping the $t$-independent 1-loop part in the prefactor extracted from the sum, we finally arrive at the rearranged expression for the tau function $\tau^{(1)}$:
\begin{equation}\label{Zak_tau1_rearr}
\tau^{(1)}(a,\eta|t)=\mathcal{Z}_{cl}(a;\epsilon{-}\kappa,\kappa|t)  \sum_{n\in\mathbb{Z}}e^{\ri n\eta(t)}\cdot
C_n^{[N_f]}(a;\kappa)\, t^{n^2}\widetilde{\mathcal{Z}}^{[N_f]}_{inst}(a{+}n\kappa;\epsilon{-}\kappa,\kappa|t),
\end{equation}
where the new rational "1-loop part" is
\begin{equation}\label{C_n_weak}
C_n^{[N_f]}(a;\kappa)=\prod_{\pm}\frac{\prod\limits_{f=1}^{N_f}P_{(\pm n)}\big(m_f\pm a-\epsilon/2;\kappa,\epsilon\big)}{P_{(\pm2n)}(\pm2a;\kappa,\epsilon)}.    
\end{equation}
Using the involutive antiautomorphism $^\mathrm{T}$ defined by \eqref{aut_T}, we immediately obtain the corresponding rearranged form of \eqref{Zak_tau2}:
\begin{equation}\label{Zak_tau2_rearr}
\tau^{(2)}(a,\eta|t)=  \left(\sum_{n\in\mathbb{Z}}
C_{-n}^{[N_f]}(a;-\kappa)\,t^{n^2}\widetilde{\mathcal{Z}}^{[N_f]}_{inst}(a{+}n\kappa;-\kappa,\epsilon{+}\kappa|t)\cdot e^{-\ri n\eta(t)^\mathrm{T}} \right) \mathcal{Z}_{cl}(a;-\kappa,\epsilon{+}\kappa|t),
\end{equation}
where
$e^{-\ri \eta(t)^\mathrm{T}}$ has the same structure as $e^{\ri\eta(t)}$ in \eqref{eta_t}, but with the gamma function arguments replaced accordingly.
These rearranged expression for the tau functions
can be considered as asymptotic expansions in $t$, 
provided that $\mathrm{Re}(\epsilon/\kappa)>-1$. This condition can be seen by collecting the powers of $t$ from \eqref{eta_t}.

\paragraph{Hamiltonian expansions and fulfillment of the Hamiltonian form equations.}
We now solve \eqref{tau_universal} for the Hamiltonian function, with the tau functions given by the  rearranged Zak transforms above. Namely, taking $\tau^{(1)}$ (for definiteness) in the form \eqref{Zak_tau1_rearr}, we solve
\begin{equation}\label{tau_ln_H}
-2\epsilon_1(\epsilon_2{-}\epsilon_1)\frac{d\tau^{(1)}}{d\ln t}=\tau^{(1)}\,H(a,\eta;\kappa,\epsilon|\ln t),   
\end{equation}
where, for uniformity, we use the Hamiltonians with respect to time $\ln t$ also in the QPVI case. Guided by the generic classical case (with $\mathrm{Re}(a/\kappa)\neq\mathbb{Z}+\frac12$), we expect the Hamiltonian to admit an expansion of the form
\begin{equation}\label{Ham_weak_ansatz}
H(a,\eta;\kappa,\epsilon|\ln t)=\sum_{n\in\mathbb{Z}}e^{\ri n\eta(t)}t^{|n|}H_n(t),\qquad H_n(t)\in \mathbb{C}[[t]].
\end{equation}
We can regard this as an asymptotic expansion as $t\to 0$ provided $\mathrm{Re}(\epsilon/\kappa)>0$.
Collecting in \eqref{tau_ln_H} the terms that are multiplied by $e^{\ri N\eta}$ moved to the left, we obtain an equation for the coefficients $H_n$, namely
\begin{multline}\label{Ham_weak_eq}
\big((a{+}N\kappa)^2-\epsilon^2/4\big) \mathcal{Z}^{[N_f]}_{inst}(a{+}N\kappa;\epsilon{-}\kappa,\kappa|t)-\kappa(\epsilon{-}\kappa)\frac{d}{d\ln t} \mathcal{Z}^{[N_f]}_{inst}(a{+}N\kappa;\epsilon{-}\kappa,\kappa|t)\\=
\left(C_N^{[N_f]}(a;\kappa)\right)^{-1}\sum_{n\in\mathbb{Z}}C_{N-n}^{[N_f]}(a{+}n\epsilon;\kappa)\mathcal{Z}^{[N_f]}_{inst}(a{+}(N{-}n)\kappa{+}n\epsilon;\epsilon{-}\kappa,\kappa|t)t^{n(n{-}2N){+}|n|}H_n(t).
\end{multline}
For each fixed $N$, this should be understood as a system of recursive equations for the coefficients of the power series in $t$.
It turns out that, to check the Hamiltonian form equations for QPVI, QPV, and QPIII's around the regular singularities, it suffices to know $H_0(t)$ up to order $t^1$ and $H_{\pm1}(t)$ up to order $t^0$.
Considering \eqref{Ham_weak_eq} for $N=0$ and the powers $t^0, t^1$, and for $N=\pm1$ and the power $t^0$, we obtain
\begin{align}\label{H0_weak}
H_0&=a^2{-}\frac{\epsilon^2}4+\sum_{\pm}\frac{\prod_{f=1}^{N_f}(m_f\pm a+\epsilon/2)}{\pm2a(\epsilon\pm2a)}\cdot t+O(t^2),\\ \label{Hpm1_weak}
H_{\pm1}&=-\frac{\kappa\prod_{f=1}^{N_f}(m_f\mp a-\epsilon/2)}{\pm2a(\epsilon\pm2a)(\kappa+\epsilon\pm2a)}+O(t).
\end{align}
Substituting the Hamiltonian expansion into the Hamiltonian form equations, we expect to obtain the additional $t$-constant operator $C$ in place of one of the mass parameters. On the other hand, by the general structure of the ansatz \eqref{Ham_weak_ansatz}, any such output $C$ must be of the form $\sum_{n\in\mathbb{Z}}e^{\ri n\eta(t)}c_n(t)$ with $c_n\in\mathbb{C}[[t]]$. The $t$-independence of $C$ implies immediately $c_n=0,\, n\neq0$, and thus it is enough to determine the constant term $c_0$. 
Using \eqref{H0_weak}, \eqref{Hpm1_weak}, we finally find that the Hamiltonian form equations for QPVI  \eqref{Ham_VI}, QPV \eqref{Ham_V},  QPIII$_1$ \eqref{Ham_III1}, QPIII$_2$ \eqref{Ham_III2}, and QPIII$_3$ \eqref{Ham_III3} are satisfied. Hence the integration-constant freedom for the regular-type solutions was fixed correctly in \cite{BST25}.

\subsection{Quantum tau and Hamiltonian functions irregular type expansions}
\label{ssec:large}
\paragraph{Ansatz of \cite{BST25}.}
The noncommutative Zak transform \eqref{Zak_ncmt_weak}, together with the commutation relation \eqref{as_comm_rel}, was used
in \cite[Sec. 4]{BST25} to construct the quantum Painlev\'e tau functions around the irregular singularity $t=\infty$ as well:
\begin{equation}\label{Zak_ncmt_comm_strong}
\tau^{[\mathbf{th}]}(a_{\D},\eta_{\D};\epsilon_1,\epsilon_2|s)=\sum_{n\in\mathbb{Z}}e^{\ri n \eta_{\D} }\mathcal{Z}^{[\mathbf{th}]}(a_{\D}+n\epsilon_2;\epsilon_1,\epsilon_2|s), \qquad a_{\D} e^{\ri \eta_{\D}}=e^{\ri \eta_{\D}} (a_{\D}+\epsilon).
\end{equation}
Here the pair $(a,\eta)$ is replaced by its dual $(a_{\D},\eta_{\D})$, and
the weak-coupling partition functions $\mathcal{Z}^{[N_f]}$ are replaced by the strong-coupling partition functions $\mathcal{Z}^{[\mathbf{th}]}$ of the gauge theory $\mathbf{th}$ from Table~\ref{tab:theories}.
To distinguish the strong-coupling regime, we write the theory superscript in boldface.
The time dependence of the strongly coupled partition function is through the dimension-$1$ variable $s=\varkappa\, t^{1/[t]}, \, \varkappa\in\mathbb{C}$. 
The solutions of the quantum Painlev\'e tau forms are then given by the direct strong-coupling analogs of \eqref{tau_factor}. Recall that the tau forms for QPIV, QPII, QPI in the present paper coincide with those in \cite{BST25}, so the corresponding prefactors $f$ are trivial.

In contrast to the weak-coupling regime, there are no closed (combinatorial) formulas for the strong-coupling partition functions as (asymptotic) expansions in $s^{-1}$. Nevertheless, they imply a general structure, used in \cite{BST25} as an ansatz to extract several leading terms of these expansions from the tau forms of the corresponding quantum Painlev\'e equations.  The resulting expansions are quantum deformations of the classical ones of \cite{BLMST16}. Based on the classical case, we have that QPV, QPIV, QPII admit two distinct asymptotic expansions, whereas for QPIII's and QPI there is a single expansion in each case. The ansatz of \cite{BST25} assumes that $\mathcal{Z}^{[\mathbf{th}]}$ also factorizes into three parts, as in \eqref{Zstr}. We use the same terminology for these factors, reflecting the similarity of their functional form. Their dependence on $a_{\D}$, $s$, and $\epsilon_1,\epsilon_2$ is fixed as a deformation of the structure in \cite{BLMST16}. The ansatz involves several parameters of appropriate dimension, which are assumed to be symmetric under the exchange $\epsilon_1\leftrightarrow\epsilon_2$. The explicit expressions are as follows:
 \begin{enumerate}
    \item A classical term of the form
    \begin{equation}\label{Zcl_anz}
    \mathcal{Z}_{cl}^{[\mathbf{th}]}(a_{\D};\epsilon_1,\epsilon_2|s)=s^{-\frac{\xi_2-N_p a_{\D}^2/2}{\epsilon_1\epsilon_2}} e^{-\frac{\beta s^2+\xi_1 s+\delta a_{\D} s}{\epsilon_1\epsilon_2}},
    \end{equation}   
    with dimensionless parameters $N_p\in\mathbb{N}$, $\beta,\delta\in\mathbb{C}$, and parameters $\xi_1$ and $\xi_2$ with $[\xi_1]=1, [\xi_2]=2$. The number $N_p$ distinguishes the two expansions in the applicable cases. 
 \item A 1-loop part of the form   \begin{equation}\label{Z1loop_anz}
     \mathcal{Z}_{1-loop}^{[\mathbf{th}]}(a_{\D};\epsilon_1,\epsilon_2)=e^{-\frac{\chi a_{\D}^2}{2\epsilon_1\epsilon_2}}\prod_{i=1}^{N_p} \exp\gamma_{\epsilon_1,\epsilon_2}(a_{\D}+\mu_i-\epsilon/2), \qquad \sum_{i=1}^{N_p}\mu_i=0,
     \end{equation}
     with a parameter $\chi\in\mathbb{C}$ and dimension-$1$ parameters $\mu_i$.
    \item An instanton part, written as an asymptotic power series in $s^{-1}$, namely 
    \begin{equation}\label{Zinst_anz}
\mathcal{Z}_{inst}^{[\mathbf{th}]}(a_{\D};\epsilon_1,\epsilon_2|s)=1+\sum_{k=1}^{K} Q_{3k}(a_{\D}) (\epsilon_1\epsilon_2s)^{-k}+O(s^{-K-1}).    
    \end{equation}
    Here $Q_{3k}(a_{\D})$ is a polynomial of degree $3k$ in $a_{\D}$ (and of the same dimension), whose coefficients are additional ansatz parameters of appropriate dimension.
\end{enumerate} 

For each quantum-deformed expansion, the coefficients in the classical and one-loop parts, together with several leading terms of the instanton part, were determined in \cite[Sec.~4]{BST25}. The resulting coefficients depend not only on the mass parameters but also on the additional integration constant that appears in the previous sections upon integrating the precursor equations. More precisely, excluding the parameter-free cases QPIII$_3$ and QPI (which we comment on below), this dependence is through the mass invariants of the corresponding finite Weyl group, with one of these invariants replaced by the finite Weyl group invariant integration constant. This statement holds literally for the instanton parts and for the classical parts (after including the connection prefactors, given by \eqref{prefactors_weak} for QPVI, QPV, and QPIII's). By contrast, the one-loop parts are expressed in terms of the tilded masses defined from the modified set of invariants; nevertheless, the full one-loop contribution remains invariant under the finite Weyl group. As before, QPIII$_3$ can be viewed as a special case of QPIII$_2$, while the integration-constant freedom for QPI is trivial (c.f. Sec.~\ref{ssec:equivalence_I}).
 
Our goal is to ensure that these bilinear tau form solutions also yield solutions of the quantum Hamiltonian form equations, with the Hamiltonian defined by \eqref{tau_universal}. In the remainder of this subsection we obtain the leading terms of the Hamiltonian function expansion for the above ansatz. The explicit identifications of the extra integration constants with the corresponding mass invariants (as well as the checks in the QPIII$_3$ and QPI cases) are then presented case by case in the next subsection.

\paragraph{Symmetries and rearrangement of the Zak transforms.}
For the irregular-type expansions we use direct analogs of formulas \eqref{Zak_tau1}, \eqref{Zak_tau2}, with $(a,\eta)$ replaced by its dual $(a_{\D},\eta_{\D})$ and the weak-coupling partition functions replaced by their strong-coupling counterparts. Accordingly, these tau functions satisfy the analog of the shift relation
\eqref{weak_shift}. However, the symmetry $a_{\D}\mapsto-a_{\D}$ is not automatic: it requires an additional transformation of the remaining parameters  (see Sec.~\ref{ssec:aut_symm}). 
The irregular-type tau function expansions are also  exchanged by the involutive antiautomorphism $^\mathrm{T}$. Its action on the formal variables is given by \eqref{aut_T}, with $(a,\eta)$ replaced by $(a_{\D},\eta_{\D})$, and similarly on the mass parameters of the Argyres-Douglas theories. We again postpone further discussion of tau function symmetries to Sec.~\ref{ssec:aut_symm}.

As in the regular type case, we rearrange the tau function $\tau^{(1)}$ (c.f. \eqref{rearr}) to facilitate extracting the Hamiltonian from \eqref{tau_universal}. Attaching the connection prefactors \eqref{prefactors_weak} (recall that $f^{[\mathrm{H_2}]}=f^{\mathrm{[H_1]}}=f^{\mathrm{[H_0]}}=1$) to the classical part by denoting $\widetilde{\mathcal{Z}}_{cl}^{\mathbf{[th]}}=f^{[\mathrm{th}]}\mathcal{Z}_{cl}^{\mathbf{[th]}}$, we obtain (c.f. \eqref{Zak_tau1_rearr})
\begin{equation}\label{Zak_tau1_rearr_strong}
\tau^{(1)}(a_{\D},\eta_{\D}|s)=\widetilde{\mathcal{Z}}^{\mathbf{[th]}}_{cl}(a_{\D};\epsilon{-}\kappa,\kappa|s) \sum_{n\in\mathbb{Z}}e^{\ri n\eta_{\D}(s)}\cdot C_n^{\mathbf{[th]}}(a_{\D};\kappa)s^{-N_p n^2/2}\mathcal{Z}^{\mathbf{[th]}}_{inst}(a_{\D}{+}n\kappa;\epsilon{-}\kappa,\kappa|s),
\end{equation}
with the modified operator $e^{\ri\eta_{\D}}$
\begin{equation}\label{eta_s}
e^{\ri\eta_{\D}(s)}=e^{\ri\eta_{\D}}\cdot\kappa^{N_p/2}\prod\limits_{i=1}^{N_p}\Gamma\big(1+\kappa^{-1}(a_{\D}+\mu_i+\epsilon/2)\big)\,e^{\delta s/\kappa}\left(e^{\chi}s^{-N_p}\right)^{\frac{2a_{\D}{+}\epsilon}{2\kappa}},
\end{equation}
and the rational "1-loop part"
\begin{equation}\label{C_n_strong}
C_n^{\mathbf{[th]}}(a_{\D};\kappa)=e^{\chi n^2/2}\prod_{i=1}^{N_p}P_{(n)}(a_{\D}+\mu_i-\epsilon/2;\kappa,\epsilon).   
\end{equation}
Applying the involutive antiautomorphism $^\mathrm{T}$ immediately yields the analogous expression for $\tau^{(2)}$ (c.f. \eqref{Zak_tau2_rearr})
\begin{equation}\label{Zak_tau2_rearr_strong}
\tau^{(2)}(a_{\D},\eta_{\D}|s)= \left(\sum_{n\in\mathbb{Z}}C_{-n}^{\mathbf{[th]}}(a_{\D};-\kappa)s^{-N_p n^2/2}\mathcal{Z}^{\mathbf{[th]}}_{inst}(a_{\D}{+}n\kappa;-\kappa,\epsilon{+}\kappa|s) \cdot e^{-\ri n\eta_{\D}(s)^\mathrm{T}}\right)\widetilde{\mathcal{Z}}^{\mathbf{[th]}}_{cl}(a_{\D};-\kappa,\epsilon{+}\kappa|s),
\end{equation}
where
$e^{-\ri \eta_{\D}(s)^\mathrm{T}}$ has the same structure as $e^{\ri\eta_{\D}(s)}$ in \eqref{eta_s}, but with the gamma function arguments (and the dimensional prefactor) replaced accordingly.
To interpret these rearranged expression 
as asymptotic expansions in $t$, we impose the same condition as in the regular-type case, i.e. $\mathrm{Re}(\epsilon/\kappa)>-1$. In the irregular-type case, however, one may also encounter an exponential growth coming from the factor $e^{\delta s/\kappa}$ in \eqref{eta_s}. We therefore require $\delta s/\kappa$ to be purely imaginary. This restricts the expansion to specific radial rays in the $t$-plane; after the rescaling $s\mapsto s/\kappa$, these rays coincide with the Stokes rays in the classical case \cite{BLMST16}. The relevant rays for all expansions are presented in \cite[Fig. 3, Tab. 3]{BST25}, where the rays corresponding to the two distinct expansions of a given (quantum) Painlev\'e equation are indicated by different colors.  

\paragraph{Hamiltonian expansions: equation.}
As in the regular type case, we substitute the rearranged Zak transform into \eqref{tau_universal} viewed as an equation for the Hamiltonian function. Taking $\tau^{(1)}$ (for definiteness) in the form \eqref{Zak_tau1_rearr_strong}, we solve
\begin{equation}\label{tau_s_H}
-2\epsilon_1(\epsilon_2{-}\epsilon_1)\frac{d\tau^{(1)}}{ds}=\tau^{(1)}\,H(a_{\D},\eta_{\D};\kappa,\epsilon|s),
\end{equation}
where we use the Hamiltonians with respect to time $s$ in all cases.
Guided by the generic classical case (with $\mathrm{Re}(a_{\D}/\kappa)\neq \mathbb{Z}+
\frac12$), we expect the Hamiltonian to admit an expansion of the form (c.f. \eqref{Ham_weak_ansatz})
\begin{equation}\label{Ham_strong_ansatz}
H(a_{\D},\eta_{\D};\kappa,\epsilon|s)=2\beta s+\xi_1+\xi_2 s^{-1}+\sum_{n\in\mathbb{Z}}e^{\ri n\eta_{\D}(s)}\left(e^{\chi}s^{-N_p}\right)^{|n|/2}H_n(s),\quad H_n(s)\in \mathbb{C}[[s^{-1}]].
\end{equation}
As in the regular-type case, we can regard this as an asymptotic expansion as $t\to\infty$ on the appropriate canonical rays provided $\mathrm{Re}(\epsilon/\kappa)>0$.
Collecting in \eqref{Ham_strong_ansatz} the terms that are multiplied by $e^{\ri N\eta}$ moved to the left, we obtain an equation for the coefficients $H_n$, namely (c.f. \eqref{Ham_weak_eq})
\begin{multline}\label{Ham_strong_eq}
\big(\delta(a_{\D}{+}N\kappa) -N_p/2(a_{\D}{+}N\kappa)^2s^{-1}\big) \mathcal{Z}^{\mathbf{[th]}}_{inst}(a_{\D}{+}N\kappa;\epsilon{-}\kappa,\kappa|s)-\kappa(\epsilon{-}\kappa)\frac{d}{ds} \mathcal{Z}^{\mathbf{[th]}}_{inst}(a_{\D}{+}N\kappa;\epsilon{-}\kappa,\kappa|s)\\=
\left(C_N^{\mathbf{[th]}}(a_{\D})\right)^{-1}\sum_{n\in\mathbb{Z}}C_{N-n}^{\mathbf{[th]}}(a_{\D}{+}n\epsilon)\mathcal{Z}^{\mathbf{[th]}}_{inst}(a_{\D}{+}(N{-}n)\kappa{+}n\epsilon;\epsilon{-}\kappa,\kappa|s)\left(e^{\chi}s^{-N_p}\right)^{\frac12|n|(|n|{+}1)-Nn}H_n(s).
\end{multline}
Again, for each fixed $N$, this should be understood as a system of recursive equations for the coefficients of the power series in $s^{-1}$. Let us denote these coefficients for $H_n(s)$ by
\begin{equation}\label{H_terms}
H_n(s)=\sum_{k=0}^{+\infty}H_{n}^{(k)}\,s^{-k}.  
\end{equation}
Then, considering \eqref{Ham_strong_eq} for $N=0,\pm1$ at order $s^0$, we obtain
\begin{equation}\label{H_leading}
H_{0}^{(0)}=\delta a_{\D}, \qquad H_{1}^{(0)}=\delta\kappa, \qquad H_{-1}^{(0)}=-\delta \kappa C_n^{[\mathbf{th}]}.  
\end{equation}
Higher-order terms start to depend on $N_p$ and on the successive coefficients of the instanton part. To verify the Hamiltonian-form equations in the case $N_p=1$, we need all coefficients $H^{(k)}_n$ with $k+n\leq 4$, except for $H_0^{(4)}$. We computed these terms for general values of the ansatz parameters, but the resulting expressions are too lengthy to include in the paper. For the cases $N_p>1$, besides \eqref{H_leading}, it is only necessary to use $H_0^{(1)}=-\frac12N_pa_{\D}^2$, which follows immediately from \eqref{Ham_strong_eq} with $N=0$ at order $s^{-1}$.

\subsection{Fixing asymptotics for the irregular-type expansions}
\label{ssec:fixing}

\begin{itemize}
    \item {\bf QPV ($N_p=4$, linear).}
For the QPV case there are two distinct irregular-type expansions. First one, with $N_p=4$ (called also linear $\mathbf{L}$, due to $\beta=0$) is defined by the following asymptotical behavior of the classical part 
\begin{equation}
\widetilde{\mathcal{Z}}^{[\mathbf{3_L}]}_{cl}(a_{\D},w_2^{[3]};\epsilon_1,\epsilon_2|t)=t^{\frac{4a_{\D}^2-w_2^{[3]}+\epsilon^2}{2\epsilon_1\epsilon_2}}e^{-\frac{a_{\D}t}{\epsilon_1\epsilon_2}},  
\end{equation}
where $\widetilde{\mathcal{Z}}_{cl}^{[\mathbf{3_L}]}$ differs from the classical part \cite[(4.9)]{BST25} by multiplication on the prefactor $f^{[3]}$ in \eqref{prefactors_weak}, thus eliminating its dependence on $e_1^{[3]}$. Then the 1-loop part is given by \cite[(4.18)]{BST25}:
    \begin{equation}\label{Z1loop_3L}
\mathcal{Z}_{1-loop}^{[\mathbf{3_L}]}(a_{\D},\tilde{m}_{1,2,3};\epsilon_1,\epsilon_2)= \prod\limits_{\varsigma_1,\varsigma_2=\pm1}  \exp \gamma_{\epsilon_1,\epsilon_2} \left(a_{\D}{+}\frac12(\varsigma_1 \tilde{m}_1{+}\varsigma_2 \tilde{m}_2{+}\varsigma_1 \varsigma_2 \tilde{m}_3{-}\epsilon)\right),
\end{equation}
where the tilded masses are defined via the $W(D_3)$ basic invariants with $w_4^{[3]}$ replaced by the $W(D_3)$-invariant integration constant $\tilde{w}_4^{[3]}$, i.e.
\begin{equation}
w_2^{[3]}=\tilde{m}_1^2+\tilde{m}_2^2+\tilde{m}_3^2, \qquad e_3^{[3]}=\tilde{m}_1\tilde{m}_2\tilde{m}_3, \qquad \tilde{w}_4^{[3]}=\tilde{m}_1^2 \tilde{m}_2^2+\tilde{m}_1^2 \tilde{m}_3^2+\tilde{m}_2^2 \tilde{m}_3^2\, .
\end{equation}
The instanton part \cite[(4.14)]{BST25} depends on the integration constant $\tilde{w}_4^{[3]}$ starting from order $t^{-2}$: 
\begin{multline}
-\epsilon_1\epsilon_2 \ln \mathcal{Z}_{inst}^{[\mathbf{3_L}]}(a_{\D},w_2^{[3]},e_3^{[3]},\tilde{w}_4^{[3]};\epsilon_1,\epsilon_2|t)=\left(4a_{\D}^3-(w_2^{[3]}{-}\epsilon^2)a_{\D}+e_3^{[3]}\right)\cdot\frac1{t}\\+\left(10 a_{\D}^4-(3w_2^{[3]}{-}5\epsilon^2)a_{\D}^2+4e_3^{[3]} a_{\D} +\frac{(w_2^{[3]}{-}\epsilon^2)^2}8-\frac{\tilde{w}_4^{[3]}}2\right)\cdot\frac1{t^2}+O\left(\frac1{t^3}\right)
\end{multline}
To fix $\tilde{w}_4^{[3]}$ from the Hamiltonian form equation \eqref{Ham_V} it is enough to use only the universal leading terms \eqref{H_leading} together with $H_0^{(1)}=-2a_{\D}^2$, also written there. It gives $\tilde{w}_4^{[3]}=w_4^{[3]}$, accordingly to the holomorphic anomaly computations in \cite[Sec. 5.1]{BST25}.
    \item {\bf QPV ($N_p=1$, square).}
The other QPV irregular-type expansion, with $N_p=1$ (called also square $\mathbf{S}$, due to $\beta\neq0$) is defined by the following asymptotical behavior of the classical part 
\begin{equation}
\widetilde{\mathcal{Z}}_{cl}^{[\mathbf{3_S}]}(a_{\D},w_2^{[3]};\epsilon_1,\epsilon_2|t)=t^{\frac{a_{\D}^2-2w_2^{[3]}+5\epsilon^2/4}{2\epsilon_1\epsilon_2}-\frac14}e^{-\frac{t^2/16+\ri a_{\D}t}{2\epsilon_1\epsilon_2}},
   \end{equation}  
which is the prefactor modified expression \cite[(4.25)]{BST25}. 
Then the mass-independent 1-loop part is given by \cite[(4.34)]{BST25}:    
\begin{equation}
\mathcal{Z}_{1-loop}^{[\mathbf{3_S}]}(a_{\D};\epsilon_1,\epsilon_2)=(2\ri)^{\frac{a_{\D}^2}{2\epsilon_1\epsilon_2}}  \exp \gamma_{\epsilon_1,\epsilon_2} (a_{\D}-\epsilon/2). 
\end{equation}
The instanton part \cite[(4.30)]{BST25} depends on the integration constant $\check{w}_4^{[3]}$ starting from order $t^{-2}$: 
\begin{multline}
-\epsilon_1\epsilon_2 \ln \mathcal{Z}_{inst}^{[\mathbf{3_S}]}(a_{\D},w_2^{[3]},e_3^{[3]},\check{w}_4^{[3]};\epsilon_1,\epsilon_2|t)=\Bigg(\frac{\alpha_{\D}^3}2+\left(4w_2^{[3]}{+}\frac74\epsilon_1\epsilon_2{-}\frac{11}8\epsilon^2\right)\alpha_{\D}+8e_3^{[3]}\Bigg)\cdot\frac1{t}\\+\Bigg(-\frac58\alpha_{\D}^4-\left(12w_2^{[3]}{+}\frac{45}8\epsilon_1\epsilon_2{-}\frac{65}{16}\epsilon^2\right)\alpha_{\D}^2-64e_3^{[3]}\alpha_{\D}-8\check{w}_4^{[3]}{-}2w_2^{[3]}\epsilon_1\epsilon_2{-}\frac12(\epsilon_1\epsilon_2)^2{+}\frac{61}{32}\epsilon_1\epsilon_2\epsilon^2\Bigg)\cdot\frac1{t^2}\\+\Bigg(\frac{11}8\alpha_{\D}^5+\left(\frac{148}3w_2^{[3]}{+}\frac{191}8\epsilon_1\epsilon_2{-}\frac{797}{48}\epsilon^2\right)\alpha_{\D}^3+448e_3^{[3]}\alpha_{\D}^2 \\ +\frac13\left(448\check{w}_4^{[3]}{+}16(w_2^{[3]})^2{+}(166\epsilon_1\epsilon_2{+}\epsilon^2)w_2^{[3]}{+}\frac{813}{16}(\epsilon_1\epsilon_2)^2{-}\frac{3797}{32}\epsilon_1\epsilon_2\epsilon^2{-}\frac{259}{128}\epsilon^4\right)\alpha_{\D}\\+\frac{16e_3^{[3]}}{3}(8w_2^{[3]}{+}10\epsilon_1\epsilon_2{-}17\epsilon^2)\Bigg)\cdot\frac1{t^3}+O\left(\frac1{t^4}\right), \qquad \quad \alpha_{\D}\equiv\ri a_{\D}.  
\end{multline}
In contrast to the $N_p=4$ case, to fix $\check{w}_4^{[3]}$ from the Hamiltonian form equation \eqref{Ham_V} it is necessary to use substantially more terms \eqref{H_terms} of the Hamiltonian expansion: $H_0(t)$ up to $t^{-3}$, $H_{\pm1}(t)$ up to $t^{-3}$, $H_{\pm2}(t)$ up to $t^{-2}$, $H_{\pm3}(t)$ up to $t^{-1}$, $H_{\pm4}(t)$ up to $t^0$. Their derivation requires the terms of the above instanton part up to $t^{-3}$. 
Finally, this derivation gives 
\begin{equation}
    \check{w}_4^{[3]}=w_4^{[3]}-\frac38w_2^{[3]}\epsilon^2+\frac{105}{1024}\epsilon^4,
     \end{equation}
which explicitly reproduces the holomorphic anomaly result \cite[(5.27)]{BST25}.
     
    \item {\bf QPIII$_1$.} For the QPIII$_1$ case there is a single irregular-type expansion, with $N_p=2$, which is defined by the following asymptotical behavior of the classical part
    \begin{equation}
\widetilde{\mathcal{Z}}^{[\mathbf{2}]}_{cl}(a_{\D},e_2^{[2]},\tilde{w}_2^{[2]};\epsilon_1,\epsilon_2|s)= s^{\frac{2a_{D}^2-e_2^{[2]}-\frac32\tilde{w}_2^{[2]}{+}\frac32\epsilon^2}{2\epsilon_1\epsilon_2}}e^{\frac{s^2/64-a_{\D} s}{2\epsilon_1\epsilon_2}},    
    \end{equation}
    where $s=8\ri\, t^{1/2}$ and $\widetilde{\mathcal{Z}}_{cl}^{[\mathbf{2}]}$ differs from the classical part \cite[(4.38)]{BST25} by multiplication on the prefactor $f^{[2]}$ in \eqref{prefactors_weak}. This classical part already depends on the $W(D_2)$-invariant integration constant $\tilde{w}_2^{[2]}$. Then the 1-loop part is given by \cite[(4.45)]{BST25}:
    \begin{equation}\label{Z1loop_2}
    \mathcal{Z}_{1-loop}^{[\mathbf{2}]}(a_{\D},\tilde{m}_{1,2};\epsilon_1,\epsilon_2)=\prod\limits_{\varsigma=\pm1}  \exp \gamma_{\epsilon_1,\epsilon_2} \left(a_{\D}{+}\frac{\varsigma}2(\tilde{m}_1{-}\tilde{m}_2)-\epsilon/2\right), 
\end{equation}
where the tilded masses are defined via the $W(D_2)$ basic invariants with $w_2^{[2]}$ replaced by $\tilde{w}^{[2]}_2$, i.e.
\begin{equation}
e_2^{[2]}=\tilde{m}_1\tilde{m}_2, \qquad \tilde{w}_2^{[2]}=\tilde{m}_1^2+\tilde{m}_2^2.
\end{equation}
The instanton part \cite[(4.41)]{BST25} depends on the integration constant $\tilde{w}_2^{[2]}$ right away from order $s^{-1}$.
To fix $\tilde{w}_2^{[2]}$ from the Hamiltonian form equation \eqref{Ham_III1} it is enough to use only the universal leading terms \eqref{H_leading} together with $H_0^{(1)}=-a_{\D}^2$, also written there. It gives $\tilde{w}_2^{[2]}=w_2^{[2]}$, accordingly to the holomorphic anomaly computations in \cite[Sec. 5.2]{BST25}.
    \item {\bf QPIII$_2$.}
    For the QPIII$_2$ case there is a single irregular-type expansion, with $N_p=1$, which is defined by the following asymptotical behavior of the classical part \cite[(4.53)]{BST25}
    \begin{equation}
\mathcal{Z}_{cl}^{[\mathbf{1}]}(a_{\D},\tilde{m}_1;\epsilon_1,\epsilon_2|\tilde{s})=\tilde{s}^{\frac{a_{\D}^2+23\epsilon^2/12-2\tilde{m}_1^2}{2\epsilon_1\epsilon_2}+\frac1{12}}e^{\frac{\tilde{s}^2/8+\tilde{m}_1 \tilde{s}-\sqrt3 a_{\D} \tilde{s}}{\epsilon_1\epsilon_2}},    
\end{equation}
where $\tilde{s}=\big(\frac{m_1}{\tilde{m}_1}\big)^{1/3}s=\big(54\frac{m_1}{\tilde{m}_1}t\big)^{1/3}$. It already depends on the integration constant $\tilde{m}_1$. Then the mass-independent 1-loop part is given by \cite[(4.61)]{BST25}, i.e. \eqref{Z1loop_anz} with $N_p=1$ and $\chi=-\ln (-12\sqrt{3})$.
The instanton part \cite[(4.55)]{BST25} depends on the integration constant $\tilde{m}_1$ right away from order $\tilde{s}^{-1}$.
To fix $\tilde{m}_1$ from the Hamiltonian form equation \eqref{Ham_III2} it is enough to use only the leading term $2\beta s=-\frac14\big(\frac{m_1}{\tilde{m}_1}\big)^{1/3}\tilde{s}$ of the expansion \eqref{Ham_strong_ansatz}. It gives $\tilde{m}_1^2=m_1^2$, accordingly to the holomorphic anomaly computations in \cite[Sec. 5.3]{BST25}.
    \item {\bf QPIII$_3$.}
    For the QPIII$_3$ case there is a single irregular-type expansion, with $N_p=1$, which is defined by the following asymptotical behavior of the classical part \cite[(4.67)]{BST25}
    \begin{equation}
    \label{Zcl_0} \mathcal{Z}_{cl}^{[\mathbf{0}]}(a_{\D};\epsilon_1,\epsilon_2|s)=s^{\frac{a_{\D}^2+9/4\epsilon^2}{2\epsilon_1\epsilon_2}+\frac14}e^{\frac{s^2/64+a_{\D} s}{4\epsilon_1\epsilon_2}},   
    \end{equation}
    where $s=-32\ri\, t^{1/4}$. Then the 1-loop part is given by \cite[(4.70)]{BST25}, i.e. \eqref{Z1loop_anz} with $N_p=1$ and $\chi=0$.
The instanton part is \cite[(4.69)]{BST25}. The Hamiltonian form equation \eqref{Ham_III3} is fulfilled simply by the structure of expansion \eqref{Ham_strong_ansatz}, accordingly to the holomorphic anomaly computations in \cite[Sec. 5.4]{BST25}.
   
    \item {\bf QPIV ($N_p=3$, linear).}
    For the QPIV case there are two distinct irregular-type expansions. First one, with $N_p=3$ (called also linear $\mathbf{L}$, due to $\beta=0$) is defined by the following asymptotical behavior of the classical part \cite[(4.80)]{BST25}
    \begin{equation}
\mathcal{Z}^{[\mathbf{H_{2,L}}]}_{cl}(a_{\D},\boldsymbol{e}_2;\epsilon_1,\epsilon_2|s)=s^{\frac{3a_{\D}^2+\boldsymbol{e}_2+\epsilon^2/4}{2\epsilon_1\epsilon_2}}e^{-\frac{a_{\D}s}{2\epsilon_1\epsilon_2}},
    \end{equation}
    where $s=t^2$.
    Then the 1-loop part is given by \cite[(4.90)]{BST25}:
\begin{equation}\label{Z1loop_H2L}
\mathcal{Z}_{1-loop}^{[\mathbf{H_{2,L}}]}(a_{\D},(\tilde{\boldsymbol{m}}_i)_{i=1}^3;\epsilon_1,\epsilon_2)= \prod\limits_{i=1}^3  \exp \gamma_{\epsilon_1,\epsilon_2} \left(a_{\D}+\tilde{\boldsymbol{m}}_i-\epsilon/2\right), 
\end{equation}
where the tilded masses are defined via the $W(A_2)$ basic invariants with $\boldsymbol{e}_3$ replaced by the $W(A_2)$-invariant integration constant $\tilde{\boldsymbol{e}}_3$, i.e.
\begin{equation}
\boldsymbol{e}_1=\tilde{\boldsymbol{m}}_1+\tilde{\boldsymbol{m}}_2+\tilde{\boldsymbol{m}}_3\equiv0, \qquad \boldsymbol{e}_2=\tilde{\boldsymbol{m}}_1\tilde{\boldsymbol{m}}_2+\tilde{\boldsymbol{m}}_2\tilde{\boldsymbol{m}}_3+\tilde{\boldsymbol{m}}_3\tilde{\boldsymbol{m}}_1, \qquad \tilde{\boldsymbol{e}}_3=\tilde{\boldsymbol{m}}_1\tilde{\boldsymbol{m}}_2\tilde{\boldsymbol{m}}_3\, .
\end{equation}
The instanton part \cite[(4.85)]{BST25} depends on the integration constant $\tilde{\boldsymbol{e}}_3$ right away from order $s^{-1}$.
To fix $\tilde{\boldsymbol{e}}_3$ from the Hamiltonian form equation \eqref{Ham_IV} it is enough to use only the universal leading terms \eqref{H_leading}. It gives $\tilde{\boldsymbol{e}}_3=\boldsymbol{e}_3$, accordingly to the holomorphic anomaly computations in \cite[Sec. 5.5]{BST25}.
    
    \item {\bf QPIV ($N_p=1$, square).}
    The other QPIV irregular-type expansion, with $N_p=1$ (called also square $\mathbf{S}$, due to $\beta\neq0$) is defined by the following asymptotical behavior of the classical part \cite[(4.97)]{BST25}
    \begin{equation}
    \mathcal{Z}_{cl}^{[\mathbf{H_{2,S}}]}(a_{\D},\boldsymbol{e}_2;\epsilon_1,\epsilon_2|s)= s^{\frac{a_{\D}^2+5\epsilon^2/12+3\boldsymbol{e}_2}{2\epsilon_1\epsilon_2}-\frac16}e^{-\frac{s^2/18+\ri \sqrt3 a_{\D} s}{6\epsilon_1\epsilon_2}}, 
    \end{equation}
    where $s=t^2$.
    Then the mass-independent 1-loop part is given by \cite[(4.107)]{BST25}, i.e. \eqref{Z1loop_anz} with $N_p=1$ and $\chi=-\ln(\sqrt3 \ri)$. The instanton part \cite[(4.102)]{BST25} depends on the integration constant $\check{\boldsymbol{e}}_3$ right away from order $s^{-1}$. In contrast to the $N_p=3$ case, to fix $\check{\boldsymbol{e}}_3$ from the Hamiltonian form equation \eqref{Ham_IV} it is necessary to use substantially more terms \eqref{H_terms} of the Hamiltonian expansion: $H_0(t)$ up to $t^{-2}$, $H_{\pm1}(t)$ up to $t^{-2}$, $H_{\pm2}(t)$ up to $t^{-1}$, $H_{\pm3}(t)$ up to $t^0$. Their derivation requires the terms of the instanton part up to $s^{-2}$. 
    Finally, this derivation gives $\check{\boldsymbol{e}}_3=\boldsymbol{e}_3$,  accordingly to the holomorphic anomaly computations in \cite[Sec. 5.5]{BST25}.
    
    \item {\bf QPII ($N_p=2$, linear).} 
    For the QPII case there are two distinct irregular-type expansions. First one, with $N_p=2$ (called also linear $\mathbf{L}$, due to $\beta=0$) is defined by the following asymptotical behavior of the classical part \cite[(4.111)]{BST25}
    \begin{equation}
\mathcal{Z}^{[\mathbf{H_{1,L}}]}_{cl}(a_{\D},\tilde{\boldsymbol{m}}^2;\epsilon_1,\epsilon_2|s)=s^{\frac{2a_{\D}^2-\frac23\tilde{\boldsymbol{m}}^2+\epsilon^2/6}{2\epsilon_1\epsilon_2}}e^{-\frac{a_{\D}s}{3\epsilon_1\epsilon_2}},
    \end{equation}
    where $s=2\sqrt2\ri\,t^{3/2}$.
    It already depends on the $W(A_1)$-invariant integration constant $\tilde{\boldsymbol{m}}^2$. Then the 1-loop part is given by \cite[(4.117)]{BST25}:
\begin{equation}\label{Z1loop_H1L}
\mathcal{Z}_{1-loop}^{[\mathbf{H_{1,L}}]}(a_{\D},\tilde{\boldsymbol{m}};\epsilon_1,\epsilon_2)= 2^{\frac{a_{\D}^2}{\epsilon_1\epsilon_2}}\prod\limits_{\lambda=\pm1}  \exp \gamma_{\epsilon_1,\epsilon_2} \left(a_{\D}+\lambda\tilde{\boldsymbol{m}}-\epsilon/2\right). 
\end{equation}
The instanton part \cite[(4.113)]{BST25} depends on the integration constant $\tilde{\boldsymbol{m}}^2$ right away from order $s^{-1}$.
To fix $\tilde{\boldsymbol{m}}^2$ from the Hamiltonian form equation \eqref{Ham_II} it is enough to use only the universal leading terms \eqref{H_leading}. It gives $\tilde{\boldsymbol{m}}^2=\boldsymbol{m}^2$,  accordingly to the holomorphic anomaly computations in \cite[Sec. 5.6]{BST25}.
    
    \item {\bf QPII ($N_p=1$, square).}
    The other QPII irregular-type expansion, with $N_p=1$ (called also square $\mathbf{S}$, due to $\beta\neq0$) is defined by the following asymptotical behavior of the classical part \cite[(4.123)]{BST25}
    \begin{equation}
    \mathcal{Z}_{cl}^{[\mathbf{H_{1,S}}]}(a_{\D},\check{\boldsymbol{m}}^2;\epsilon_1,\epsilon_2|s)= s^{\frac{a_{\D}^2-\frac43\check{\boldsymbol{m}}^2}{2\epsilon_1\epsilon_2}-\frac1{12}}e^{-\frac{s^2/32-\ri \sqrt2 a_{\D} s}{6\epsilon_1\epsilon_2}}, 
    \end{equation}
    where $s=2\sqrt2\ri\,t^{3/2}$. It already depends on the $W(A_1)$-invariant integration constant $\check{\boldsymbol{m}}^2$.
    Then the mass-independent 1-loop part is given by \cite[(4.129)]{BST25}, i.e. \eqref{Z1loop_anz} with $N_p=1$ and $\chi=-\ln(-2\sqrt2 \ri)$. The instanton part \cite[(4.125)]{BST25} depends on the integration constant $\check{\boldsymbol{m}}^2$ right away from order $s^{-1}$. In contrast to the $N_p=2$ case, to fix $\check{\boldsymbol{m}}^2$ from the Hamiltonian form \eqref{Ham_II} it is necessary to use substantially more terms \eqref{H_terms} of the Hamiltonian expansion: $H_0(t)$ up to $t^{-1}$, $H_{\pm1}(t)$ up to $t^{-1}$, $H_{\pm2}(t)$ up to $t^{-0}$. Their derivation requires the terms of the instanton part up to $s^{-1}$. 
    Finally, this derivation gives
     \begin{equation}
     \check{\boldsymbol{m}}^2=\boldsymbol{m}^2-\frac3{32}\epsilon^2,   
\end{equation} 
which explicitly reproduces the holomorphic anomaly result \cite[(5.76)]{BST25}.
   
    \item {\bf QPI.}
    For the QPI case there is a single irregular type expansion, with $N_p=1$, which is defined by the following asymptotical behavior of the classical part \cite[(4.133)]{BST25}
    \begin{equation}
    \mathcal{Z}_{cl}^{[\mathbf{H_0}]}(\epsilon_1,\epsilon_2|s)= s^{\frac{a_{\D}^2+7\epsilon^2/60}{2\epsilon_1\epsilon_2}-\frac1{60}}e^{-\frac{-s^2/12^3+a_{\D} s}{60\epsilon_1\epsilon_2}},   
    \end{equation}
    where $s=-8(\ri{+}1)\, (6t)^{5/4}$. Then the 1-loop part is given by \cite[(4.136)]{BST25}, i.e. \eqref{Z1loop_anz} with $N_p=1$ and $\chi=0$.
The instanton part is \cite[(4.135)]{BST25}. The Hamiltonian form equation \eqref{Ham_I} is fulfilled from the leading term $2\beta s=-\frac{s}{30\cdot 12^3}$ of the expansion \eqref{Ham_strong_ansatz}, accordingly to the holomorphic anomaly computations in \cite[Sec. 5.7]{BST25}.
\end{itemize}

\section{Symmetries of tau functions as Zak transforms}
\label{sec:tau_symm}

\setcounter{subsection}{-1}

\subsection{Overview}
As already mentioned in Secs. \ref{sec:QPVI}, \ref{sec:coalescence}, the tau functions $\tau^{(1)}$ and $\tau^{(2)}$ are invariant under the autonomous transformations, represented by the corresponding extended finite Weyl groups. This invariance follows from the Hamiltonian symmetry \eqref{Ham_inv} via the tau function definition \eqref{tau_universal}. We start this section from lifting these symmetries to the level of the Zak transform solutions. These symmetries actually correspond to the symmetries of the instanton and the classical parts in the weak- and the strong-coupling regime, while
transformations of the 1-loop part in some cases yield simple transformations of $(a,\eta)$ or $(a_{\D},\eta_{\D})$.

The action of the nonautonomous transformations on the Hamiltonian can be expressed in terms of the Hamiltonian function itself and its total time derivatives. The simplest example is provided by the QPIII$_3$ nonautonomous transformation of order $2$. We obtain two bilinear $(\epsilon_1,\epsilon_2)$- Hirota equations relating the initial and transformed tau functions; following \cite{BS16b}, we call them Okamoto-like. Based on an asymptotic analysis, we propose that, at the level of the Zak transform solutions, this transformation acts as shift $a\mapsto a+\kappa/2$ in the weak-coupling regime and as sign change $e^{\ri\eta_{\D}}\mapsto -e^{\ri\eta_{\D}}$ in the strong-coupling regime. 
In the weak-coupling regime, the resulting bilinear $(\epsilon_1,\epsilon_2)$- Hirota equations for the Zak transform solutions are, via the AGT correspondence \cite{AGT09}, equivalent to those obtained in \cite{BS16b} from the representation theory of the $\mathcal{N}=1$ super Virasoro algebra. Such 
bilinear relations can be interpreted as $\mathbb{C}^2/\mathbb{Z}_2$ blowup relations in a nontrivial holonomy sector. We also present their analogs in the strong-coupling regime, complementing the quantum Toda-like equations of \cite[Sec.~7]{GMS20}.

As we discussed in Sec.~\ref{sec:QPVI}, \ref{sec:coalescence}, in the other cases (except trivial QPI), the whole symmetry group is a semidirect product of the autonomous symmetry group and the weight lattice (translation) group $P$ (recall the latter equality of \eqref{Pdecomp}). We derive quantum Okamoto-like equations relating tau function $\tau^{(1)}$ and tau function $\tau^{(2)}$ transformed under shifts $m_f\mapsto m_f+\kappa/2$, $f=1,\ldots,N_f$, for QPVI, QPV, and QPIII$_{1,2}$. Also we derive them for QPIV, QPII with the transformations under shifts $(\boldsymbol{m}_1,\boldsymbol{m}_2,\boldsymbol{m}_3)\mapsto\big(\boldsymbol{m}_1{-}\kappa/3,\boldsymbol{m}_2{-}\kappa/3,\boldsymbol{m}_3{+}2\kappa/3\big)$ and $\boldsymbol{m}\mapsto\boldsymbol{m}{+}\kappa/2$, respectively. At the level of the Zak transform solutions in the weak- and strong-coupling regimes, these equations determine the action on the Zak transform parameters and further lead to several $\mathbb{C}^2/\mathbb{Z}_2$ blowup relations. For each quantum Painlev\'e equation, these translations together with the autonomous symmetries generate the whole symmetry group. Thus we find its action on the Zak transform parameters in the both regimes.

This section is organized as follows. In Sec.~\ref{ssec:aut_symm} we discuss the tau function invariance under the autonomous symmetries at the level of the Zak transforms from Sec.~\ref{sec:asymptotics}, and determine the corresponding transformations of pairs $(a,\eta)$ and $(a_{\D},\eta_{\D})$. In addition, we discuss $(\epsilon_1,\epsilon_2)$-symmetries; in particular, we introduce the hermitian conjugation operator $^\dagger$. In Sec.~\ref{ssec:C2} we treat the toy example of the $C_2$ nonautonomous symmetry of QPIII$_3$ and derive the corresponding quantum Okamoto-like bilinear equations. We then show that, in the weak- and the strong-coupling regimes, these equations can be interpreted as $\mathbb{C}^2/\mathbb{Z}_2$ blowup relations. Finally, in Sec.~\ref{ssec:Okamoto-like}
we present quantum Okamoto-like equations for all the other cases (except trivial QPI) considered for certain symmetries
together with the corresponding $\mathbb{C}^2/\mathbb{Z}_2$ blowup relations amd the action on the Zak transform parameters.

\subsection{Autonomous symmetries}
\label{ssec:aut_symm}

\paragraph{Symmetries of $\mathcal{Z}^{[N_f]}_{inst}$.}
We start describing the weak-coupling instanton part \eqref{Zinst} symmetries from the case of $N_f=4$. Via the AGT correspondence \cite{AGT09}, these symmetries follow from the symmetries of the $4$-point Virasoro conformal blocks.
In our notations the AGT relation is given by formula \cite[(C.23)]{BST25} and the dictionary just below it. This immediately gives invariance of $\mathcal{Z}^{[4]}_{inst}$ under exchange  $\epsilon_1\leftrightarrow\epsilon_2$. Moreover, product $f^{[4]}\mathcal{Z}_{inst}^{[4]}$ (with $\epsilon_1\leftrightarrow\epsilon_2$-invariant prefactor \eqref{prefactors_weak}) is explicitly invariant under the sign changes $(\epsilon_1,\epsilon_2)\mapsto(-\epsilon_1,-\epsilon_2)$, $m_1\leftrightarrow-m_3$, and $m_2\leftrightarrow-m_4$. Together with the mass permutation invariance of the Nekrasov formula \eqref{Zinst}, the latter two symmetries generate the full $W(D_4)$ mass symmetry (of Sec. \ref{ssec:QPVI_symm}) of product $f^{[4]}\mathcal{Z}_{inst}^{[4]}$. Finally, let us consider element $\sigma_{13}=\sigma_{14}\sigma_{34}\sigma_{14}$ (recall Table~\ref{table:QPVI_Backlund}), which acts on $t$ by $t\mapsto \frac{t}{t{-}1}$, and thus generates the stabilizer of $t=0$ in $S_3=\mathrm{Aut}(D_4)$; as mentioned in Sec.~\ref{ssec:QPVI_symm} it acts
on the masses simply by $m_2\mapsto -m_2$. Then this element preserves (up to a numerical complex phase) product $\mathcal{Z}_{cl}^{[4]}f^{[4]}\mathcal{Z}_{inst}^{[4]}$, according to \cite[(2.39)]{BST25}.

The corresponding instanton part symmetries for $N_f<4$ can be obtained following the successive limits of the first row of the quantum Painlev\'e coalescence diagram~\ref{fig:QPainleve_coalescence}, via formula \cite[(3.24)]{BST25}. This procedure implies that products $f^{[N_f]}\mathcal{Z}_{inst}^{[N_f]}$ (with $\epsilon_1\leftrightarrow\epsilon_2$-invariant prefactors \eqref{prefactors_weak}) for $N_f=3,2,1,0$ are automatically invariant under the sign change $(\epsilon_1,\epsilon_2)\mapsto(-\epsilon_1,-\epsilon_2)$ and exchange $\epsilon_1\leftrightarrow\epsilon_2$. Moreover, these products obey the corresponding extended finite Weyl group symmetries:
$C_2\ltimes W(D_3)$ for $N_f=3$ (see Sec. \ref{ssec:QPVI_Ham_tau}), $C_2\langle\sigma\rangle\ltimes W(D_2)$ for $N_f=2$ (see Sec. \ref{ssec:QPIII1}), and $W(A_1)$ for $N_f=1$ (see Sec. \ref{ssec:QPIII2}). 

\paragraph{Symmetries of $\mathcal{Z}^{[\mathbf{th}]}_{inst}$.} The strong-coupling instanton part \eqref{Zinst_anz} is invariant under exchange $\epsilon_1\leftrightarrow\epsilon_2$ simply by an assumption of the ansatz. All the strong coupling expansions derived in \cite{BST25}, and further specified in the previous section, are expressed in terms of $\epsilon_1\epsilon_2, \epsilon^2$ and the finite Weyl group basic mass invariants, in accordance with the parameter dependence of the yielding equations.
Thus they are automatically invariant under the sign change
$(\epsilon_1,\epsilon_2)\mapsto(-\epsilon_1,-\epsilon_2)$ and the finite Weyl group action. They are also symmetric under the finite Weyl group automorphisms, but with the additional sign change $a_{\D}\mapsto-a_{\D}$. Specifically:
\begin{itemize}
    \item In the QPV case, $\mathcal{Z}_{inst}^{[\mathbf{3_L}]}$ \cite[(4.14)]{BST25} and $\mathcal{Z}_{inst}^{[\mathbf{3_S}]}$ \cite[(4.30)]{BST25} are invariant under $(a_{\D},e_3|t)\mapsto {(-a_{\D},-e_3|-t)}$, which corresponds to the action of $\sigma_{13}$ (recall Table~\ref{table:QPV_Backlund}). This symmetry relates the expansions along the canonical rays $\mathrm{Arg}\, t=\pm\pi/2$ (for $\mathbf{3_L}$) or $\mathrm{Arg}\, t=0,\pi$ (for $\mathbf{3_S}$).
    
    \item In the QPIII$_1$ case, $\mathcal{Z}_{inst}^{[\mathbf{2}]}$ \cite[(4.41)]{BST25} is invariant under $(a_{\D}|s)\mapsto (-a_{\D}|-s)$, which corresponds to the branching of variable $s=8\ri\, t^{1/2}$.
    At the same time, the finite Weyl group automorphism $\sigma$ of Table~\ref{table:QPIII1_Backlund} relates the expansions along two canonical rays $\mathrm{Arg}\, t=0,\pi$.

    \item In the QPIII$_2$ case, $\mathcal{Z}_{inst}^{[\mathbf{1}]}$ \cite[(4.55)]{BST25} is invariant under $(a_{\D},m_1|s)\mapsto (-a_{\D},-m_1|-s)$, which corresponds to the action of $s_1$ (recall Table~\ref{table:QPIII2_Backlund}). This symmetry relates the expansions along two canonical rays $\mathrm{Arg}\, t=\pm\pi/2$.

   \item In the QPIII$_3$ case, $\mathcal{Z}_{inst}^{[\mathbf{0}]}$ \cite[(4.69)]{BST25} is invariant under $(a_{\D}|s)\mapsto (-a_{\D}|-s)$, which corresponds to the branching of variable $s=-32\ri\, t^{1/4}$.

   \item In the QPIV case, $\mathcal{Z}_{inst}^{[\mathbf{H_{2,L}}]}$ \cite[(4.85)]{BST25} and $\mathcal{Z}_{inst}^{[\mathbf{H_{2,S}}]}$ \cite[(4.102)]{BST25} are invariant under $(a_{\D},\boldsymbol{e}_3|s)\mapsto (-a_{\D},-\boldsymbol{e}_3|-s)$, which corresponds to the action of $\sigma_{12}\in C_4$ (recall Table~\ref{table:QPIV_Backlund}). This symmetry relates the expansions along the canonical rays $\mathrm{Arg}\, t=\pm \frac\pi4,\pm\frac{3\pi}4$ (for $\mathbf{H_{2,L}}$) or $\mathrm{Arg}\, t=0,\pm \frac\pi2,\pi$ (for $\mathbf{H_{2,S}}$).

   \item In the QPII case, $\mathcal{Z}_{inst}^{[\mathbf{H_{1,L}}]}$ \cite[(4.113)]{BST25} and $\mathcal{Z}_{inst}^{[\mathbf{H_{1,S}}]}$ \cite[(4.125)]{BST25} are invariant under $(a_{\D}|s)\mapsto {(-a_{\D}|-s)}$, which corresponds to the action of $\sigma\in C_3$ (recall Table~\ref{table:QPII_Backlund}) and to the branching of variable $s=2\sqrt2\ri\,t^{3/2}$, simultaneously. This symmetry relates the expansions along the canonical rays $\mathrm{Arg}\, t=0, \pm\frac{2\pi}3$ (for $\mathbf{H_{1,L}}$) or $\mathrm{Arg}\, t=\pi,\pm\frac\pi3$ (for $\mathbf{H_{1,S}}$).

   \item In the QPI case, $\mathcal{Z}_{inst}^{[\mathbf{H_{0}}]}$ \cite[(4.135)]{BST25} is invariant under $(a_{\D}|s)\mapsto (-a_{\D}|-s)$, which corresponds to the action of the $C_5$ group \eqref{C5_I} and to the branching of variable $s=-8(\ri{+}1)\, (6t)^{5/4}$, simultaneously. It relates the expansions along five canonical rays $\mathrm{Arg}\, t=\pi, \pm\frac{3\pi}5, \pm\frac{\pi}5$.
\end{itemize}

\paragraph{Symmetries of the tau functions.}
In the weak-coupling case, we now lift the instanton part symmetries to those of the tau functions $\tau^{(1)}$ and $\tau^{(2)}$ given by the Zak transforms \eqref{Zak_tau1}, \eqref{Zak_tau2}, respectively. The mass independent classical part \eqref{Zcl} is automatically invariant under the finite Weyl groups, while the 1-loop parts \eqref{Z1loop} are explicitly invariant only under the mass permutation subgroups. Nevertheless, the transformations of the 1-loop parts under the mass sign changes can be absorbed by an appropriate shift of $\eta$. Indeed, the rearranged Zak transforms \eqref{Zak_tau1_rearr}, \eqref{Zak_tau2_rearr} with \eqref{eta_t}, \eqref{C_n_weak} under the mass sign change $m_f\mapsto -m_f$ obtain a multiplication of $e^{\ri\eta}$ by a $\kappa$-periodic function in $a$, namely
\begin{equation}
e^{\ri\eta}\mapsto -e^{\ri\eta}\prod_{\pm}\sin^{\pm1}\Big(\pi\kappa^{-1}\big(m_f\pm(a{+}\epsilon/2)\big)\Big).     
\end{equation}
To obtain this formula it is necessary to use the symmetry property of polynomial $P_{(n)}(x;\kappa,\epsilon)$ 
\begin{equation}
P_{(n)}(-x;\kappa,\epsilon)=(-1)^{\frac{n(n{-}1)}2}\prod_{k=\frac12}^{|n|-\frac12}\left(x-\frac12\epsilon-\mathrm{sgn}(n)k\epsilon\right)^{-\mathrm{sgn}(n)} P_{(-n)}(x-\epsilon;\kappa,\epsilon),  
\end{equation}
which immediately follows from its definition \eqref{Pn}. 
With this result we lift to the level of the tau functions not only the whole finite Weyl groups, but also their automorphisms, except those in the QPVI case that do not preserve $t=0$. These QPVI automorphisms map the expansion around $t=0$ to  the expansions around the remaining regular singularities $t=1$ and $t=\infty$. 

In contrast to the weak-coupling regime, the strong-coupling 1-loop parts \eqref{Z1loop_anz} are invariant under the whole finite Weyl group. This is automatically true for the mass-independent 1-loop parts with $N_p=1$, while for the 1-loop parts from \eqref{Z1loop_3L}, \eqref{Z1loop_2}, \eqref{Z1loop_H2L}, \eqref{Z1loop_H1L} with $N_p>1$ it is visible case by case; the tilded masses there coincide with the non-tilded masses, as fixed in Sec.~\ref{ssec:fixing}. The automorphisms of the finite Weyl groups, together with the branchings of the strong-coupling variable $s$, already described above for the instanton part, preserve the classical parts up to a numerical complex phases (with an exception for QPIII$_1$, where automorphism $\sigma$ relates the expansions on two canonical rays). The corresponding transformation of the 1-loop parts can be absorbed by an appropriate shift of $\eta_{\D}$, as in the weak-coupling case. Indeed, all these finite Weyl groups automorphisms and the branchings act on the 1-loop part \eqref{Z1loop_anz} arguments by $a_{\D}\mapsto-a_{\D}$ and $\mu_i\mapsto-\mu_i, {\scriptstyle i=1,\ldots N_p}$. So, the rearranged Zak transforms \eqref{Zak_tau1_rearr_strong}, \eqref{Zak_tau2_rearr_strong} with \eqref{eta_s}, \eqref{C_n_strong} under these symmetries obtain a multiplication of $e^{\ri\eta_{\D}}$ by a $\kappa$-periodic function in $a$, namely
\begin{equation}\label{eta_D_periodic_factor}
e^{\ri\eta_{\D}}\mapsto (-1)^{-N_p/2}e^{-\ri\eta_{\D}}(-1)^{-\frac{2a_{\D}{+}\epsilon}{2\kappa}N_p}\prod_{i=1}^{N_p}\pi^{-1}\sin\Big(\pi\kappa^{-1}\big(a_{\D}+\mu_i+\epsilon/2\big)\Big).
\end{equation}

Finally, we consider the sign change transformation $(\epsilon_1,\epsilon_2)\mapsto(-\epsilon_1,-\epsilon_2)$ at the level of the tau functions. The weak-coupling \eqref{Zcl} and the strong-coupling  \eqref{Zcl_anz} classical parts multiplied by prefactors \eqref{prefactors_weak} (recall that $f^{[\mathrm{H}_k]}=1,{\scriptstyle k=2,1,0}$) are automatically invariant under this sign change. 
The strong-coupling 1-loop part \eqref{Z1loop_anz} and the fundamental part (numerator) of the weak-coupling 1-loop part \eqref{Z1loop}
are also invariant provided by the double gamma function symmetry $\gamma_{-\epsilon_1,-\epsilon_2}(x)=\gamma_{\epsilon_1,\epsilon_2}(x+\epsilon)$, which follows from its definition \eqref{gammaNYdef}. Thus the $(\epsilon_1,\epsilon_2)$- sign change for the strong-coupling tau function expansions immediately leads to the sign change $\eta_{\D}\mapsto -\eta_{\D}$ Although the vector part (denominator) of \eqref{Z1loop} is not invariant, the $(\epsilon_1,\epsilon_2)$- sign change in the weak-coupling case leads to the sign change $\eta\mapsto -\eta$ as well, which can be seen from the rearranged form \eqref{Zak_tau1_rearr}, \eqref{Zak_tau2_rearr} with \eqref{eta_t}, \eqref{C_n_weak}. The composition of the obtained $(\epsilon_1,\epsilon_2)$- sign change with the involutive antiautomorphism $^\mathrm{T}$, given by \eqref{aut_T} yields an involutive antiautomorphism, which we denote by $^\dagger$. It can be lifted to the level of the Hamiltonian system imposing $q^\dagger=q,\,p^\dagger=p$. Then, in the case of a real Planck's constant $\ri\epsilon$ and a real scaling $\kappa$ this involutive antiautomorphism can be considered as the standard QM hermitian conjugation. Notice that these realness restrictions correspond via the AGT correspondence \cite{AGT09} to the Virasoro central charge $c\in [1,\infty)$. The physical sense of the $(\epsilon_1,\epsilon_2)$- sign change, however, remains unclear for us.

\subsection{QPIII$_3$ symmetry and $\mathbb{C}^2/\mathbb{Z}_2$ blowup relations}
\label{ssec:C2}

\paragraph{Action on the Hamiltonian and bilinear equations on the tau functions.}

Actions of the nonautonomous symmetries on the Hamiltonians and, via definition \eqref{tau_universal}, on the tau functions are rather sophisticated. We start discussing them on the toy example of QPIII$_3$, which has $C_2$ symmetry generated by the nonautonomous transformation $\pi$ (see Table~\ref{table:QPIII3_Backlund}). The corresponding transformations of $H, H', H''$ follow from \eqref{Hder_III3}:
\begin{equation}\label{pi_H_III3}
\pi(H)=H+\frac{\kappa^2}2(H')^{-1}H''{-}\frac{\kappa}2\epsilon{-}\frac{\kappa^2}4, \qquad \pi(H')=t(H')^{-1}, \qquad \pi(H'')=t (H')^{-1}-t(H')^{-1}H''(H')^{-1}.
\end{equation}
These transformations of the Hamiltonian were used in \cite[Sec. 2.2]{BS16b} in the classical ($\epsilon=0$) case to obtain the so-called Okamoto-like \cite[(2.14, 2.15)]{BS16b} and Toda-like \cite[(2.24)]{BS16b} bilinear equations relating the tau function $\tau$ and its B\"acklund transformation $\pi(\tau)$. We obtain the quantized versions of the Okamoto-like equations, slightly modifying our previous derivation of the bilinear tau forms. Namely, we obtain
\begin{equation}\label{Okamoto_III3}
\mathfrak{D}^{k;\pi}_{\epsilon_1,\epsilon_2}\left(t^{\frac{\epsilon}{8\kappa}}\tau^{(1)},t^{-\frac{\epsilon}{8\kappa}}\pi\big(\tau^{(2)}\big)\right)=0,  
\end{equation}
where $\mathfrak{D}^{k;\pi}_{\epsilon_1,\epsilon_2}$ are bilinear differential operators of order $k=2,3$ in $\ln t$, given by
\begin{equation}\label{Okamoto_operators_III3}
\mathfrak{D}^{k;\pi}_{\epsilon_1,\epsilon_2}=D^k_{\epsilon_1,\epsilon_2}+\frac12\left(\epsilon_1\epsilon_2\frac{d}{d\ln t}\right)D^{k-2}_{\epsilon_1,\epsilon_2}-\frac1{16}\left(\epsilon_1\epsilon_2{+}\epsilon^2\right)D^{k-2}_{\epsilon_1,\epsilon_2}.
\end{equation}
Concretely, these equations are ensured by substituting definition \eqref{tau_universal} for the tau functions to their $(\epsilon_1,\epsilon_2)$- Hirota derivatives and checking the fulfillment of the corresponding relation in $H$ and $\pi(H)$ by using their expressions as polynomials in $q,q^{-1},p,t,\kappa,\epsilon$. Notice that, in addition to these equations, we immediately obtain the same pair but with subscript $\pi$ on $\tau^{(1)}$ instead of $\tau^{(2)}$. Obtained two pairs of equations can be viewed as an implicit description of the action of symmetry $\pi$ on the tau functions. Unfortunately, we have not found a way to derive the quantized version of the Toda-like equations from the QPIII$_3$ Hamiltonian system, despite their actual existence \cite[Sec. 7]{GMS20} as $\mathbb{C}^2/\mathbb{Z}_2$ blowup equations. For comparison, we present these Toda-like equations below.

\paragraph{$\mathbb{C}^2/\mathbb{Z}_2$ blowup relations of \cite{BS16b}.}
As already mentioned, 
in \cite{BST25} the tau form \eqref{D13_VI}, \eqref{D4_VI} of the QPVI equation was obtained from the $\mathbb{C}^2/\mathbb{Z}_2$ blowup relations. Via the AGT correspondence \cite{AGT09}, these relations were derived in \cite{BS14} as the bilinear relations on the irregular Virasoro conformal blocks, using the representation theory of the $\mathcal{N}=1$ super Virasoro algebra in the Neveu--Schwarz sector. Paper \cite{BS16b} presents the analogous derivation but in the Ramond sector of the $\mathcal{N}=1$ super Virasoro algebra. The Okamoto-like equations (which are \eqref{Okamoto_III3} with \eqref{Okamoto_operators_III3} for $\epsilon=0$) were obtained there from relations \cite[(4.47)]{BS16b} via \cite[(4.44), (4.41)]{BS16b} in the special case of the central charge $c=1$ for the irregular Virasoro conformal blocks. Via the AGT correspondence, these relations are simply translated into the bilinear relations on the pure gauge $\mathcal{N}=2$ $D=4$ SUSY $SU(2)$ partition function. Namely, this partition function is related with the irregular Virasoro conformal block $\mathcal{F}_c(\Delta|\mathrm{t})$ (defined by \cite[(3.6)]{BS16b}) by the irregular analog of the general AGT relation \cite[(C.23)]{BST25} under the AGT dictionary \cite[(C.24)]{BST25}, namely
\begin{equation}\label{AGT_dct}
\mathrm{t}^{-\Delta}\mathcal{F}_c(\Delta|\mathrm{t})=\mathcal{Z}^{[0]}_{inst}\big(a;\epsilon_1,\epsilon_2\big|\epsilon_1^2\epsilon_2^2\mathrm{t}\big), \qquad \textrm{under} \qquad c=1+6\frac{\epsilon^2}{\epsilon_1\epsilon_2}, \qquad   \Delta=\frac{\epsilon
^2-4a^2}{4\epsilon_1\epsilon_2}.  
\end{equation}
Also, under dictionary \cite[(C.31)]{BST25} factor $(l_n^{+,+})^2$ from \cite[(4.42)]{BS16b} becomes
\begin{equation}\label{l_n_++}
(l_n^{+,+})^2=\frac12(\epsilon_1\epsilon_2)^{\frac{a^2{-}\epsilon^2/4}{\epsilon_1\epsilon_2}-\frac14}\frac{\big(2\epsilon_1(\epsilon_2{-}\epsilon_1)\big)^{\frac{\epsilon^2/4{-}(a{+}2n\epsilon_1)^2}{\epsilon_1(\epsilon_2{-}\epsilon_1)}} \big(2\epsilon_2(\epsilon_1{-}\epsilon_2)\big)^{\frac{\epsilon^2/4{-}(a{+}2n\epsilon_2)^2}{\epsilon_2(\epsilon_1{-}\epsilon_2)}}}{\prod\limits_{\mathrm{reg}(4n)}\Big(\epsilon^2/4- \big(2a+\mathrm{sgn}(n)(i\epsilon_1+j\epsilon_2)\big)^2\Big)},   
\end{equation}
where in the denominator we see exactly that of the blowup factor \cite[(2.21)]{BST25}, which comes from the vector part (denominator) of \eqref{Z1loop} via \cite[(B.9)]{BST25}. In addition to the bilinear relations \cite[(4.47)]{BS16b}, according to and in the notation of \cite[Sec. 4.4]{BS16b}, one immediately has relations $\widehat{\mathcal{F}}_0^{'-}=\widehat{\mathcal{F}}_1^{'+}=\widehat{\mathcal{F}}_2^{'-}=\widehat{\mathcal{F}}_3^{'+}=0$ (which are trivial in the $c=1$ case).
Altogether, these relations lead to relations on the partition functions with operators \eqref{Okamoto_operators_III3}, namely
\begin{equation}\label{blowup_odd}
\sum_{n\in\mathbb{Z}+\frac{\mathfrak{p}}2+\frac14}\mathfrak{D}^{k;\pi}_{\epsilon_1,\epsilon_2}\left(t^{\frac{\epsilon/8}{\epsilon_2{-}\epsilon_1}}\mathcal{Z}^{[0]}(a+2n\epsilon_1;2\epsilon_1,\epsilon_2{-}\epsilon_1|t),t^{\frac{\epsilon/8}{\epsilon_1{-}\epsilon_2}}\mathcal{Z}^{[0]}(a+2n\epsilon_2;2\epsilon_2,\epsilon_1{-}\epsilon_2|t)\right)=0,
\end{equation}
with $\mathfrak{p}=0,1$.
Note that, in the obtained blowup relations for $\mathfrak{p}=0,1$ the sum runs over $n\in\mathbb{Z}+\frac{\mathfrak{p}}2+\frac14$, while for the blowup relations \cite[(2.12)]{BST25} it runs over $n\in\mathbb{Z}+\frac{\mathfrak{p}}2$.
We can refer to the former blowup relations as to the $\mathbb{C}^2/\mathbb{Z}_2$ blowup relations in a nontrivial holonomy sector of the gauge theory.

\paragraph{Action of symmetry $\pi$ on the Zak transform.}
Analogously to \cite[Sec. 2.3]{BST25} bilinear relations \eqref{blowup_odd} can be rewritten in terms of the Zak transforms \eqref{Zak_tau1}, \eqref{Zak_tau2}, namely 
\begin{equation}\label{Okamoto_blowup_III3}
\mathfrak{D}^{k;\pi}_{\epsilon_1,\epsilon_2}\left(t^{\frac{\epsilon}{8\kappa}}\tau^{(1)}(a,\eta|t), t^{-\frac{\epsilon}{8\kappa}}\tau^{(2)}(a+\kappa/2,\eta|t)\right)=0,\qquad k=2,3.   
\end{equation}
These equations are just the above \eqref{Okamoto_III3} for the Zak transform solutions if $\pi(a,\eta)=(a+\kappa/2,\eta)$. For the classical case this action is due to \cite[Prop. 3.1]{BS16b}, which proof is based on the asymptotic analysis. Proceeding analogously in the noncommutative case, from \eqref{Hder_III3} under \eqref{eta_t}, \eqref{Ham_weak_ansatz}, \eqref{Hpm1_weak} we obtain
\begin{equation}\label{q_as_III3}
q=-\dot{H}^{-1}\sim e^{\ri\eta}\,\frac{\Gamma\big(1-2a/\kappa\big)\Gamma\big(1-(2a{+}\epsilon)/\kappa\big)}{\Gamma\big(2a{+}\epsilon)/\kappa\big)\Gamma\big(2a{+}2\epsilon)/\kappa\big)}\,t^{\frac{2a+\epsilon}{\kappa}}, \qquad t\rightarrow 0,
\end{equation}
under assumptions $\mathrm{Re}(\epsilon/\kappa)>0$ and $0<\mathrm{Re}\,\frac{a}{\kappa}<\frac12$ of Sec.~\ref{ssec:small}\footnote{Non-general lines $\mathrm{Re}\,\frac{a}{k}=0,\frac12$ are excluded.}. Assuming that the transformed solution $\pi(q)=t/q$ also belongs to the Zak transform family yields $\pi(a,\eta)=(\kappa/2-a,-\eta)$. This tuple is equivalent to $(a+\kappa/2,\eta)$ by symmetry $(a,\eta)\mapsto(-a,-\eta)$ 
of the Zak transforms \eqref{Zak_tau1}, \eqref{Zak_tau2}, mentioned in Sec.~\ref{ssec:small}. The crucial argument of the proof in the classical case was that any Zak transform solution is unambiguously recovered from its asymptotics and these solutions form a general family (\cite[Prop. 2.1]{BS16b}).
In the quantized version, we just assume that the Zak transform solution is general enough to be, as a rule, closed under the symmetry group. 

For the strong-coupling solutions we reverse the logic: obtain the $\mathbb{C}^2/\mathbb{Z}_2$ blowup equations from the quantum Okamoto-like equations and the asymptotic analysis. Namely, from  \eqref{Hder_III3} under \eqref{eta_s}, \eqref{Ham_strong_ansatz}, \eqref{H_leading} of ansatz \eqref{Zcl_anz}, \eqref{Z1loop_anz} with $\beta=-\frac1{256}$, $\delta=-\frac14$, $\chi=0$, $s=-32\ri t^{1/4}$ (due to \eqref{Zcl_0} and discussion there) we have that
\begin{equation}\label{QPIII3_strong_as}
-tq^{-1}=\frac{s^2}{2^{10}}\left(1-4\kappa^{1/2}e^{\ri\eta_{\D}}\Gamma\big(1{+}\kappa^{-1}(a_{\D}{+}\epsilon/2)\big)e^{-\frac{s}{4\kappa}}s^{-\frac12-\frac{2a_{\D}+\epsilon}{2\kappa}}+O\big(s^{-1}\big)\right), \qquad t/\kappa^4\rightarrow +\infty    
\end{equation}
under assumptions $\mathrm{Re}(\epsilon/\kappa)>0$ and $-\frac12<\mathrm{Re}\,\frac{a_{\D}}{\kappa}<0$. The latter region of $a_{\D}$ (including the boundary points) can be considered without loss of generality due to the strong-coupling analog of the shift symmetry \eqref{weak_shift} and branching $(a_{\D},s)\mapsto(-a_{\D},-s)$ of Sec.~\ref{ssec:aut_symm}. Assuming that the transformed solution $\pi(q)=t/q$ also belongs to the Zak transform family, we obtain that $\pi(a_{\D},e^{\ri\eta_{\D}})=(a_{\D},-e^{\ri\eta_{\D}})$
in agreement with \cite[Sec. 7.2]{GMS20} for the Toda-like equations (see below). Then, analogously to the weak-coupling case in the trivial holonomy sector, the quantum Okamoto-like equations \eqref{Okamoto_III3} for the Zak transform solutions in the strong-coupling regime are equivalent to the $\mathbb{C}^2/\mathbb{Z}_2$ blowup relations with operators \eqref{Okamoto_operators_III3} (c.f. weak-coupling case \eqref{blowup_odd})
\begin{equation}\label{blowup_odd_strong}
\sum_{n\in\mathbb{Z}+\frac{\mathfrak{p}}2}(-1)^n\,\mathfrak{D}^{k;\pi}_{\epsilon_1,\epsilon_2}\left(s^{\frac{\epsilon/2}{\epsilon_2{-}\epsilon_1}}\mathcal{Z}^{[\mathbf{0}]}(a_{\D}+2n\epsilon_1;2\epsilon_1,\epsilon_2{-}\epsilon_1|s),s^{\frac{\epsilon/2}{\epsilon_1{-}\epsilon_2}}\mathcal{Z}^{[\mathbf{0}]}(a_{\D}+2n\epsilon_2;2\epsilon_2,\epsilon_1{-}\epsilon_2|s)\right)=0, 
\end{equation}
with $\mathfrak{p}=0,1$.
We checked these relations, using the successive terms of the instanton expansion \cite[(4.69)]{BST25} up to order $s^{-8}$.

\paragraph{Toda-like equations.}
Besides the blowup relations \cite[(3.1),(3.6)]{BST25} of the tau form, the representation theory of the $\mathcal{N}=1$ super Virasoro algebra in the Neveu-Schwarz sector provide the second-order blowup relation \cite[(4.38)]{BS16b}, that in terms of the Zak transforms \eqref{Zak_tau1}, \eqref{Zak_tau2} reads 
\begin{equation}
D^2_{\epsilon_1,\epsilon_2}\big(\tau^{(1)}(a,\eta|t),\tau^{(2)}(a,\eta|t)\big)=\frac12t^{1/2}\tau^{(1)}(a+\epsilon_1,\eta|t)\tau^{(2)}(a+\epsilon_2,\eta|t).
\end{equation}
It can be obtained via the AGT dictionary \eqref{AGT_dct} analogously to the blowup equation \eqref{Okamoto_blowup_III3} using \cite[(4.31)]{BS16b} to reconstruct the 1-loop parts. 
Its right-hand side is equal to the product of $e^{-\ri\eta/2}\tau^{(1)}(a-\kappa/2)$ and $\tau^{(2)}(a+\kappa/2)e^{\ri\eta/2}$, which are claimed to be the $\pi$-transformed tau functions. Notice that these Toda-like equations (more precisely, their $q$-deformed version) gave the initial example of the Zak transform quantization in \cite[Sec. 4]{BGM17}. These equations was also considered in the strong-coupling regime in \cite[Sec. 7]{GMS20}, where there was also detected action $\pi(e^{\ri\eta_{\D}})=-e^{\ri\eta_{\D}}$. That gave a first example of the $\mathbb{C}^2/\mathbb{Z}_2$ blowup relations in the strong-coupling regime. In our notations these blowup relations \cite[(7.18)]{GMS20} read
\begin{multline}\label{blowup_Toda_strong}
\sum_{n\in\mathbb{Z}+\frac{\mathfrak{p}}2}D^2_{\epsilon_1,\epsilon_2}\left(\mathcal{Z}^{[\mathbf{0}]}(a_{\D}+2n\epsilon_1;2\epsilon_1,\epsilon_2{-}\epsilon_1|s),\mathcal{Z}^{[\mathbf{0}]}(a_{\D}+2n\epsilon_2;2\epsilon_2,\epsilon_1{-}\epsilon_2|s)\right)\\=(1/2-\mathfrak{p})\sum_{n\in\mathbb{Z}+\frac{\mathfrak{p}}2}\mathcal{Z}^{[\mathbf{0}]}(a_{\D}+2n\epsilon_1;2\epsilon_1,\epsilon_2{-}\epsilon_1|s)\mathcal{Z}^{[\mathbf{0}]}(a_{\D}+2n\epsilon_2;2\epsilon_2,\epsilon_1{-}\epsilon_2|s).
\end{multline}
These equations together with the $D^1$-equation of \eqref{D13_V} were used in \cite[Sec. 7.2]{GMS20} to get the strong-coupling expansion of $\mathcal{Z}^{[\mathbf{0}]}$, that gave the idea that we followed in \cite{BST25}. Thus the Toda-like equations, together with the $D^1$-equation can be also regarded as the tau form of the QPIII$_3$ equation. 

\subsection{Nonautonomous symmetries and $\mathbb{C}^2/\mathbb{Z}_2$ blowup relations}
\label{ssec:Okamoto-like}

Let us consider translations $T^{[N_f]}(m_f)= m_f+\kappa/2,\, f=1,\ldots N_f$ from the weight lattice symmetry subgroup $P$ of the corresponding equation for $N_f=4,3,2,1$. Following the QPIII$_3$ derivation, for these translations we derive the quantum Okamoto-like equations of the form (c.f. \eqref{Okamoto_III3} for the QPIII$_3$)
\begin{equation}\label{Okamoto_weak_eq}
\mathfrak{D}^{k;T^{[N_f]}}_{\epsilon_1,\epsilon_2}\left(\frac{t^{\frac{\epsilon}{8\kappa}}\tau^{(1)}}{f^{[N_f]}(\epsilon{-}\kappa,\kappa|t)},\frac{t^{-\frac{\epsilon}{8\kappa}}T^{[N_f]}(\tau^{(2)})}{f^{[N_f]}(-\kappa,\epsilon{+}\kappa|t)}\right)=0, \qquad k=2,3, 
\end{equation}
where the concrete expressions for the bilinear differential operator $\mathfrak{D}^{k}$ of order $k$ are presented below\footnote{We divided the tau functions by prefactors \eqref{prefactors_weak} to simplify formulas for the operators.}. For QPIV, QPII we derive the quantum Okamoto-like equations for translations $T_1^{[\mathrm{H_2}]}(\boldsymbol{m}_1,\boldsymbol{m}_2,\boldsymbol{m}_3)=\big(\boldsymbol{m}_1-\kappa/3,\boldsymbol{m}_2-\kappa/3,\boldsymbol{m}_3+2\kappa/3\big)$ and $T_1^{[\mathrm{H_1}]}(\boldsymbol{m})=\boldsymbol{m}+\kappa/2$, respectively.

\paragraph{Okamoto-like equations for QPVI.}
According to \eqref{PVI_masses}, we have that $T^{[4]}=T^{[4]}_3$ (recall definition \eqref{elem_transl_root}). As an element of $P/Q\ltimes W\left(D_4^{(1)}\right)$ this translation is given by $T_3^{[4]}=\pi_1s_3s_2s_4s_1s_2s_3$
(see Table~\ref{table:QPVI_Backlund} and definition of $\pi_1$ below it). It is a product of $\pi_1$ and the autonomous symmetries, so for us it was enough to find the Okamoto-like equations for $\pi_1$. Namely, the operators \eqref{Okamoto_weak_eq} are given by (c.f. \eqref{Okamoto_operators_III3})
\begin{equation}\label{Okamoto_op_VI_2}
\mathfrak{D}^{2;T_3^{[4]}}_{\epsilon_1,\epsilon_2}=(1{-}t)D^2_{\epsilon_1,\epsilon_2[\ln t]}+\frac{1{+}t}2\left(\epsilon_1\epsilon_2\frac{d}{d\ln t}\right)-\frac{\tilde{e}_1^{[4]}t}2 D^1_{\epsilon_1,\epsilon_2[\ln t]}-\frac{\epsilon_1\epsilon_2{+}\epsilon^2+4\tilde{e}_2^{[4]}t}{16}D^0_{\epsilon_1,\epsilon_2[\ln t]},
\end{equation}

\vspace{-0.7cm}

\begin{multline}\label{Okamoto_op_VI_3}
\mathfrak{D}^{3;T_3^{[4]}}_{\epsilon_1,\epsilon_2}=(1{-}t)^2D^3_{\epsilon_1,\epsilon_2[\ln t]}+\frac{1{-}t^2}2\left(\epsilon_1\epsilon_2\frac{d}{d\ln t}\right)D^1_{\epsilon_1,\epsilon_2[\ln t]}+\frac{\tilde{e}_1^{[4]}t(t{+}3)}4\left(\epsilon_1\epsilon_2\frac{d}{d\ln t}\right)D^0_{\epsilon_1,\epsilon_2[\ln t]}\\
-\frac{\epsilon_1\epsilon_2(1{-}17t)+\big(4\tilde{e}_2^{[4]} t{+}\epsilon^2\big)(1{-}t)+4\tilde{e}_1^{[4]}t(\tilde{e}_1^{[4]}t{+}2\epsilon)}{16}D^1_{\epsilon_1,\epsilon_2[\ln t]}
-t\frac{2\tilde{e}_3^{[4]}+\tilde{e}_2^{[4]}\tilde{e}_1^{[4]}t}8 D^0_{\epsilon_1,\epsilon_2[\ln t]},
\end{multline}
where for convenience of presentation we used $(\epsilon_1,\epsilon_2)$- Hirota derivatives with respect to time $\ln t$ (instead of QPVI time $\ln (1{-}1/t)$) and introduced mass parameter abbreviations
\begin{multline}
\tilde{e}_1^{[4]}=e_1^{[4]}+\kappa+3\epsilon, \qquad \tilde{e}_2^{[4]}=e_2^{[4]}+\frac14(3\kappa{+}7\epsilon)\big(e_1^{[4]}{+}\kappa\big)+\frac34(\epsilon_1\epsilon_2{+}3\epsilon^2), \\   \tilde{e}_3^{[4]}=e_3^{[4]}+\frac12(\kappa{+}3\epsilon)\Big(e_2^{[4]}+\frac34\big(e_1^{[4]}{+}\kappa{+}3\epsilon\big)\kappa-\frac14(\kappa{+}3\epsilon)^2\Big)+\frac18(\epsilon_1\epsilon_2{+}17\epsilon^2)\big(e_1^{[4]}{+}\kappa{+}3\epsilon\big).
\end{multline}

For the QPIII$_3$ case we used the asymptotics \eqref{q_as_III3} to claim that $\pi$ acts on the Zak transform parameters by shift $a\mapsto a+\kappa/2$. In general case it is much more tricky to follow such asymptotic analysis: in particular, for the QPVI case we should derive such an asymptotics from \eqref{q_VI}. However, the same shift is expected also for $4\geq N_f>0$ because these relations are of the representation-theoretic origin, where it appears naturally. Under this conjecture, the Okamoto-like equations \eqref{Okamoto_weak_eq} for the Zak transform solutions \eqref{Zak_tau1}, \eqref{Zak_tau2} are equivalent to the $\mathbb{C}^2/\mathbb{Z}_2$ blowup relations of the form (analogously to the equivalence between \eqref{Okamoto_III3} and \eqref{Okamoto_blowup_III3})
\begin{equation}\label{blowup_odd_weak}
\sum_{n\in\mathbb{Z}+\frac{\mathfrak{p}}2+\frac14}\mathfrak{D}^{k;T^{[N_f]}}_{\epsilon_1,\epsilon_2}\left(t^{\frac{\epsilon}{8\kappa}}\mathcal{Z}^{[N_f]}\big(a{+}2n\epsilon_1;(m_f)_{f=1}^{N_f};\epsilon{-}\kappa,\kappa|t\big),t^{-\frac{\epsilon}{8\kappa}}\mathcal{Z}^{[N_f]}\big(a{+}2n\epsilon_2;\big(m_f{+}\frac{\kappa}2\big)_{f=1}^{N_f};\epsilon{+}\kappa,-\kappa|t\big)\right)=0
\end{equation}
with $\mathfrak{p}=0,1$.
At the other hand, the Okamoto-like equations \eqref{Okamoto_weak_eq} are actually certain relations on the Hamiltonian and its  transformation,
so the fulfillment of the corresponding $\mathbb{C}^2/\mathbb{Z}_2$ blowup relations in the leading orders in $t$ can be considered as an asymptotic argument. Anyway, we made a computational check of the first $5$ orders of \eqref{blowup_odd_weak} with operators \eqref{Okamoto_op_VI_2}, \eqref{Okamoto_op_VI_3}. 

The action of translation $T_3^{[4]}$ on the Zak transform parameters yields the action of the whole weight lattice $P_{D_4}$ since the whole QPVI symmetry group is generated by $\pi_1$ and the extended finite Weyl group of the autonomous symmetries. Namely we have
\begin{equation}
T_i^{[4]}(a,\eta)=(a+\kappa/2,\eta), \quad i=1,3,4, \qquad T_2^{[4]}(a,\eta)=(a,\eta).    
\end{equation}
Thus we find the explicit action of the whole QPVI symmetry group on the Zak transform solution.

\paragraph{Okamoto-like equations for QPV, QPIII$_1$, QPIII$_2$.}
We can derive $\mathbb{C}^2/\mathbb{Z}_2$ blowup relations \eqref{blowup_odd_weak} in the weak-coupling regime for $N_f=3,2,1$ by the straightforward coalescence limits along the first row of the Diagram~\ref{fig:QPainleve_coalescence} in accordance with the results of \cite[Sec. 3.1, 3.3]{BST25} in the trivial holonomy sector:
\begin{itemize}
\item
Sending $m_4\rightarrow\infty$ under scaling $t\mapsto t/m_4$, we obtain from the $\mathbb{C}^2/\mathbb{Z}_2$ blowup relation \eqref{blowup_odd_weak} with operators \eqref{Okamoto_op_VI_2}, \eqref{Okamoto_op_VI_3} for $N_f=4$ those for $N_f=3$ with translation $T_3^{[3]}=\pi^{-1}s_1s_2s_3$ (see Table~\ref{table:QPV_Backlund}): 
\begin{equation}\label{Okamoto_op_V_2}
\mathfrak{D}^{2;T_3^{[3]}}_{\epsilon_1,\epsilon_2}=D^2_{\epsilon_1,\epsilon_2}+\frac12\left(\epsilon_1\epsilon_2\frac{d}{d\ln t}\right)-\frac{t}2 D^1_{\epsilon_1,\epsilon_2}-\frac{\epsilon_1\epsilon_2{+}\epsilon^2+\big(4e_1^{[3]}+(3\kappa{+}7\epsilon)\big)t}{16}D^0_{\epsilon_1,\epsilon_2},
\end{equation}

\vspace{-0.7cm}

\begin{multline}\label{Okamoto_op_V_3}
\mspace{-50mu}\mathfrak{D}^{3;T_3^{[3]}}_{\epsilon_1,\epsilon_2}=D^3_{\epsilon_1,\epsilon_2}+\frac12\left(\epsilon_1\epsilon_2\frac{d}{d\ln t}\right)D^1_{\epsilon_1,\epsilon_2}+\frac{3t}4\left(\epsilon_1\epsilon_2\frac{d}{d\ln t}\right)D^0_{\epsilon_1,\epsilon_2}
-\frac{\epsilon_1\epsilon_2{+}\epsilon^2+\big(4e_1^{[3]}+(3\kappa{+}15\epsilon)\big)t+4t^2}{16}D^1_{\epsilon_1,\epsilon_2}\\
-t\frac{8e_2^{[3]}+(\kappa{+}3\epsilon{+}t)\big(4e_1^{[3]}{+}3\kappa\big)+7\epsilon t+\epsilon_1\epsilon_2{+}17\epsilon^2}{32}D^0_{\epsilon_1,\epsilon_2}.
\end{multline}

\item Sending $m_3\rightarrow\infty$ under scaling $t\mapsto t/m_3$, we obtain from the $\mathbb{C}^2/\mathbb{Z}_2$ blowup relation \eqref{blowup_odd_weak} with operators \eqref{Okamoto_op_V_2}, \eqref{Okamoto_op_V_3} for $N_f=3$ those for $N_f=2$ with translation $T_1^{+[2]}=\pi^+s_1^+$ (see Table~\ref{table:QPIII1_Backlund}): 
\begin{align}\label{Okamoto_op_III1_2}
&\mathfrak{D}^{2;T_1^+}_{\epsilon_1,\epsilon_2}=D^2_{\epsilon_1,\epsilon_2}+\frac12\left(\epsilon_1\epsilon_2\frac{d}{d\ln t}\right)-\frac{\epsilon_1\epsilon_2{+}\epsilon^2+4t}{16}D^0_{\epsilon_1,\epsilon_2},\\ \label{Okamoto_op_III1_3}
&\mathfrak{D}^{3;T_1^+}_{\epsilon_1,\epsilon_2}=D^3_{\epsilon_1,\epsilon_2}+\frac12\left(\epsilon_1\epsilon_2\frac{d}{d\ln t}\right)D^1_{\epsilon_1,\epsilon_2}
-\frac{\epsilon_1\epsilon_2{+}\epsilon^2+4t}{16}D^1_{\epsilon_1,\epsilon_2}
-t\frac{2e_1^{[2]}{+}\kappa{+}3\epsilon}8 D^0_{\epsilon_1,\epsilon_2}.
\end{align}

\item
Sending $m_2\rightarrow\infty$ under scaling $t\mapsto t/m_2$, we obtain from the $\mathbb{C}^2/\mathbb{Z}_2$ blowup relation \eqref{blowup_odd_weak} with operators \eqref{Okamoto_op_III1_2}, \eqref{Okamoto_op_III1_3} for $N_f=2$ those for $N_f=1$ with translation $T_1^{[1]}=\pi s_1$ (see Table~\ref{table:QPIII2_Backlund}): 
\begin{align}\label{Okamoto_op_III2_2}
&\mathfrak{D}^{2;T_1^{[1]}}_{\epsilon_1,\epsilon_2}=D^2_{\epsilon_1,\epsilon_2}+\frac12\left(\epsilon_1\epsilon_2\frac{d}{d\ln t}\right)-\frac{\epsilon_1\epsilon_2{+}\epsilon^2}{16}D^0_{\epsilon_1,\epsilon_2},\\ \label{Okamoto_op_III2_3}
&\mathfrak{D}^{3;T_1^{[1]}}_{\epsilon_1,\epsilon_2}=D^3_{\epsilon_1,\epsilon_2}+\frac12\left(\epsilon_1\epsilon_2\frac{d}{d\ln t}\right)D^1_{\epsilon_1,\epsilon_2}
-\frac{\epsilon_1\epsilon_2{+}\epsilon^2}{16}D^1_{\epsilon_1,\epsilon_2}
-\frac{t}4D^0_{\epsilon_1,\epsilon_2}.
\end{align}

\item Finally, sending $m_1\rightarrow\infty$ under scaling $t\mapsto t/m_1$, we obtain from the $\mathbb{C}^2/\mathbb{Z}_2$ blowup relation \eqref{blowup_odd_weak} with operators \eqref{Okamoto_op_III2_2}, \eqref{Okamoto_op_III2_3} for $N_f=1$ those for $N_f=0$, i.e. the QPIII$_3$ operators \eqref{Okamoto_operators_III3}.
\end{itemize}
Naturally, direct derivation of the Okamoto-like equations gives the same operators as the coalescence. Note that the limiting procedure keeps the Zak transform parameters $(a,\eta)$ uninvolved until the final $N_f=0$. This further supports that  $T^{[N_f]}(a,\eta)=(a+\kappa/2,\eta)$. As for the QPVI case, this gives the action of the corresponding weight lattices on the weak-coupling Zak transform parameters (and therefore of the whole symmetry groups).

\paragraph{Okamoto-like equations and strong-coupling relations.}
The same way as before, we derive the QPIV Okamoto-like equations with translation $T_1^{[\mathrm{H_2}]}=\pi^{-1}s_2s_1$ (see Table~\ref{table:QPIV_Backlund})
\begin{equation}
D^2_{\epsilon_1,\epsilon_2}\left(\tau^{(1)},T_1^{[\mathrm{H_2}]}\big(\tau^{(2)}\big)\right)+\frac{t}6D^1_{\epsilon_1,\epsilon_2}\left(\tau^{(1)},T_1^{[\mathrm{H_2}]}\big(\tau^{(2)}\big)\right)-\frac16\left(\frac{t^2}3+\frac{3\boldsymbol{m}_3+\kappa+\epsilon}{2}\right)\tau^{(1)}T_1^{[\mathrm{H_2}]}\big(\tau^{(2)}\big)=0,
\end{equation}

\vspace{-0.7cm}

\begin{multline}
D^3_{\epsilon_1,\epsilon_2}\left(\tau^{(1)},T_1^{[\mathrm{H_2}]}\big(\tau^{(2)}\big)\right)+\frac12\epsilon_1\epsilon_2\frac{d}{dt}\left(\tau^{(1)}T_1^{[\mathrm{H_2}]}\big(\tau^{(2)}\big)\right)-\frac14\left(\frac{t^2}3-3\boldsymbol{m}_3-\kappa-\epsilon\right)D^1_{\epsilon_1,\epsilon_2}\left(\tau^{(1)},T_1^{[\mathrm{H_2}]}\big(\tau^{(2)}\big)\right)\\+\frac1{12}\left(\frac{t^2}9-\frac{3\boldsymbol{m}_3+\kappa+3\epsilon}2\right)t\,\tau^{(1)}T_1^{[\mathrm{H_2}]}\big(\tau^{(2)}\big)=0,
\end{multline}
and the QPII Okamoto-like equations with translation $T_1^{[\mathrm{H_1}]}=\pi s_1$ (see Table~\ref{table:QPII_Backlund})
\begin{align}
&D^2_{\epsilon_1,\epsilon_2}\left(\tau^{(1)},T_1^{[\mathrm{H_1}]}\big(\tau^{(2)}\big)\right)+\frac{t}8\tau^{(1)}T_1^{[\mathrm{H_1}]}\big(\tau^{(2)}\big)=0,\\
&D^3_{\epsilon_1,\epsilon_2}\left(\tau^{(1)},T_1^{[\mathrm{H_1}]}\big(\tau^{(2)}\big)\right)+\frac{t}8D^1_{\epsilon_1,\epsilon_2}\left(\tau^{(1)},T_1^{[\mathrm{H_1}]}\big(\tau^{(2)}\big)\right)+\frac{4\boldsymbol{m}+\kappa+3\epsilon}{16}\tau^{(1)}T_1^{[\mathrm{H_1}]}\big(\tau^{(2)}\big)=0.
\end{align}

We can consider these, as well as all the previous Okamoto-like equations for the Zak transform solutions in the strong coupling regime. Analogously to the weak-coupling case, this let us to conjecture for the chosen symmetry group actions on the Zak transform parameters $(a_{\D},\eta_{\D})$. Namely, we checked several (at least 5) orders of the $\mathbb{C}^2/\mathbb{Z}_2$ blowup equations that arise from all the previous blowup equations for the Zak transform solutions under the symmetry group action on $(a_{\D},\eta_{\D})$ of Table~\ref{tab:nonaut_Zak_strong}. As for the weak-coupling cases, this gives the action of the corresponding weight lattices on the strong-coupling Zak transform parameters (and therefore of the whole symmetry groups).

\begin{table}[ht]
    \centering
    \begin{tabular}{|c||c|c|c|c|c|}
      \hline Eq./gen. & \multicolumn{2}{|c|}{QPV / $\pi$} & QPIII$_1$ / $\pi^+$ & QPIII$_2$ / $\pi s_1$ & QPIII$_3$ / $\pi$ \\
      \hline
      Exp. sing. & linear & square &  \multicolumn{3}{|c|}{square} \\   
      \hline
      Action & $(a_{\D}{-}\frac{\kappa}4,e^{\ri\eta_{\D}})$ & $(a_{\D},\ri e^{\ri\eta_{\D}})$ & $(a_{\D},-e^{\ri\eta_{\D}})$ & $(a_{\D},-(2{+}\sqrt3)e^{\ri\eta_{\D}})$ & $(a_{\D},-e^{\ri\eta_{\D}})$ \\
      \hline
    \end{tabular}

   \vspace{0.7cm}
    
    \begin{tabular}{|c||c|c|c|c|}
      \hline Eq./gen. & \multicolumn{2}{|c|}{QPIV/ $\pi$} & \multicolumn{2}{|c|}{QPII / $\pi$} \\
      \hline
      Exp. sing. & linear & square & linear & square \\   
      \hline
      Action & $(a_{\D}{-}\frac{\kappa}3,e^{\ri\eta_{\D}})$ & $(a_{\D},e^{2\pi\ri/3} e^{\ri\eta_{\D}})$ & $(a_{\D}{-}\frac{\kappa}2,e^{\ri\eta_{\D}})$ & $(a_{\D},- e^{\ri\eta_{\D}})$\\
      \hline
    \end{tabular}
    \caption{Action of $P_X/Q_X$ on the Zak transform parameters}
    \label{tab:nonaut_Zak_strong}
\end{table}

Actions, presented in Table~\ref{tab:nonaut_Zak_strong} require several comments:
\begin{itemize}
\item Actions of the group $P_X/Q_X$ on the Zak transform parameters are effective, unlike the weak-coupling case. In the QPIII$_2$ case we presented action of translation $\pi s_1=T_1^{[1]}$ instead of $\pi$ for convenience; recall that $s_1$ inverts sign of $a_{\D}$ and acts on $e^{\ri\eta_{\D}}$ by \eqref{eta_D_periodic_factor} for $N_p=1$. Factor of infinite order (i.e. $2{+}\sqrt3$) under translations appears only in this special case.

\item For the classical parts $\mathcal{Z}_{cl}$ with the square exponent singularities we have shifts of $\eta_{\D}$, while for the classical parts with the linear exponent singularities --- shifts of $a_{\D}$. We have already seen the role of the square exponent singularity for the asympotics in the QPIII$_3$ case, namely in \eqref{QPIII3_strong_as}.

\item For the cases of the linear exponent singularities for QPV, QPIV we face $\mathbb{C}^2/\mathbb{Z}_2$ blowup relations in  nontrivial holonomy sectors of the gauge theory, other than for QPIII$_3$. Namely, they resemble structure of \eqref{blowup_odd}, \eqref{blowup_odd_weak} (without factors $t^{\pm \frac{\epsilon}{8\kappa}}$ for QPIV) but with the sum running over $n\in\mathbb{Z}+\frac{\mathfrak{p}}2+\frac18$ for QPV and $n\in\mathbb{Z}+\frac{\mathfrak{p}}2+\frac16$ for QPIV. The special structure of the 1-loop parts \eqref{Z1loop_3L}, \eqref{Z1loop_H2L} ensure the polynomiality (in masses, $a_{\D},\epsilon_1,\epsilon_2$) of the corresponding blowup factors. 

\end{itemize}

\section{Further directions}
\label{sec:further}
\begin{itemize}
    \item We derive the Hamiltonian forms and show their equivalence with the Heisenberg dynamics by a rather ad hoc and computational approach.
    At the same time, the Hamiltonian forms appear to reflect a rich underlying geometric structure, in particular through the remarkable factorizations in \eqref{Ham_V_twisted} and \eqref{Ham_IV_twisted}.
    This naturally leads to the question of a noncommutative generalization of the Okamoto--Sakai space of initial conditions approach \cite{S01}.

    \item Quantum Painlev\'e dynamics is expected to be rooted in the formulation of the quantum isomonodromic deformation theory. In this realm the analytic properties of the quantum tau functions in the variables describing the initial conditions of the quantum mechanical problem should become much clearer. Relatedly, it would be interesting to find isomonodromic definitions of the quantum tau functions, guided by the relation between quantum Painlev\'e equations and the KZ equations in \cite{JNS08,gaiur2023}.
    This relation has been developed in the gauge theory perspective in \cite{Nekrasov:2020qcq}.
    
    \item 
     The definition of the tau functions as solutions of the first order quantum bilinear equation  \eqref{tau_universal} is reminiscent of the Schr\"odinger equation, the tau functions playing the role of left and right evolution kernels.
    As the BPZ equation is indeed the quantum Painlev\'e equation in the Schr\"odinger representation, it stays as an open problem how to {\it explicitly} rebuild its bilocal evolution kernel out of the solution of the quantum bilinear equations.

    \item We would like to understand better the action of the nonautonomous symmetries. Namely, we lack a rigorous derivation of the action of the symmetries on the Zak transform parameters, especially in the strong-coupling regime. Related to this issue, we would like to characterize the zoo of the quantum bilinear equations on the tau functions.
 
    \item It would be interesting to further investigate the relation between our results and cluster integrable systems.
    Concretely, the noncommutative Zak transform approach originally appeared in \cite{BGM17} in the context of cluster tau functions for the $q$-difference quantum Painlev\'e III$_3$. 
    Applied to our approach, this should make systematic both the link between the quantum bilinear tau form and Toda-like equations of the cluster system and
    the symmetry properties of the Hamiltonian quantum Painlev\'e we formulate.
    Indeed, according to \cite{BGM17, GMS20}, such equations hold for the Zak transform representations of the tau function, but we do not yet know how to reproduce them within our approach.


    \item 
     It would be interesting to study the existence of an integer Hurwitz-like expansion for the quantum Painlev\'e tau functions by generalizing the results in \cite{Bonelli:2024wha,Bonelli:2025juz} (see also \cite{Fucito:2023txg,Francois:2023trm} for related aspects on surface operators).
     Also, it should be possible to derive bilinear relations for the tau function in elliptic form by studying the blowup equations of $\mathcal{N}=2^*$ gauge theory \cite{Bershtein:2021uts,Bonelli:2025bmt}. The related isomonodromic deformation problem is formulated on the torus with one puncture, whose tau function is described in gauge theoretical terms in \cite{Bonelli:2019boe}

     \item Isomonodromic deformation problems for flat connections of general Lie groups were studied in \cite{Guest2012IsomonodromyAO} in the context of $tt^*$-equations and in \cite{Bonelli:2021rrg,Bonelli:2022iob} in relation with supersymmetric gauge theories. It would be interesting to study their quantization along the lines of this paper. 
     
     \item Study more in general the relation between the quantum hamiltonian form of the dynamics and blowup equations in gauge theory, both for blowup of higher singularities $\widehat{\mathbb{C}^2/\mathbb{Z}_p}$ and of multiple successive blowups. These take a multi-linear form which we conjecture to be  geometrical pictures of Painlev\'e hierarchies (see, e.g., \cite{Mazzocco_2007}), the basic idea being to identify the volumes of the successive blown up $\mathbb{P}^1$'s with the multiple times of the hierarchy. For example, it would be interesting to concretely link to \cite{Mazzocco_2007,Bobrova_2021}
     for the Painlev\'e II hierarchy case.
    
\end{itemize}

\printbibliography[
heading=bibintoc,
title={References}
]

\end{document}